\documentclass{aa}  
\usepackage{graphicx,float,subfigure}
\usepackage{amsmath,amssymb}
\usepackage{txfonts}
\usepackage{natbib}

\newcommand{\ideal}[1]{\ensuremath{\overset{\infty}{#1}}}
\newcommand{\vect}[1]{\ensuremath{\boldsymbol{#1}}}
\newcommand{\dd}{\ensuremath{\mathrm{d}}}

\newcommand{\unit}[1]{\ensuremath{\,\mathrm{pc}}}
\begin{document}
   \title{A Fluid-Dynamical Subgrid Scale Model for\\ Highly Compressible Astrophysical Turbulence}
   
   \titlerunning{An SGS Model for Highly Compressible Turbulence}

   \author{W. Schmidt\inst{1}
          \and
          C. Federrath\inst{2,3,4}
   }
   

   \institute{Institut f\"ur Astrophysik, Universit\"at G\"ottingen, Friedrich-Hund-Platz 1,
	D-37077 G\"ottingen, Germany\\
              	\email{schmidt@astro.physik.uni-goettingen.de}
	\and
	Zentrum f\"ur Astronomie der Universit\"at Heidelberg, Institut f\"ur Theoretische Astrophysik, Albert-Ueberle-Str. 2, D--69120 Heidelberg, Germany\\
            	\email{chfeder@ita.uni-heidelberg.de}
	\and
	Max-Planck-Institut f\"ur Astronomie, K\"onigstuhl 17, D--69117 Heidelberg, Germany
	\and
	Ecole Normale Sup\'{e}rieure de Lyon, CRAL, 69364 Lyon Cedex 07, France
  }

   \date{}

 
  \abstract
   {Compressible turbulence influences the dynamics of the interstellar and the intergalactic medium over a vast range of length scales. In numerical simulations, phenomenological subgrid scale (SGS) models are used to describe particular physical processes below the grid scale. In most cases, these models do not cover fluid-dynamical interactions between resolved and unresolved scales, or the employed SGS model is not applicable to turbulence in the
highly compressible regime.}
   {We formulate and implement the Euler equations with SGS dynamics and provide numerical tests of an SGS turbulence energy model that predicts the turbulent pressure of unresolved velocity fluctuations and the rate of dissipation for highly compressible turbulence.}
   {We test closures for the turbulence energy cascade by filtering data from high-resolution simulations of forced isothermal and adiabatic turbulence. Optimal properties and an excellent correlation are found for a linear combination of the eddy-viscosity closure that is employed in LES of weakly compressible turbulence and a term that is non-linear in the Jacobian matrix of the velocity. Using this mixed closure, the SGS turbulence energy model is validated in LES of turbulence with stochastic forcing.}
   {It is found that the SGS model satisfies several important requirements: 1.\ The mean SGS turbulence energy follows a power law for varying grid scale. 2.\ The root mean square (RMS) Mach number of the unresolved velocity fluctuations is proportional to the RMS Mach number
of the resolved turbulence, independent of the forcing. 3.\ The rate of dissipation and the turbulence energy flux are constant. 
Moreover, we discuss difficulties with direct estimates of the turbulent pressure and the dissipation rate on the basis of resolved flow quantities that have recently been proposed. }
   {In combination with the energy injection by stellar feedback and other unresolved processes, the proposed SGS model is applicable to a variety of problems in computational astrophysics. Computing the SGS turbulence energy, the treatment of star formation and stellar feedback in galaxy simulations can be improved. Further, we expect that the turbulent pressure on the grid scale affects the stability of gas against gravitational collapse. The influence of small-scale turbulence on emission line broadening, e.~g., of \ion{O}{VI}, in the intergalactic medium is another potential application.}


   \maketitle
%

\section{Introduction}

The effects of numerically unresolved turbulence have recently met increasing attention in a variety of astrophysical simulations \citep[see, for instance,][]{ScannBruegg08,MaierIap09,Joung09,OppDav09}. Some approaches comprise subgrid scale (SGS) models, although these models are basically phenomenological parameterizations of astrophysical processes on length scales smaller than the grid scale. The full multi-scale dynamics of turbulence, however, is not embraced. The essence of an SGS model in the fluid-dynamical sense is that, at high Reynolds numbers, energy is transported through a turbulent cascade from larger, numerically resolved length scales to the subgrid scales. The energy of the unresolved turbulent velocity fluctuations is eventually dissipated into heat. Numerical simulations, in which an explicit closure for the turbulence cascade is applied on the grid scale, are called large eddy simulations (LES). A closure is an approximation to an SGS quantity in terms of resolved flow quantities.

If the unresolved turbulent velocity fluctuations reach a non-negligible fraction of the speed of sound, they give rise to a turbulent pressure in addition to the thermal pressure of the gas. This contribution to the pressure is proportional to the energy density of the SGS turbulence.
Turbulent pressure effects in the baryonic gas component of star-forming galaxies are discussed in
\citet{BurkGenz09}. In contemporary numerical simulations of disk galaxies \citep[e.~g.,][]{DobbsGlov08,TaskTan09,AgerLake09,AgerTey10} or in galactic-scale simulations of the interstellar medium \citep[e.~g.,][]{Joung06,Joung09}, the minimal grid scale (or the SPH smoothing length) $\Delta\gtrsim 1\,\mathrm{pc}$. Since this length scale is comparable to the size of molecular clouds, the unresolved turbulent velocity fluctuations can exceed the speed of sound in the cold gas phase. Consequently, it can be expected that a significant turbulent pressure is caused by turbulence below the grid scale. To compute the turbulent pressure, which has an impact on the star formation rate through the stability of the gas against gravitational collapse, a model for the highly compressible regime is indispensable. \citet{BonaHey87} and \citet{BonaPer92} formulate an analytic theory to calculate the turbulent pressure of isotropic compressible turbulence on the integral length scale. Applying an SGS model, on the other hand, the turbulent pressure can be computed for any length scale within the inertial subrange. \citet{Joung09} present an SGS model that is based on the equation for the kinetic energy of the unresolved turbulent velocity fluctuations, the so-called SGS turbulence energy, where energy is solely supplied by supernova feedback. Since the non-diagonal SGS turbulence stresses are neglected, the model of \citet{Joung09} reduces the effects of SGS turbulence to the turbulent pressure alone, and the turbulence energy cascade, i.~e., the production of SGS turbulence by the shear of the numerically resolved flow, is not considered. 

A further example for this type of SGS models is the model of \citet{ScannBruegg08} for the simulation of Rayleigh-Taylor-driven turbulence in active galactic nuclei, where it is assumed that SGS turbulence is produced by buoyancy processes on unresolved length scales only. These processes are modelled by an equation for the characteristic length scale of the Rayleigh-Taylor instability. In contrast, \citet{SchmNie06c} incorporate unresolved buoyancy effects into an SGS model that includes the production by shear for the treatment of turbulent combustion in thermonuclear supernovae.

In the cosmological simulations of galaxy clusters by \citet{MaierIap09}, the role of SGS turbulence has been explored with a numerical technique that combines adaptive mesh refinement and LES. They apply the SGS turbulence energy model of \citet{SchmNie06b}. The main effect of the SGS model is an enhancement of the turbulent heating in the cluster core. The SGS turbulence energy also serves as a tracer of turbulence production in the intergalactic medium \citep{IapSchm10}. However, a deficiency in these simulations is that the employed SGS model is only applicable to moderately compressible turbulence. Shocks are treated tentatively, i.~e., SGS turbulence production is suppressed in the vicinity of shock fronts. While this is not a severe constraint for the bulk of the intracluster medium, in which the Mach numbers of the turbulent flow are small compared to unity, an erroneous production of SGS turbulence energy is likely to occur near accretion shocks in the outer regions of the cluster. 

In this article, we improve on the previous approaches to SGS modelling by addressing the closure problem for highly compressible turbulence. In Sect.~\ref{sc:euler_sgs}, we discuss the meaning of the compressible Euler equations in the context of computational fluid dynamics. The verification of the proposed closure and the calibration of the closure coefficients are presented in Sect.~\ref{sc:test}. Data from several high-resolution simulations of forced turbulence \citep{SchmNie07,SchmFeder09a,FederRom10} allow us to compute the rate at which energy is transferred from length scales greater than the filter length to smaller length scales and to test the correlation with different closures. As a result, we propose a combination of the eddy-viscosity closure, which has successfully been used in LES of incompressible turbulence, and a non-linear closure that is put forward by \citet{WoodPort06}. Then we show that physically reasonable statistics of the SGS turbulence energy and the rate of dissipation are obtained for varying grid resolutions and forcing in LES of supersonic turbulence (Sect.~\ref{sc:les2}). Furthermore, we investigate correlations of the SGS quantities with quantities derived from the numerically resolved flow. We demonstrate that the turbulent pressure and energy dissipation cannot be predicted in a straight-forward way on the basis of the resolved turbulent flow, as proposed, for instance, by \citet{PanPad09a} and \citet{ZhuFeng10}. Instead, a full SGS model is needed to estimate unresolved turbulence effects. In the last Section, we summarize the results and discuss potential astrophysical applications of the closure for the highly compressible turbulence cascade in combination with the phenomenological
approaches described above.


\section{The compressible Euler equations with subgrid-scale dynamics}
\label{sc:euler_sgs}

The Reynolds number of turbulent flows in astrophysics is usually considered to be high enough so that the approximation of an inviscid fluid can be applied on numerically resolvable length scales. For a physically complete picture, we begin with the compressible Navier-Stokes equations encompassing all physical length scales. This acknowledges the fact that perfect fluids do not exist in nature and that the notion of viscous dissipation is essential for turbulence. The fluid-dynamical variables determined by this set of equations are denoted by $\ideal{\varrho}$ for the mass density of the gas, $\ideal{\vec{u}}$ for the velocity of the flow, etc.\footnote{A mathematical proof for the existence of a solution is still on the agenda of the Clay Mathematics Institute Millennium Prize Problems, even in the case of incompressible turbulence.}
The resolution of a numerical simulation is given by the size of the grid cells $\Delta$,
which is called the cutoff scale or the grid scale. A consistent formulation of the equations of fluid dynamics with a cutoff scale $\Delta$ can by derived from the Navier-Stokes equations by means of the filter formalism introduced by \citet{Germano92}. Generalizing this formalism to compressible fluid dynamics is straightforward \citep[see][]{SchmNie06b}. The basic idea is to identify the numerically computed solution with filtered variables $\varrho:=\langle\ideal{\varrho}\rangle_{\Delta}$, $\vec{u}:=\langle\ideal{\varrho}\ideal{\vec{u}}\rangle_{\Delta}/\varrho$, etc. The filter operator $\langle\cdot\rangle_{\Delta}$ smoothes the physical variables that are given by the Navier-Stokes equations over the length scale $\Delta$. In LES, the filtering corresponds to the discretization of the equations of fluid dynamics. The dynamical equations for the computable, filtered quantities are similar to the unfiltered equations, with additional terms that are related to the subgrid-scale dynamics on length scales $\ell<\Delta$. 

Let us consider the dynamical equation for the momentum density of the fluid, which is given by the partial differential equation (PDE)
\begin{equation}
  \label{eq:momt}
  \frac{\partial}{\partial t}\left(\ideal{\varrho}\ideal{\vec{u}}\right) + 
  \vec{\nabla}\cdot\left(\ideal{\varrho}\ideal{\vec{u}}\otimes\ideal{\vec{u}}\right) =
  \ideal{\rho}\left(\ideal{\vect{g}} + \ideal{\vect{f}}\right) -\vec{\nabla}\ideal{P} + 
  \vec{\nabla}\cdot\ideal{\boldsymbol{\sigma}},
\end{equation}
where $\ideal{g}$ and $\ideal{f}$ are the accelerations due to gravity and other mechanical forces acting on the fluid, and $\ideal{P}$ is the thermal pressure. The viscous dissipation tensor $\ideal{\tens{\sigma}}$ is defined by 
\begin{equation}
        \label{eq:visc_diss_tens2}
        \ideal{\sigma}_{ij} =
        2\nu\ideal{\varrho}\left(\ideal{S}_{\! ij} - \frac{1}{3}\ideal{d}\delta_{ij}\right),
\end{equation}
where $\nu$ is the microscopic viscosity of the fluid\footnote{
Although we consider compressible fluid dynamics, for brevity, we neglect the second viscosity that is
related to the divergence of the flow. This does not affect subsequent arguments.
}, the rate of strain $\ideal{S}_{\! ij}$ is the symmetic part of the Jacobian matrix $\ideal{u}_{i,j}=\partial_j\ideal{u}_i$, and $\ideal{d}=u_{i,i}$. Applying a homogeneous filter operator that is uniform in time, Eq.~(\ref{eq:momt}) is converted into an equation for the filtered momentum, $\varrho\vec{u}=\langle\ideal{\varrho}\ideal{\vec{u}}\rangle$. This equation has the same form as the original equation, except for one term. Because of the non-linear advection term, the filtering introduces a stress term that accounts for the interaction between the numerically resolved flow and velocity fluctuations on the subgrid scales:
\begin{equation}
  \label{eq:momt_flt}
  \frac{\partial}{\partial t}(\varrho\vec{u}) + \vec{\nabla}\cdot(\varrho\vec{u}\otimes\vec{u}) =
  \rho(\vect{g} + \vect{f}) -\vec{\nabla}P + 
  \vec{\nabla}\cdot\left(\boldsymbol{\sigma} + \boldsymbol{\tau}_{\mathrm{sgs}}\right),
\end{equation}
where the SGS turbulence stress tensor is defined by
\begin{equation}
        \label{eq:tau_sgs_def}
        \boldsymbol{\tau}_{\mathrm{sgs}} =
        -\langle\ideal{\varrho}\ideal{\vec{u}}\otimes\ideal{\vec{u}}\rangle_{\Delta} +
        \varrho\vec{u}\otimes\vec{u}.
\end{equation}
In the following, the components of $\boldsymbol{\tau}_{\mathrm{sgs}}$ are simply denoted by $\tau_{ij}$.
The second-order moment $\langle\ideal{\varrho}\ideal{\vec{u}}\otimes\ideal{\vec{u}}\rangle_{\Delta}$ is not explicitly computable in LES because
the variations of the mass density $\ideal{\varrho}$ and the velocity $\ideal{\vec{u}}$ below the grid scale are unknown. For this reason, an approximation
in terms of filtered quantities has to be devised. This is the closure problem.\footnote{Althernatively, $\langle\ideal{\varrho}\ideal{\vec{u}}\otimes\ideal{\vec{u}}\rangle_{\Delta}$ can be expressed in terms of higher-order moments. But this merely shifts the closure problem to the higher-order moments.}

The SGS turbulence energy density is defined by the difference between the resolved 
kinetic energy and the filtered kinetic energy:
\begin{equation}
	\label{eq:trace_tau}
	K_{\mathrm{sgs}}:=
		\frac{1}{2}\langle\ideal{\varrho}\ideal{\vec{u}}\cdot\ideal{\vec{u}}\rangle_{\Delta} - \frac{1}{2}\varrho|\vec{u}|^2
		=-\frac{1}{2}\mathrm{tr}\,\boldsymbol{\tau}_{\mathrm{sgs}},
\end{equation}
where $\mathrm{tr}\,\boldsymbol{\tau}_{\mathrm{sgs}}=\tau_{ii}$ is the trace of the SGS turbulence stress tensor.
One can see that the trace of $\boldsymbol{\tau}_{\mathrm{sgs}}$ gives rise to the term $-\frac{2}{3}\vec{\nabla}K_{\mathrm{sgs}}$ on the right-hand side of Eq.~(\ref{eq:momt_flt}). This term
can be absorbed into the pressure gradient if the thermal pressure $P$ is replaced by the effective pressure
\begin{equation}
	\label{eq:press_eff}
	P_{\mathrm{eff}} = P + \frac{2}{3}K_{\mathrm{sgs}} =
	P - \frac{1}{3}\mathrm{tr}\,\boldsymbol{\tau}_{\mathrm{sgs}}.
\end{equation}
The relative contribution of the turbulent pressure $P_{\mathrm{sgs}}=\frac{2}{3}K_{\mathrm{sgs}}$ compared to the thermal pressure $P$ is characterized by the SGS turbulence Mach number $\mathcal{M}_{\rm sgs}=(2K_{\rm sgs}/\rho c_{\rm s}^2)^{1/2}$, where $c_{\rm s}$
is the thermal speed of sound. $\mathcal{M}_{\rm sgs}$ depends on the temperature of the fluid and the cutoff scale $\Delta$. The dependence on $\Delta$ is investigated in Section~\ref{sc:cutoff}.
\citet{Joung09} define the turbulent pressure by $P_{\mathrm{sgs}}=(\gamma-1)K_{\rm sgs}$,
where $\gamma$ is the adiabatic coefficient of the gas. We emphasize that, except for $\gamma=5/3$, this definition is inconsistent with the decomposition of the fluid-dynamical equations, which fixes the coefficient to be $2/3$ \citep[see also][]{Chandra51}. This is reasonable because the turbulent pressure is solely a property of the turbulent flow of a gas at a certain length scale, whereas $\gamma$ is a microscopic property of the gas that is related to thermal motions of the atoms or molecules.

The SGS turbulence energy is an intermediate reservoir of energy that exchanges energy with the resolved flow and loses energy by dissipation into heat. For the computation of $K_{\mathrm{sgs}}$, a PDE has to be solved in addition to the filtered equations for the resolved gas dynamics:
\begin{equation}
  \label{eq:energy_sgs2}
  \frac{\partial}{\partial t}K_{\mathrm{sgs}} + \vec{\nabla}\cdot(\vec{u}
  K_{\mathrm{sgs}}) = \Gamma + \Sigma - \rho(\epsilon + \lambda) + \mathfrak{D}.
\end{equation}
While $\Sigma=\tau_{ij}S_{ij}$ is the rate of SGS turbulence energy production by the turbulent cascade through the cutoff scale $\Delta$ (also called the turbulence energy flux) and $\rho\epsilon$ is the viscous dissipation rate smoothed over $\Delta$, effects caused by SGS fluctuations of the gravitational potential and the thermal pressure are given by $\Gamma$ and $\rho\lambda$, respectively. The term $\mathfrak{D}$ accounts for SGS transport effects. We refer to \citet{SchmNie06b}, Eqs.\ (33)--(37), for the exact definitions of these terms. For our purpose it is sufficient to discuss the closures of these terms, which are approximations in terms of the numerically resolved variables and $K_{\mathrm{sgs}}$.

To compute the SGS turbulence stress tensor~(\ref{eq:tau_sgs_def}), we propose the following closure for the highly compressible regime:
\begin{equation}
 \label{eq:tau_nonlin}
  \tau_{ij}= 2C_{1}\Delta(2\varrho K_{\mathrm{sgs}})^{1/2}S_{\! ij}^{\ast}
  -2C_{2}K_{\mathrm{sgs}}\frac{2u_{i,k}u_{j,k}}{|\vec{\nabla}\otimes\vec{u}|^{2}}
  -\frac{2}{3}(1-C_{2})K_{\mathrm{sgs}}\delta_{ij}.
\end{equation}
where $|\vec{\nabla}\otimes\vec{u}|:=(2u_{i,k}u_{i,k})^{1/2}$ is the norm of the resolved velocity derivative, $S_{\! ij}^{\ast}=S_{ij}-\frac{1}{3}d\delta_{ij}$ is the trace-free part of $S_{ij}=\frac{1}{2}(u_{i,j}+u_{j,i})$, and $d=u_{i,i}$. While the first term in Eq.~(\ref{eq:tau_nonlin}) corresponds to the eddy-viscosity closure that is commonly used in incompressible LES, the second, non-linear term was investigated by \citet{WoodPort06} for transonic decaying turbulence. The standard eddy-viscosity closure follows if $C_{2}=0$. In general, the linear eddy-viscosity term dominates if $(K_{\mathrm{sgs}}/\rho)^{1/2}$ is small compared to $\Delta|S^{\ast}|\lesssim\Delta|\nabla\otimes\vect{u}|$. On the other hand, for strong turbulence intensity, i.~e., $(K_{\mathrm{sgs}}/\rho)^{1/2}\gtrsim \Delta|\vect{\nabla}\otimes\vect{u}|$, the non-linear term contributes significantly. This particularly applies to intermittent events in supersonic turbulence, for which $\Delta|\vect{\nabla}\otimes\vect{u}|\gtrsim c_{\mathrm{s}}$. In moderately compressible turbulence, non-linear contributions affect the high-intermittency tails of the turbulent energy distribution. Independent of the values of $C_{1}$ and $C_{2}$, $\tau_{ii}=-2K_{\mathrm{sgs}}$, as required by the identity~(\ref{eq:trace_tau}). We denote the trace-free part of the SGS turbulence stress tensor by $\tau_{ij}^{\ast}$. The verification of the generalized closure~(\ref{eq:tau_nonlin}) for $\tau_{ij}$ and the determination of the coefficients $C_{1}$ and $C_{2}$ for supersonic turbulence is the key to the computation of the turbulent pressure $P_{\mathrm{sgs}}=\frac{2}{3}K_{\mathrm{sgs}}$, as $K_{\mathrm{sgs}}$ first and foremost depends on the production rate $\Sigma=\tau_{ij}S_{ij}$ in Eq.~(\ref{eq:energy_sgs2}).

Due to the microscopic viscosity $\nu$ of the fluid, the viscous stresses $ \ideal{\sigma}_{ij}$ dissipate kinetic energy on the smallest dynamical length scales $\ell\sim\mathrm{\eta}$ of the physical flow $\ideal{\vect{u}}$. The length scale $\eta$ is called the Kolmogorov scale. In the filtered momentum equation~(\ref{eq:momt_flt}), viscous dissipation effects are given by the divergence of the filtered tensor $\sigma_{ij}=\langle\ideal{\sigma}_{ij}\rangle_{\mathrm{\Delta}}$. The corresponding rate of energy dissipation, filtered on the grid scale, is given by
\begin{equation}
  \label{eq:diss_grid}
  \varrho\epsilon = \langle\ideal{\sigma}_{ij}\ideal{u}_{i,j}\rangle_{\Delta} =
  \langle2\nu\ideal{\varrho}\,\ideal{S}{_{\! ij}^{\,\ast}}\ideal{S}{_{\! ij}^{\,\ast}}\rangle_{\Delta} =
 \langle\nu\ideal{\varrho}|\ideal{S}{_{\! ij}^{\,\ast}}|^{2}\rangle_{\Delta}.
\end{equation}
It is important to note that $\varrho\epsilon\neq\sigma_{ij}u_{i,j}$, where $\sigma_{ij}$ and $u_{i,j}$ are the filtered viscous stress tensor and the filtered velocity gradient, respectively. 

For fully developed incompressible turbulence, the Kolmogorov scale can be related to the Reynolds number: $\eta/L\sim\mathrm{Re}^{3/4}$, where  $\mathrm{Re}:=VL/\nu$ for an integral length $L$ and characteristic velocity $V$ of the flow. As pointed out at the beginning of this section, $\mathrm{Re}$ is assumed to be very high in astrophysical systems. In this case, $\eta$ is much smaller than any feasible grid resolution $\Delta$, and simple scaling arguments show that the viscous stress term in the filtered momentum equation~(\ref{eq:momt_flt}) is negligible \citep{RoepSchm09}, i.~e., $|\sigma|\ll|\tau_{\mathrm{sgs}}|$. Consequently, the \emph{physical} energy dissipation occurs entirely on subgrid scales $\ell\ll\Delta$. As $\eta$ decreases in comparison to $\Delta$, the
velocity fluctuations on ever smaller length scales give rise to arbitrarily steep velocity gradients,
which add up to a non-vanishing product of the viscosity times the squared rate of strain on the
right-hand side of Eq.~(\ref{eq:diss_grid}), regardless of how small the viscosity is. This results in a non-zero, asymptotically constant mean rate of energy dissipation in the limit $\eta\rightarrow 0$ ($\nu\rightarrow 0$), which is supported by experimental and numerical evidences \citep[see][]{Frisch,IshiGo10}. We may reasonably conjecture that the viscous dissipation tensor is negligible in the filtered momentum equation and the energy dissipation rate does not vanish in the limit of infinite Reynolds numbers also in the case of compressible turbulence. \emph{A posteriori} tests imply that this conjecture is fulfilled for driven supersonic turbulence (see Sect.~\ref{sc:cutoff}). However, the question of energy dissipation in inhomogeneous turbulence is more difficult. For example, it is known from boundary layers of terrestrial turbulence that
viscous effects can affect the flow at relatively large scales near a wall. For this reason, the microscopic
viscosity cannot be neglected in LES of such flows. Although solid walls are not encountered in astrophysics, many relevant problems exhibit pronounced inhomogeneities, and we cannot entirely
exclude the possibility that viscous effects might become noticeable on resolved length scales in certain cases.

A closure for $\epsilon$ follows from simple dimensional reasoning:
\begin{equation}
	\label{eq:diss_sgs}
	\rho\epsilon = C_{\epsilon}\frac{K_{\mathrm{sgs}}^{3/2}}{\rho^{1/2}\Delta}.
\end{equation}
Here, it is assumed that the SGS turbulence energy is dissipated into heat at a rate proportional
to $K_{\mathrm{sgs}}$ divided by the time scale $\Delta(K_{\mathrm{sgs}}/\rho)^{-1/2}$.
For the pressure-dilatation term $\rho\lambda$ several closures have been proposed \citep[e.~g.,][]{Sarkar1992,FurTab97}. However, applying a priori tests (see Sect.~\ref{sc:test}), we find that these closures clearly fail in the case of supersonic turbulence. The simplest solution is to neglect pressure dilatation entirely \citep{WoodPort06}. 
In this article, we also set $\rho\lambda=0$, although we are aware that pressure-dilatation effects have potential significance, particularly, in the case of adiabatic turbulence. The transport term in Eq.~(\ref{eq:energy_sgs2}) can be modelled by a gradient-diffusion approximation \citep[see][]{Sagaut}:
\begin{equation}
	\mathfrak{D} = 
	\vect{\nabla}\cdot\left[\kappa_{\mathrm{sgs}}
	\vect{\nabla}\left(\frac{K_{\mathrm{sgs}}}{\rho}\right)\right],
\end{equation}
where the SGS turbulent diffusivity is approximated by $\kappa_{\mathrm{sgs}}\approx 0.65\Delta (\rho K_{\mathrm{sgs}})^{1/2}$, as shown by \citet{SchmNie06b}.

In this work, we assume that self-gravity has no significant effects on length scales $\ell\lesssim\Delta$. This corresponds to the condition that the local Jeans length $\lambda=c_{\mathrm{s}}(\pi/G\rho)^{1/2}$, where $G$ is the gravitational constant, is sufficiently large compared to the grid scale $\Delta$ \citep{TrueloveEtAl1997,FederrathBanerjeeClarkKlessen2010}. Thus, setting $\Gamma = 0$, the filtered equations resulting from the compressible Navier-Stokes equations in the limit of $\eta\ll\Delta$ \citep{SchmNie06b} read
\begin{align}
 \label{eq:euler_sgs_rho}
  \frac{\partial}{\partial t}\varrho + \vec{\nabla}\cdot(\vec{u}\varrho)\, &= 0, \\
 \label{eq:euler_sgs_momt}
  \frac{\partial}{\partial t}(\varrho\vec{u}) + 
  \vec{\nabla}\cdot\left(\varrho\vec{u}\otimes\vec{u}\right)\, &=
  \rho(\vect{g} + \vect{f}) -\vec{\nabla}(P + P_{\mathrm{sgs}}) +
  \vec{\nabla}\cdot\boldsymbol{\tau}_{\mathrm{sgs}}^{\ast}, \\
  \begin{split}
 \label{eq:euler_sgs_energy}
   \frac{\partial}{\partial t} E + \vec{\nabla}\cdot(\vec{u}E)\, &=
  -\mathcal{L} + \rho\vect{u}\cdot(\vect{g} + \vect{f})
  -\vec{\nabla}\cdot[\vec{u}(P + P_{\mathrm{sgs}})] \\
 &+ \vec{\nabla}\cdot(\vec{u}\cdot\boldsymbol{\tau}_{\mathrm{sgs}}^{\ast}) - \Sigma + \rho\epsilon,
  \end{split}
\end{align}
where $E=\frac{1}{2}\rho u^2 + E_{\mathrm{int}}$ is the sum of the resolved kinetic and internal
energy density, and $-\mathcal{L}$ accounts for sources and sinks of the internal energy due to heating and cooling, respectively. Since the resolved fluid dynamics on length scales $\ell\ge\Delta$ is unaffected by the viscosity of the fluid, the above set of equations defines the compressible \emph{Euler} equations for computational fluid dynamics in a physically meaningful and consistent way. These equations are supplemented by an equation of state, the SGS turbulence energy equation~(\ref{eq:energy_sgs2}), and the Poisson equation for the gravitational potential. The pure compressible Euler equations without SGS terms, on the other hand, do \emph{not} follow from the compressible Navier-Stokes equation in the limit of infinite Reynolds number.
In this case, there is no viscous dissipation at all, and, by definition, $\epsilon$ vanishes identically. 
This is a mathematical idealization that does not describe turbulent flows in nature.

As a special case, \emph{implicit} large eddy simulations (ILES) follow from the above approach.
ILES is the most commonly used method in astrophysical fluid dynamics. It is based on two assumptions, which are usually not stated in the literature. Firstly, the discretization of the compressible Euler equations introduces a dissipative leading error term $\mathfrak{D}_{\mathrm{num}}$ in the momentum equation~(\ref{eq:euler_sgs_rho}).
Implicitly, this term is assumed to be equivalent to the SGS turbulence stress term $\vec{\nabla}\cdot\boldsymbol{\tau}_{\mathrm{sgs}}$. The second assumption in ILES is that $\vect{u}\cdot\mathfrak{D}_{\mathrm{num}}=-\rho\epsilon$, i.~e., kinetic energy on the resolved scales is directly
dissipated into heat at a rate that approximates the viscous dissipation on unresolved length scales. This is referred to as numerical viscosity or numerical dissipation. Effectively,
the following equations are solved in ILES:
\begin{align}
 \label{eq:euler_iles_rho}
  \frac{\partial}{\partial t}\varrho + \vec{\nabla}\cdot(\vec{u}\varrho)\, &= 0, \\
 \label{eq:euler_iles_momt}
  \frac{\partial}{\partial t}(\varrho\vec{u}) + 
  \vec{\nabla}\cdot\left(\varrho\vec{u}\otimes\vec{u}\right)\, &=
  \rho(\vect{g} + \vect{f}) -\vec{\nabla}P +
  \mathfrak{D}_{\mathrm{num}}, \\
 \label{eq:euler_iles_energy}
   \frac{\partial}{\partial t} E + \vec{\nabla}\cdot(\vec{u}E)\, &=
  -\mathcal{L} + \rho\vect{u}\cdot(\vect{g} + \vect{f})
  -\vec{\nabla}\cdot(\vec{u}P).
\end{align}
Despite the lack of a mathematical justification, ILES serves as an approximation to turbulent compressible fluid dynamics that has proven its utility in numerous astrophysical applications. \citet{BenzBif08} demonstrate that the ILES approach closely reproduces two-point statistics of weakly compressible turbulence in the inertial subrange in comparison to direct numerical simulations that solve the Navier-Stokes equations. In this article, we make use of ILES to compute high-resolution data for the explicit verification of SGS closures. In contrast, LES treat the energy dissipation explicitly. However, it cannot be avoided that numerical schemes for compressible fluid dynamics such as the piecewise parabolic method (PPM, \citealt{ColWood84}) introduce some numerical dissipation. Thus, running an LES with an explicit SGS model, there is inevitably a numerical dissipation channel that competes with the transfer of energy to the subgrid scales and the subsequent dissipation of SGS turbulence energy into heat. Notwithstanding this caveat, we demonstrate in this article that physically sensible predictions can by made by an explicit SGS model, which are not possible on the basis of ILES.


\section{Closure verification}
\label{sc:test}

To test different closures for the turbulence energy flux $\Sigma$, we apply the method described in \citet{SchmNie06b}. The basic idea is to use data from ILES of non-selfgravitating isothermal and adiabatic turbulence with high numerical resolution and to apply an explicit filter to these data on a length scale that is in between the forcing and the dissipative range. The applied filters are Gaussian with filter lengths $\Delta_{\rm{G}}=24\Delta$ and $32\Delta$ for $768^{3}$ and $1024^{3}$ grids, respectively. Although the bottleneck effect has some influence on the chosen length scales \citep{SchmFeder09a,FederRom10}, we show that the results change only slightly if the filter length decreases or increases by a factor of two. Moreover, comparing to box filters, the results turn out to be rather insensitive to the filter type.

\subsection{Single-coefficient closures for supersonic isothermal turbulence}
\label{sc:test_single}

The turbulence energy flux on the filter scale $\Delta_{\rm{G}}$ can be computed explicitly from the unfiltered numerical data by the formula
\begin{equation}
\label{eq:flux_flt}
	\Sigma_{\Delta_{\mathrm{G}}} =
	\left(-\overline{\varrho u_i u_j} +
	\frac{\overline{\varrho u_i}\,\overline{\varrho u_i}}{\overline{\varrho}}
	 \right) \overline{u}_{i,j}\, ,
\end{equation}
where the first factor on the right hand side is the turbulence stress tensor on the filter scale (defined analogous to Eq.~\ref{eq:tau_sgs_def}) and the second factor is the derivative of the filtered velocity.  As a shorthand notation, we denote the explicitly filtered quantities by an overline, for instance, $\overline{\rho}=\langle\varrho\rangle_{\Delta_{\rm G}}$, and
$\overline{u}_i=\langle\varrho u_i \rangle_{\Delta_{\rm G}}/\overline{\rho}$.

The above expression for $\Sigma_{\Delta_{\mathrm{G}}}$ can be compared to
closures. For the eddy-viscosity closure, $\Sigma_{\Delta_{\mathrm{G}}}$ is given by
\begin{equation}
	\label{eq:flux_eddy_visc}
	\Sigma_{\Delta_{\mathrm{G}}}^{(\rm cls)} =
	C_{1}\Delta_{\rm G}(2\overline{\rho}K_{\Delta_{\rm G}})^{1/2}
	|\overline{S}^{\,\ast}|^{2}-\frac{2}{3}K_{\Delta_{\rm G}}\overline{d},
\end{equation}
where 
\begin{equation}
\label{eq:turb_energy_flt}
	K_{\Delta_{\rm G}} = \frac{1}{2}
	\left(\overline{\varrho u^{2}} -
	\frac{\overline{\varrho u}^{2}}{\overline{\varrho}}\right)
\end{equation}
is the turbulence energy on length scales smaller than $\Delta_{\rm G}$.
Strictly speaking, $K_{\Delta_{\rm G}}$ is the turbulence energy for the length scales
ranging from the grid resolution $\Delta$ to the smoothing length $\Delta_{\rm G}$
of the Gaussian filter. 
If $\Delta_{\rm G}$ is sufficiently large compared to $\Delta$, this distinction can be neglected \citep[see][]{SchmNie06b}.

Defining
\begin{equation}
	C_{1}f^{(\rm cls)} =
	\Sigma_{\Delta_{\mathrm{G}}}^{(\rm cls)} + \frac{2}{3}K_{\Delta_{\rm G}}\overline{d},
\end{equation}
the squared error function of the closure can be written as
\begin{equation}
	\mathrm{err}^{2}(C_{1}) =\int_{\mathcal{V}} \left|\Sigma_{\Delta_{\mathrm{G}}}+\frac{2}{3}K_{\Delta_{\rm G}}\overline{d} - C_{1}f^{(\rm cls)}\right|^{2}\dd^{3}x,
\end{equation}
where $\Sigma_{\Delta_{\mathrm{G}}}$ and $K_{\Delta_{\rm G}}$ are given by Eqs.~(\ref{eq:flux_flt}) and~(\ref{eq:turb_energy_flt}), respectively, and the volume integral extends over the whole domain $\mathcal{V}$. The minimum
of $\mathrm{err}^{2}(C_{1})$ yields the least squares error solution for the closure coefficient,
\begin{equation}
	\label{eq:coeff_lse}
	C_{1}=\frac{\int_{\mathcal{V}} f^{(\rm cls)}\left[\Sigma_{\Delta_{\mathrm{G}}}+\frac{2}{3}K_{\Delta_{\rm G}}\overline{d}\right]\,\dd^{3}x}{\int_{\mathcal{V}} |f^{(\rm cls)}|^2\dd^{3}x},
\end{equation}
where $f^{(\rm cls)}=\Delta_{\rm G}(2\overline{\rho}K_{\Delta_{\rm G}})^{1/2}|\overline{S}^{\,\ast}|^{2}$ for the eddy-viscosity closure. For statistically stationary and isotropic turbulence, the closure coefficient $C_{1}$ is independent of the filter length scale because of the local equilibrium of the transfer of turbulence energy in the inertial subrange.
Thus, the value of $C_{1}$ inferred form Eq.~(\ref{eq:coeff_lse}) is an approximation to the coefficient of the SGS closure for $\Sigma$ in LES.

\begin{figure*}[t]
\centering
\mbox{\subfigure[eddy-viscosity closure]{\includegraphics[width=80mm]{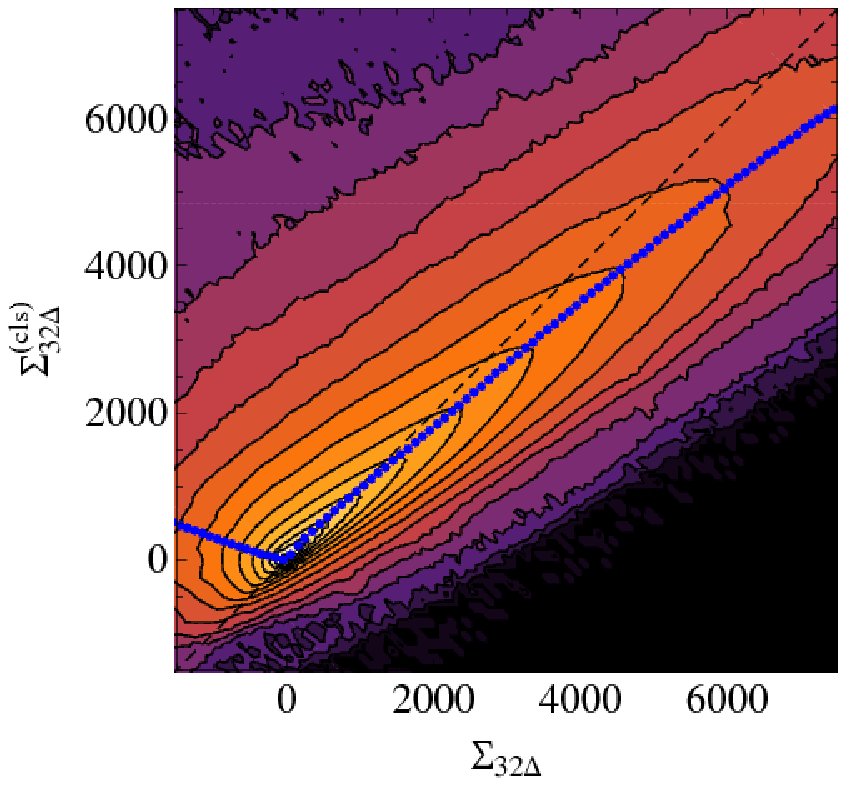}}\quad
\subfigure[eddy-viscosity closure (shocks excluded)]{\includegraphics[width=80mm]{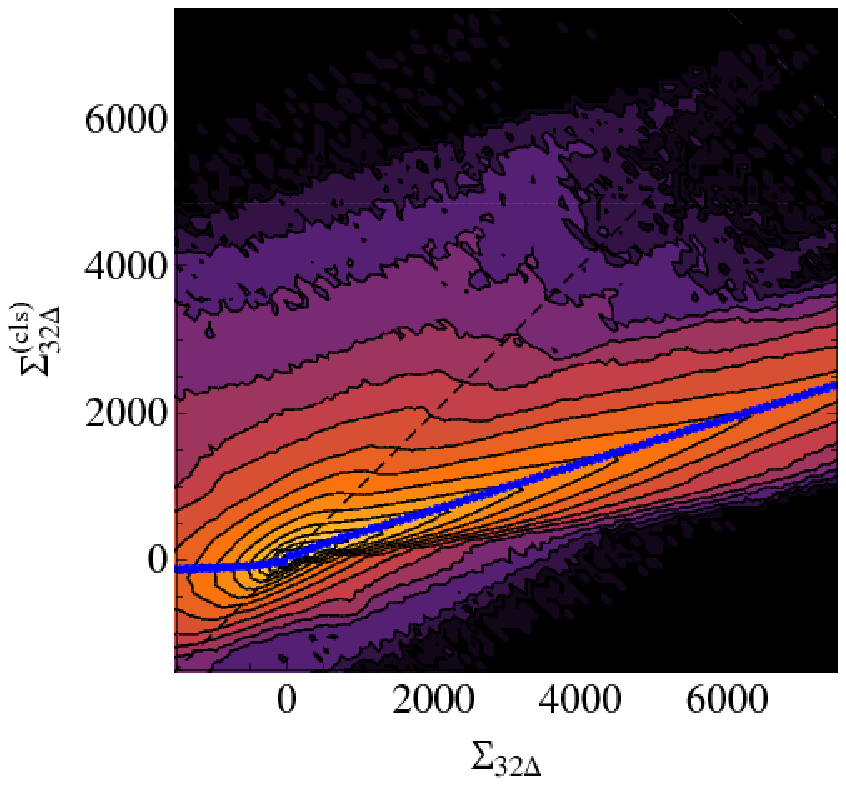}}}\\
\mbox{\subfigure[determinant closure]{\includegraphics[width=80mm]{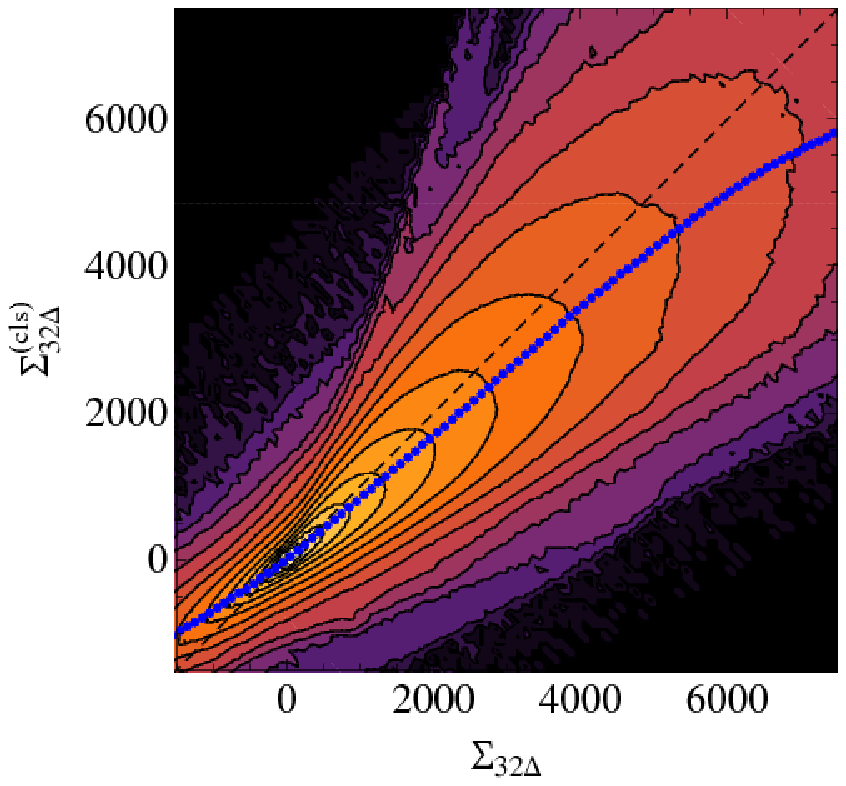}}\quad
\subfigure[non-linear closure]{\includegraphics[width=80mm]{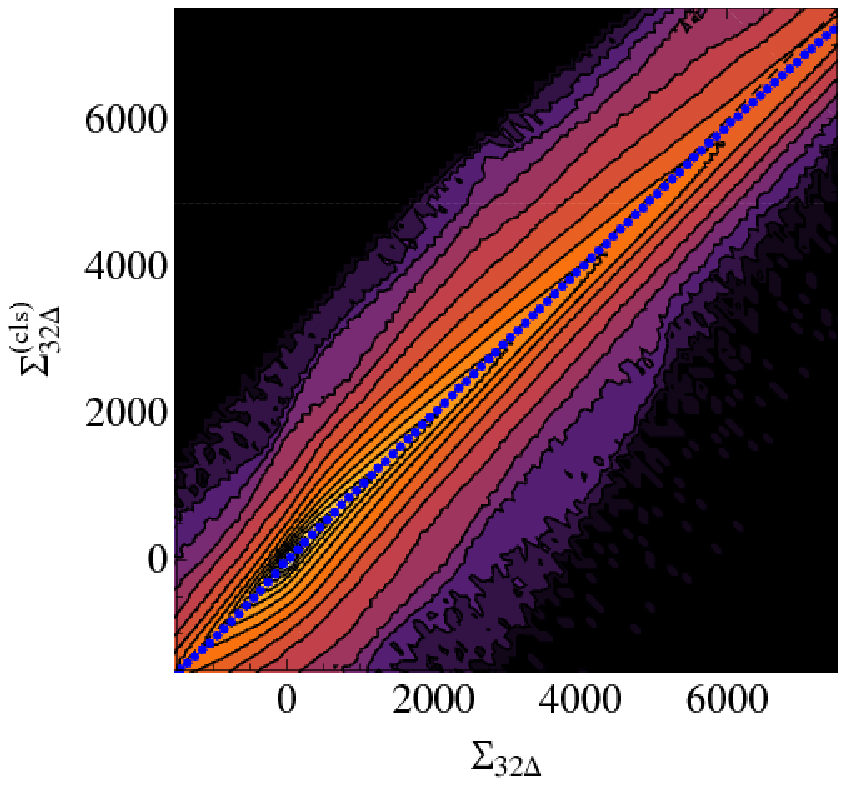}}}
\caption{Correlation diagrams for the SGS turbulence energy flux in the case of isothermal supersonic turbulence with solenoidal forcing ($\mathcal{M}_{\mathrm{rms}}\approx 5.3$).
The applied filter length is $32\Delta$. The blue dots indicate the average prediction of the closure for a given value of $\Sigma_{32\Delta}$.}
\label{fig:corrl_sol1024}
\end{figure*}

\begin{figure*}[t]
\centering
\mbox{\subfigure[eddy-viscosity closure]{\includegraphics[width=80mm]{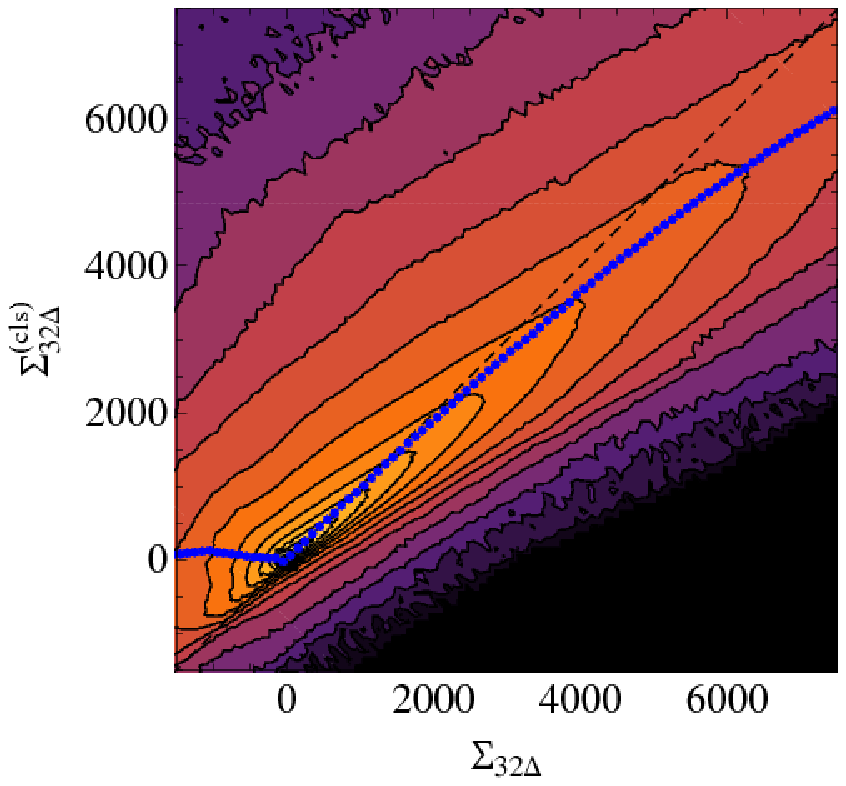}}\quad
\subfigure[shocks excluded]{\includegraphics[width=80mm]{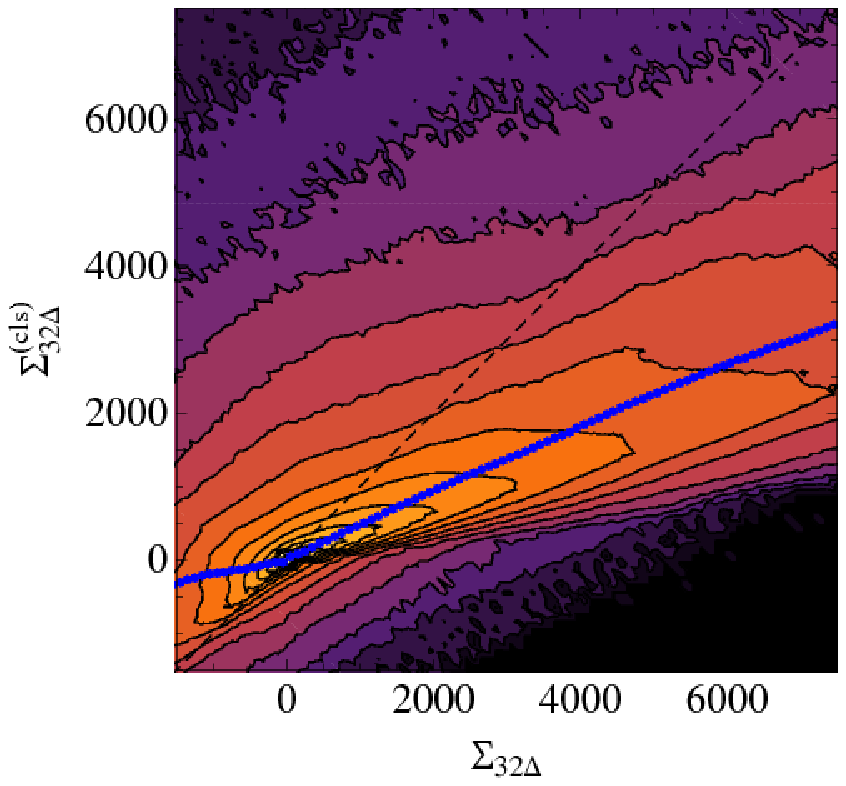}}}\\
\mbox{\subfigure[determinant closure]{\includegraphics[width=80mm]{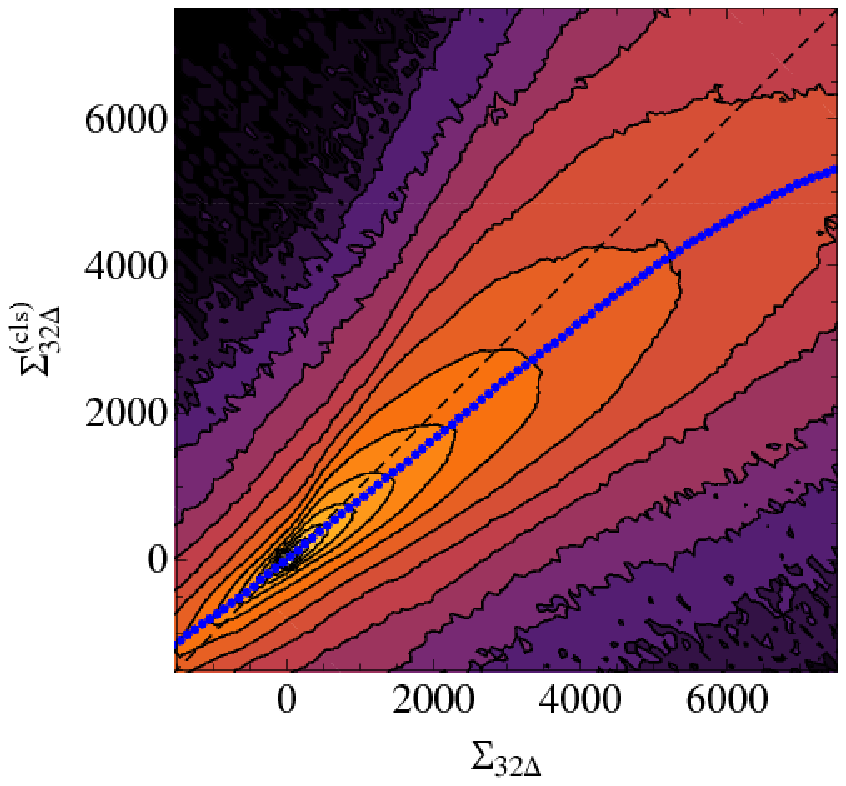}}\quad
\subfigure[non-linear closure]{\includegraphics[width=80mm]{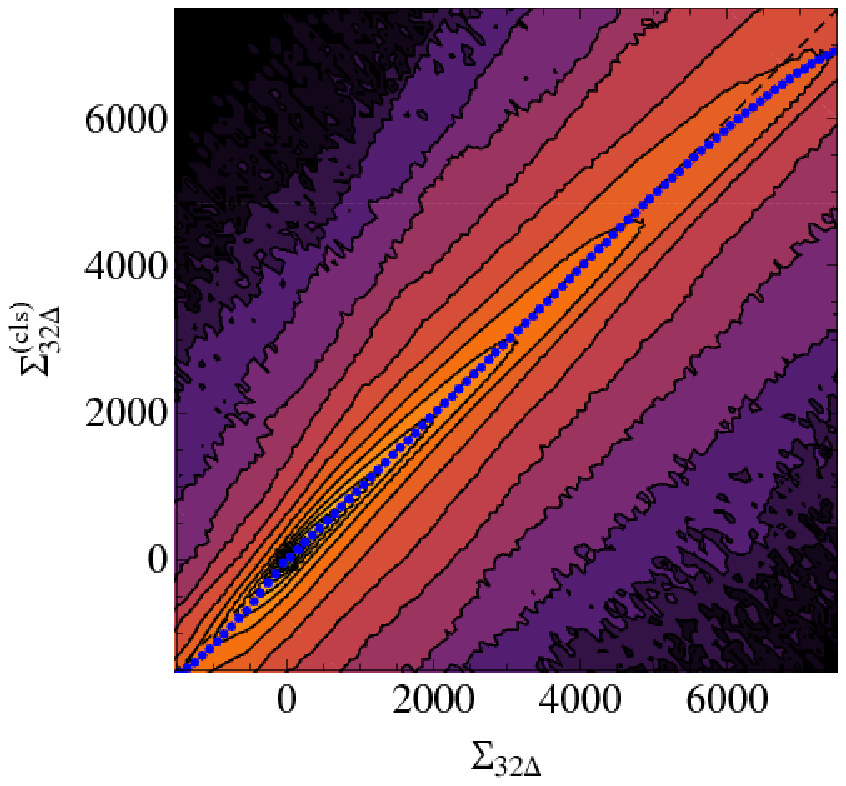}}}\caption{Correlation diagrams for isothermal supersonic turbulence with compressive forcing ($\mathcal{M}_{\mathrm{rms}}\approx 5.6$) as in Fig.~\ref{fig:corrl_sol1024}.}
\label{fig:corrl_dil1024}
\end{figure*}

To calculate $C_1$, we use data from two $1024^{3}$ simulations of supersonic isothermal turbulence with a root-mean-square (RMS) Mach number around $5.5$ \citep{FederRom10}. Statistically stationary and isotropic turbulence is produced by stochastic forcing. Solenoidal (divergence-free) forcing is applied in one simulation, while the forcing is compressive (rotation-free) in the other simulation. We
choose $\Delta_{\mathrm{G}}=32\Delta$ for the filtering of the simulation data. Figs.~\ref{fig:corrl_sol1024} and~\ref{fig:corrl_dil1024} show the correlation between $\Sigma_{\Delta_{\mathrm{G}}}^{(\rm cls)}$ and $\Sigma_{\Delta_{\mathrm{G}}}$ by means of two-dimensional probability density functions. For the eddy-viscosity closure~(\ref{eq:flux_eddy_visc}), the correlation is quite good (the spacing of the contour lines in Figs.~\ref{fig:corrl_sol1024} and~\ref{fig:corrl_dil1024} is logarithmic in the two-dimensional
probability density), but there is a problem with negative flux values. The values of the closure coefficient $C_{1}$ following from Eq.~(\ref{eq:coeff_lse}) are listed in Table~\ref{table:corrl_1024}. Also listed are the correlation coefficients
\begin{equation}
\begin{split}
   \mathrm{corr} &[\Sigma_{\Delta_{\rm G}},\Sigma_{\Delta_{\rm G}}^{(\rm cls)}] =\\
   &\frac{\int_{\mathcal{V}} \left[\Sigma_{\Delta_{\rm G}}-\langle\Sigma_{\Delta_{\rm G}}\rangle\right]
   \left[\Sigma_{\Delta_{\rm G}}^{(\rm cls)}-\langle\Sigma_{\Delta_{\rm G}}^{(\rm cls)}\rangle\right]\,\dd^{3}x}{\mathrm{std}[\Sigma_{\Delta_{\rm G}}]\mathrm{std}[\Sigma_{\Delta_{\rm G}}^{(\rm cls)}]},
   \end{split}
\end{equation}
where $\mathrm{std}[\cdot]$ denotes the standard deviation and the angle brackets indicate an average over the whole domain. 

An important question for LES of supersonic turbulence is whether shocks can be accommodated in
closures for the turbulence energy. Since the eddy-viscosity closure originates from incompressible
turbulence, \citet{MaierIap09} suggested to set $\Sigma$ equal to zero in the vicinity of shock fronts. This should suppress the spurious production of SGS turbulence energy by the large strain
at shock fronts. Thus, we tested whether excluding shocks in the computation of the eddy-viscosity closure for the turbulence energy flux would improve the correlations. However, panels (b) in Figs.~\ref{fig:corrl_sol1024} and~\ref{fig:corrl_dil1024} make clear that such a cutoff deteriorates the correlations and implies a significant underestimate of large positive fluxes. Although the applied shock detection criterion $\overline{d}<-\overline{c}_{\mathrm{s}}/\Delta_{\mathrm{G}}$ is rather crude, we interpret this trend as an indication that shocks must not be separated from the supersonic turbulent cascade.

\begin{table}[t]
\caption{Closure and correlation coefficients for the closures shown in Figs.~\ref{fig:corrl_sol1024}
and~\ref{fig:corrl_dil1024}.}
\label{table:corrl_1024}      
\centering                         
\begin{tabular}{lcc}        
\hline\hline                
closure & $C_1$ & $\mathrm{corr}[\Sigma_{32\Delta},\Sigma_{32\Delta}^{(\rm cls)}]$  \\  
\hline                        
\multicolumn{3}{c}{solenoidal forcing, $\zeta=1.0$, $\mathcal{M}_{\mathrm{rms}}\approx 5.3$}\\
\hline
eddy viscosity &  0.102 & 0.950 \\
eddy viscosity (shocks excluded) & 0.055 & 0.931\\
determinant & 0.803 & 0.950 \\
non-linear     & 0.849 & 0.991\\
\hline                             
\multicolumn{3}{c}{compressive forcing, $\zeta=0.0$, , $\mathcal{M}_{\mathrm{rms}}\approx 5.6$}\\
\hline
eddy viscosity &  0.092 & 0.930 \\
eddy viscosity (shocks excluded) & 0.059 & 0.914\\
determinant & 0.834 & 0.947 \\
non-linear & 0.833 & 0.991\\
\hline                                   
\end{tabular}
\end{table}

In addition to the conventional eddy-viscosity closure, we investigate a closure that is based on the
determinante of the velocity gradient \citep{PortWood01}. In this case,
the trace-free part of the SGS turbulence stress tensor is still given by the expression
$\tau_{ij}^{\ast}=2\rho\nu_{\mathrm{sgs}}S_{ij}^{\ast}$. The eddy viscosity, however, does not depend on $K_{\mathrm{sgs}}$. It is defined by $\nu_{\mathrm{sgs}}=-C_{1}\Delta^{2}|S^{\ast}|^{\,-2}\det\tens{S}^{\ast}$. Hence, the turbulence energy flux on the filter scale is given by
\begin{equation}
	\label{eq:flux_det}
	\Sigma_{\Delta_{\mathrm{G}}}^{(\rm cls)} =
	-C_{1}\overline{\rho}\Delta_{\mathrm{G}}^{2}\det\overline{\tens{S}}^{\,\ast}
	-\frac{2}{3}K_{\Delta_{\rm G}}\overline{d}
\end{equation}
for this closure. \citet{PortWood01} employ the same method to test the correlation between
their closure and the turbulence energy flux as we do. A particularly interesting feature of the determinant is that it switches signs and thereby describes two different flow topologies. In one case, the determinant is negative. This corresponds to the forward turbulent cascade transporting
energy from large eddies to smaller eddies. In the other case, the flow is contracting in one dimension and expanding in the other two. Then the determinant is positive, corresponding to a backscattering of energy from small eddies to larger eddies. This phenomenon can be explained by the alignment of vortices along a single stretching direction (the "tornado" topology). While an energy flux of the form~(\ref{eq:flux_eddy_visc}) fails to describe the reverse cascade, we see in panels (c) of Figs.~\ref{fig:corrl_sol1024} and~\ref{fig:corrl_dil1024} that the determinant closure yields a good correlation for negative energy flux. However, the overall correlation does not significantly improve (see Table~\ref{table:corrl_1024}), because of the relatively large scatter in the forward cascade.

In \citet{WoodPort06}, a non-linear expression for the turbulence stress tensor is investigated, which
depends on the full Jacobian $\vect{\nabla}\otimes\vect{u}$ of the velocity:
\begin{equation}
  	\label{eq:tau_nonlin_pure}
  	\tau_{ij}=
  	-2C_{1}K_{\mathrm{sgs}}\frac{2u_{i,k}u_{j,k}}{|\vect{\nabla}\otimes\vect{u}|^{2}}
	-\frac{2}{3}(1-C_{1})K_{\mathrm{sgs}}\delta_{ij}.
\end{equation}
Since $|\vect{\nabla}\otimes\vect{u}|=(2u_{i,k}u_{i,k})^{1/2}$, the above expression fulfills the identity $\tau_{ii}=-2K_{\mathrm{sgs}}$. The corresponding turbulence energy flux on the filter length scale $\Delta_{\mathrm{G}}$ is given by
\begin{equation}
	\label{eq:flux_nonlin}
	\Sigma_{\Delta_{\mathrm{G}}}^{(\rm cls)} =
	-4C_{1}K_{\Delta_{\rm G}}
	\frac{\overline{u}_{i,k}\overline{u}_{j,k}\overline{S}_{\! ij}^{\,\ast}}
	{|\vect{\nabla}\otimes\overline{\vect{u}}|^{2}}
	-\frac{2}{3}(1-C_{1})K_{\Delta_{\rm G}}\overline{d}.
\end{equation}
Figs.~\ref{fig:corrl_sol1024} (d) and~\ref{fig:corrl_dil1024} (d) show that the correlation is excellent
for the above closure, with correlation coefficients above $0.99$, as listed in Table~\ref{table:corrl_1024}. Like the determinant closure discussed above, the trace-free part of the non-linear closure for the SGS turbulence stress switches signs and, thus, allows for a backward energy cascade .

\subsection{Mixed closure for supersonic isothermal turbulence}
\label{sc:test_mixed}

Even though the correlation of the turbulence energy flux is very good, the purely non-linear closure~(\ref{eq:tau_nonlin_pure}) is generally not adequate as a model for the turbulence stress tensor for the following reasons. Most importantly, rotation invariance is violated because of the antisymmetric
part of $\vect{\nabla}\otimes\vect{u}$. As a consequence, spurious turbulence energy would be produced for a uniformly rotating fluid. Apart from that, the application of this closure in LES of forced turbulence show that the growth of turbulence energy during the transition from laminar to turbulent flow is insufficient for this closure. This is because of the linear dependence on $K_{\mathrm{sgs}}$.\footnote{The eddy-viscosity closure depends on $K_{\mathrm{sgs}}^{1/2}$. Writing $K_{\mathrm{sgs}}=\frac{1}{2}\rho q_{\mathrm{sgs}}^2$, a factor $q_{\mathrm{sgs}}$ can be cancelled from the
SGS turbulence energy Eq.~(\ref{eq:energy_sgs2}). This results in an equation for $q_{\mathrm{sgs}}$ with a non-vanishing production rate of $q_{\mathrm{sgs}}$ even starting from the
initial condition $q_{\mathrm{sgs}}=0$. For the non-linear closure, on the other hand, $q_{\mathrm{sgs}}=0$ is a fixed point of the equation.} As pointed out by \citet{WoodPort06}, a seed term has to be included in order to trigger the production of turbulence energy. If the seed term is constructed from the symmetric part of the velocity gradient, then turbulence energy production vanishes for a uniformly rotating fluid, and, consequently, the problem of rotation invariance is also resolved. \citet{WoodPort06} consider a linear combination of the closure~(\ref{eq:tau_nonlin_pure}) with the determinant closure. Because of the relatively large scatter of the determinant closure, however, we propose a combination of the non-linear closure with the linear eddy-viscosity closure. Conceptually, this combination has
the advantage that the eddy-viscosity closure, which is well established for LES of incompressible turbulence, follows as a limiting case.

\begin{figure*}[t]
\centering
\mbox{\subfigure[solenoidal, $\mathcal{M}_{\mathrm{rms}}\approx 5.3$, $\Delta_{\rm G}=32\Delta$]{\includegraphics[width=80mm]{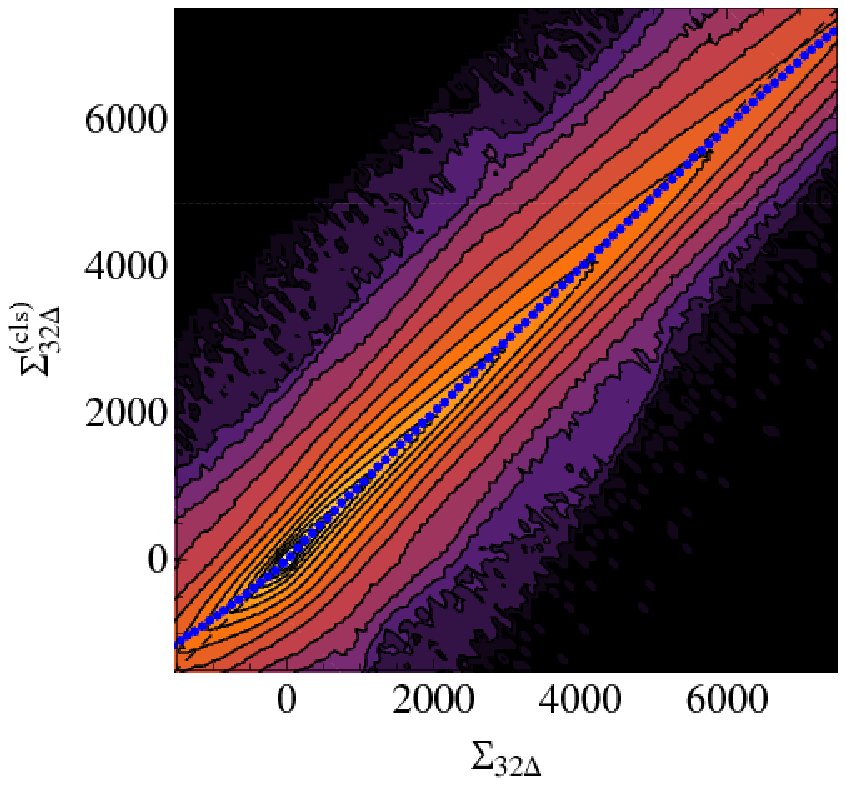}}\quad
\subfigure[compressive, $\mathcal{M}_{\mathrm{rms}}\approx 5.6$, $\Delta_{\rm G}=32\Delta$]{\includegraphics[width=80mm]{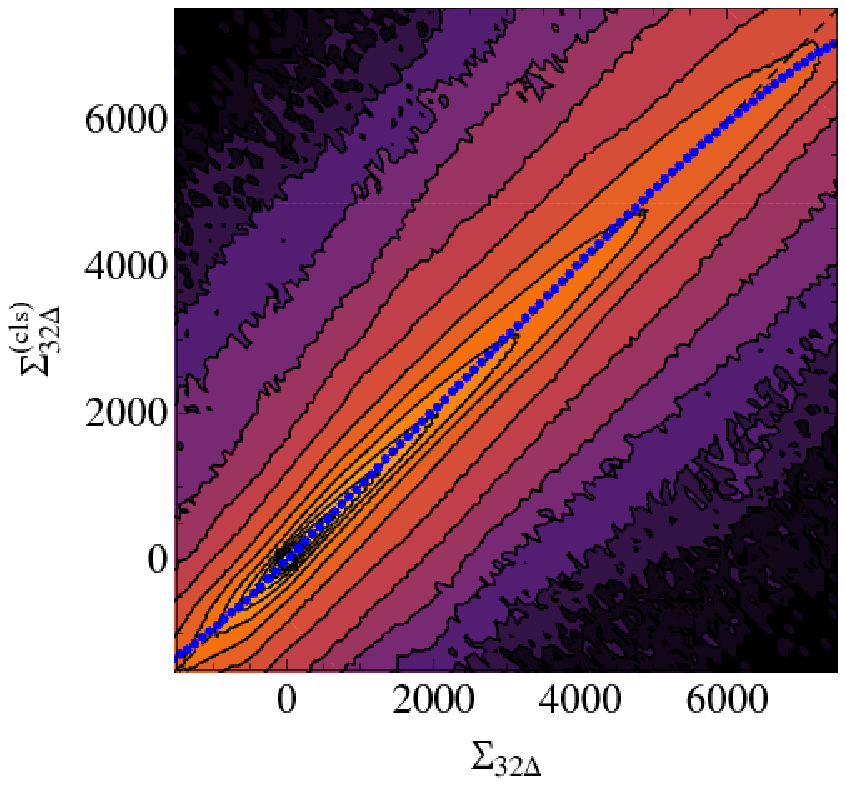}}}\\
\caption{Correlation diagrams for the mixed closures for isothermal supersonic turbulence with different forcing.
}
\label{fig:corrl_mixed1024}
\end{figure*}

\begin{table}[t]
\caption{Closure and correlation coefficients for the linear combination of the eddy-viscosity and
the non-linear closure.}
\label{table:mixed_1024}      
\centering                         
\begin{tabular}{lcccc}        
\hline\hline                
$\Delta_{\rm G}$ & $\mathcal{M}_{\Delta_{\rm G}}$ & $C_1$ & $C_2$ & $\mathrm{corr}[\Sigma_{\Delta_{\rm G}},\Sigma_{\Delta_{\rm G}}^{(\rm cls)}]$  \\  
\hline                        
\multicolumn{5}{c}{solenoidal forcing ($\zeta=1.0$), $\mathcal{M}_{\mathrm{rms}}\approx 5.3$}\\
\hline
$16\Delta$ & 0.96 & 0.0204 & 0.749 & 0.991\\
$32\Delta$ & 1.48 & 0.0229 & 0.723 & 0.991\\
$64\Delta$ & 2.16 & 0.0242 & 0.696 & 0.986 \\
\hline                             
\multicolumn{5}{c}{compressive forcing ($\zeta=0.0$), $\mathcal{M}_{\mathrm{rms}}\approx 5.6$}\\
\hline
$32\Delta$ &1.29 & 0.0189 & 0.698 & 0.991\\
\hline                                   
\end{tabular}
\end{table}

The least-squared-errors approach can be generalized to a mixed closure with two coefficient, $C_1$ and $C_2$. For
\begin{equation}
	C_{1}f^{(\rm cls)} +  C_{2}g^{(\rm cls)} =
	\Sigma_{\Delta_{\mathrm{G}}}^{(\rm cls)} + \frac{2}{3}K_{\Delta_{\rm G}}\overline{d},
\end{equation}
the closure coefficients are given by the linear system of equations
\begin{alignat}{2}
	\label{eq:coeff_lse_2}
	&\left(\int_{\mathcal{V}} |f^{(\rm cls)}|^2\dd^{3}x\right)\,C_1 + 
	\left(\int_{\mathcal{V}} f^{(\rm cls)}g^{(\rm cls)}\dd^{3}x\right)\,C_2\\
	&=\int_{\mathcal{V}} f^{(\rm cls)}\left[\Sigma_{\Delta_{\mathrm{G}}}+\frac{2}{3}K_{\Delta_{\rm G}}\overline{d}\right]\,\dd^{3}x, \\
	&\left(\int_{\mathcal{V}}f^{(\rm cls)}g^{(\rm cls)} \dd^{3}x\right)\,C_1 + 
	\left(\int_{\mathcal{V}} |g^{(\rm cls)}|^{2}\dd^{3}x\right)\,C_{2} \\
	&=\int_{\mathcal{V}} g^{(\rm cls)}\left[\Sigma_{\Delta_{\mathrm{G}}}+\frac{2}{3}K_{\Delta_{\rm G}}\overline{d}\right]\,\dd^{3}x,
\end{alignat}
where 
\begin{align}
	f^{(\rm cls)}&=
	\Delta_{\rm G}(2\overline{\rho}K_{\Delta_{\rm G}})^{1/2}|\overline{S}^{\,\ast}|^{2}, \\
	g^{(\rm cls)} &=-4K_{\Delta_{\rm G}}
	\frac{\overline{u}_{i,k}\overline{u}_{j,k}\overline{S}_{\! ij}^{\,\ast}}
	{|\vect{\nabla}\otimes\overline{\vect{u}}|^{2}}.
\end{align}
The solutions for $C_1$ and $C_2$ that are obtained from our numerical data are listed in Table~\ref{table:mixed_1024}. As one can see, the correlation coefficients are about as high as for the purely non-linear closure and there is only little variation with the forcing and the filtering length scale. For 
$\Delta_{\mathrm{G}}=64\Delta$ the ratio of $\Delta_{\mathrm{G}}$ to the integral scale
$L$ is 8, which is quite small. As a consequence, there might be a marginal influence of the forcing. The filter length $\Delta_{\mathrm{G}}=16\Delta$, on the other hand is significantly affected by numerical dissipation.
The correlation diagrams for the mixed closure are plotted in Fig.~\ref{fig:corrl_mixed1024}. 
Although the relation between $\Sigma_{32\Delta}^{(\rm cls)}$ and $\Sigma_{32\Delta}$ is slightly tilted for negative fluxes, the results are comparable to ~\ref{fig:corrl_sol1024} (d) and~\ref{fig:corrl_dil1024} (d). Therefore, we base our SGS model on the mixed non-linear closure~(\ref{eq:tau_nonlin}) with the averaged coefficients $C_{1}=0.02$ and $C_{2}=0.7$.

\begin{figure*}[t]
\centering
\mbox{\subfigure[isothermal, $\mathcal{M}_{\mathrm{rms}}\approx 2.2$]{\includegraphics[width=80mm]{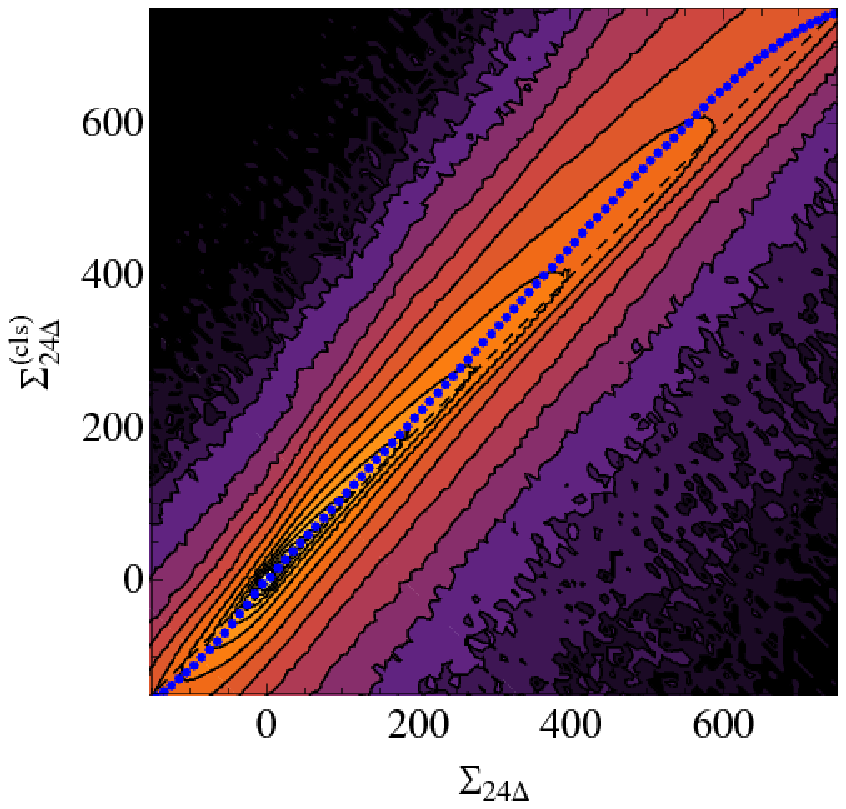}}\quad
\subfigure[adiabatic, $\mathcal{M}_{\mathrm{rms}}\approx 0.5$]{\includegraphics[width=80mm]{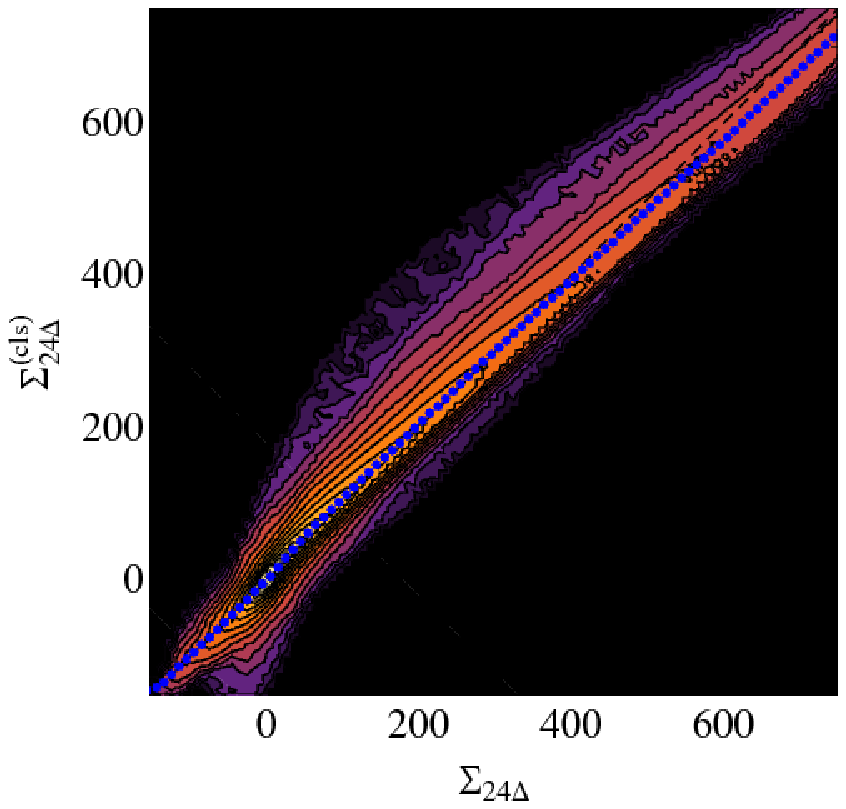}}}\\
\caption{Correlation diagrams for the mixed closure in the case of isothermal (a) and adiabatic (b) turbulence with lower RMS Mach numbers as in Fig.~\ref{fig:corrl_mixed1024}.}
\label{fig:corrl_768}
\end{figure*}

\begin{table}[t]
\caption{Correlation coefficients for the linear combination of the eddy-viscosity and the non-linear closure with $C_{1}=0.02$ and $C_{2}=0.7$ for isothermal turbulence and adiabatic
turbulence at various instants with different Mach numbers.}
\label{table:corrl_768}      
\centering                         
\begin{tabular}{rccc}        
\hline\hline                
t/T & $\mathcal{M}_{\mathrm{rms}}$ & $\langle\mathcal{M}_{24\Delta}\rangle$ & $\mathrm{corr}[\Sigma_{24\Delta},\Sigma_{24\Delta}^{(\rm cls)}]$  \\  
\hline                        
\multicolumn{4}{c}{isothermal ($\gamma = 1.01$)}\\
\hline
9.1 & 2.2 & 0.66 & 0.990 \\
\hline                             
\multicolumn{4}{c}{adiabatic ($\gamma = 1.4$)}\\
\hline
2.0 & 1.3 & 0.48 & 0.981\\
3.9 & 0.9 & 0.34 & 0.986\\ 
8.0 & 0.6 & 0.26 & 0.990 \\
15.9 & 0.5 & 0.21 & 0.990\\
\hline                                   
\end{tabular}
\end{table}

\subsection{Supplementary tests for different Mach numbers}

For the data listed in Table~\ref{table:mixed_1024}, the average Mach numbers associated with the filter scale, $\mathcal{M}_{\Delta_{\rm G}}=\langle (2K_{\Delta_{\rm G}}/\rho c_{\rm s}^2)^{1/2}\rangle $, assume values around the speed of sound. Thus, the question arises
whether the closure coefficients calculated above are applicable to subsonic velocity fluctuations. 
To test the mixed closure for a different Mach number, we calculated the turbulence energy flux 
$\Sigma_{\Delta_{\rm G}}^{(\rm cls)}$ with fixed values $C_{1}=0.02$ and $C_{2}=0.7$ for data from a simulation of isothermal turbulence with $\mathcal{M}_{\rm rms}\approx 2.2$
\citep{SchmFeder09a}.
The resulting correlation diagram is plotted in Fig.~\ref{fig:corrl_768} (a). There
is a small bias to overestimate the turbulence energy flux, but the prediction of the mixed closure is still very good. Indeed, a correlation coefficient $\mathrm{corr}[\Sigma_{\rm G},\Sigma_{\rm G}^{(\rm cls)}]=0.990$ is obtained in this case (see Table~\ref{table:corrl_768}). In addition, we investigated data from an adiabatic turbulence simulation \citep{SchmNie07}, in which the RMS Mach number gradually decreases with time because of the dissipative heating of the gas. The results are summarized in Table~\ref{table:corrl_768}, and the correlation diagram for the final snapshot of the simulation is shown in Fig.~\ref{fig:corrl_768} (b). Our results suggest that the closure coefficients are not very sensitive to the Mach number. Nevertheless, we cannot exclude that the optimal values of $C_1$ and $C_2$ differ significantly if the SGS turbulence Mach number $\mathcal{M}_{\rm sgs}$ (see Section~\ref{sc:euler_sgs}) is only a tiny fraction of the speed of sound. Answering this question is left for future studies.

\subsection{Energy dissipation}

For the turbulence energy on the length scale $\Delta_{\rm G}$, which is defined by Eq.~(\ref{eq:turb_energy_flt}), a dynamical equation analogous to Eq.~(\ref{eq:energy_sgs2}) can be formulated.
Averaging this equation over the whole periodic domain and assuming statistical equilibrium, i.~e.,
$\partial_t \langle K_{\Delta_{\rm G}}\rangle\simeq 0$, the following global balance equation is obtained:
\begin{equation}
	\langle\Sigma_{\Delta_{\rm G}}\rangle -
	\langle\overline{Pd}-\overline{P}\,\overline{d}\rangle -
	C_{\epsilon}\left\langle\frac{K_{\Delta_{\rm G}}^{3/2}}{\overline\rho^{1/2}\Delta_{\rm G}}\right\rangle \simeq 0.
\end{equation}
The first term is the mean turbulence energy flux, the second term is the mean pressure dilatation
\citep[see][]{SchmNie06b}, and the third term is the mean dissipation rate expressed in terms
of $K_{\Delta_{\rm G}}$. Substituting Eq.~(\ref{eq:flux_flt}) for $\Sigma_{\Delta_{\rm G}}$, yields
the coefficient of turbulence energy dissipation, $C_{\epsilon}$. From the supersonic isothermal
turbulence data, we find a value $C_{\epsilon}\approx 1.5$, which is somewhat higher yet still
comparable to typical values calculated for incompressible turbulence \citep[see][]{Sagaut}.


\begin{figure*}[t]
\sidecaption
\includegraphics[width=12cm]{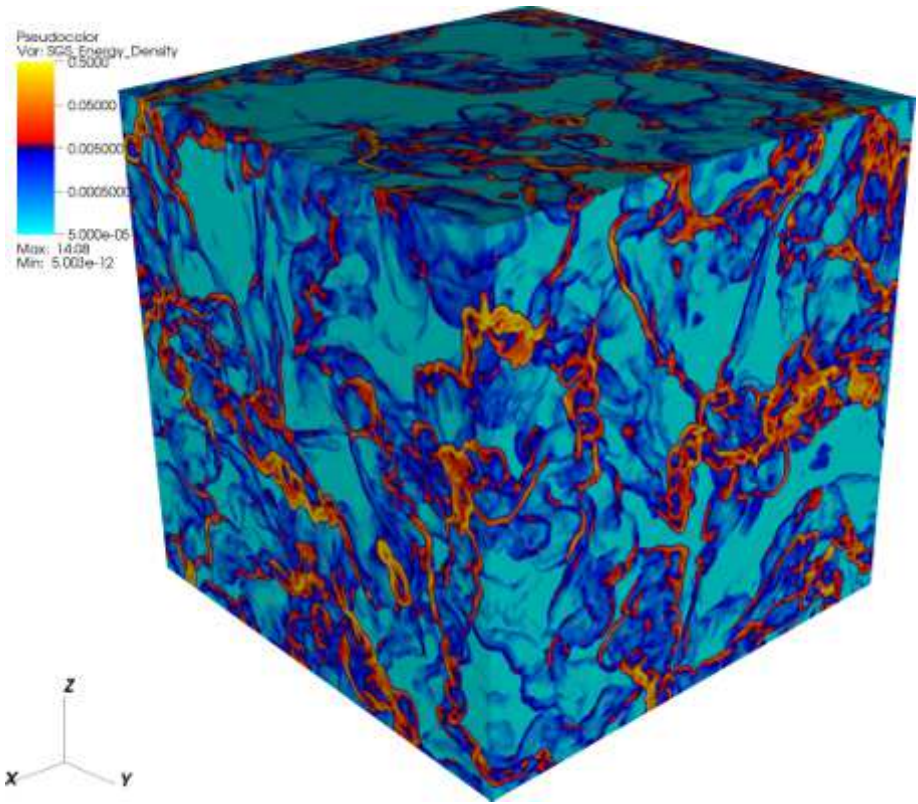}
\caption{Visualization of the SGS turbulence energy density $K_{\mathrm{sgs}}$ in a $512^{3}$ LES with solenoidal forcing.}
\label{fig:ksgs}
\end{figure*}

\section{Large eddy simulations of forced supersonic turbulence}
\label{sc:les2}

To investigate statistical properties of the SGS turbulence energy and related quantities, we run LES of forced supersonic isothermal turbulence with the SGS model defined in Sects.~\ref{sc:euler_sgs} and~\ref{sc:test}. For the implementation, we use the code Enzo 1.5 developed by the Laboratory for Computational Astrophysics at the University of California in San Diego  (http://lca.ucsd.edu). In these simulations, we apply solenoidal, compressive and mixed force fields to produce statistically stationary and homogeneous turbulence with different RMS Mach numbers \citep[see][]{SchmHille06,SchmFeder09a,FederRom10}. The forcing acts on length scales around the integral
length $L$, where $L$ is one half of the box size. The autocorrelation time of the force field is given by the time scale $T=L/V$, where the characteristic velocity $V$ specifies the magnitude of the
turbulent velocity fluctuations on the integral scale. The mixture of solenoidal (divergence-free) and compressive (rotation-free) modes of the force field is adjusted by means of a Helmholtz decomposition with weighing parameter $0\le\zeta\le 1$. Purely solenoidal forcing results for $\zeta=1$. Setting the adiabatic exponent $\gamma=1.001$, the energy dissipated per integral time is small compared to the internal energy for $\mathcal{M}_{\rm rms}$ up to
about $10$. For this reason, the gas is pseudo-isothermal. This approximate treatment of isothermality enables us to monitor energy conservation. With our implementation of the SGS model, the sum of resolved kinetic energy, SGS turbulence energy, and internal energy minus the power of the forcing integrated over time is conserved for the whole computational domain to a relative precision better than $10^{-8}$. The fraction of computational time consumed by the SGS model is in the percent range. 

\subsection{Correlations with resolved flow quantities and the effective pressure}
\label{sc:corrl_res}

\begin{figure*}[t]
\sidecaption
\includegraphics[width=12cm]{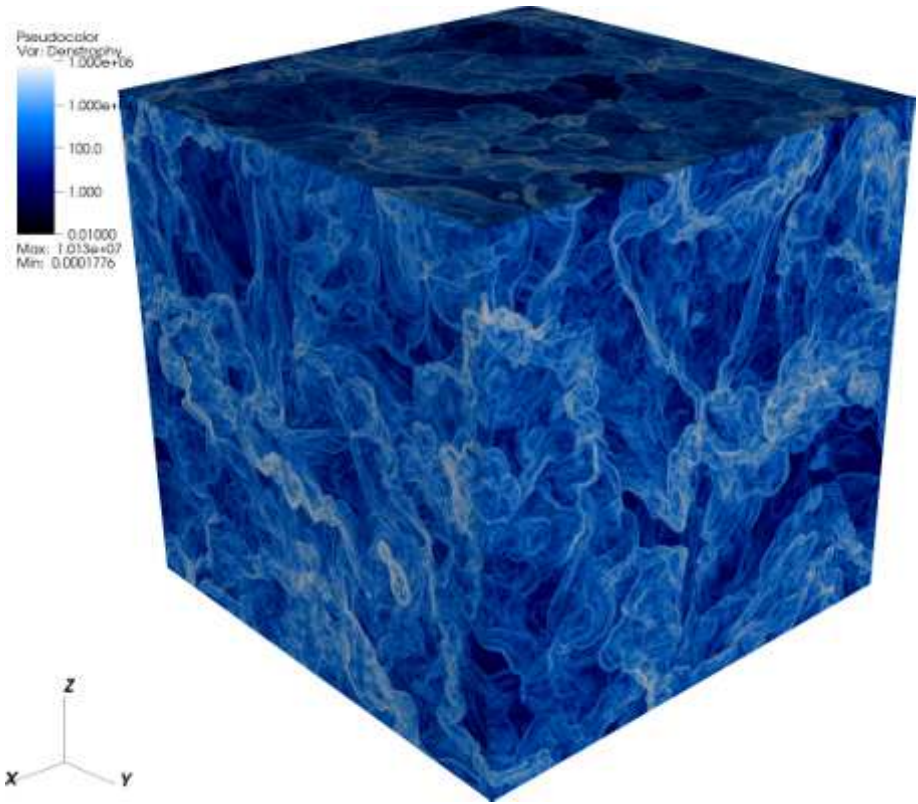}
\caption{Visualization of the denstrophy $\Omega_{1/2}$ for the same snapshot as in Fig.~\ref{fig:ksgs}.}
\label{fig:denstrophy}
\end{figure*}

\begin{figure*}[t]
\centering
\mbox{\subfigure[solenoidal]{\includegraphics[width=80mm]{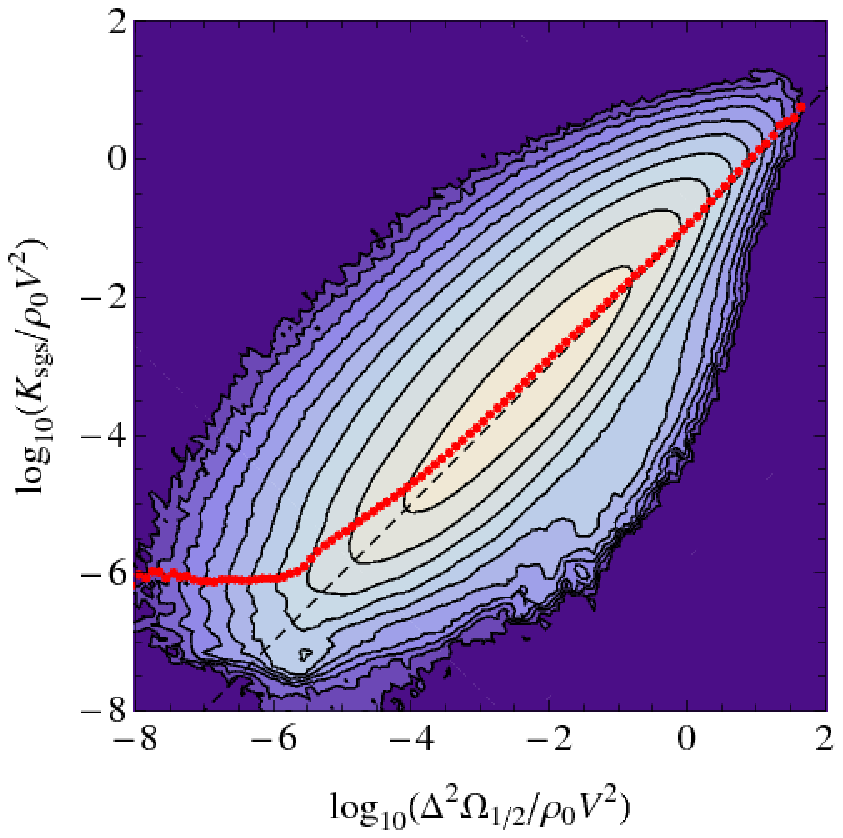}}\quad
\subfigure[compressive]{\includegraphics[width=80mm]{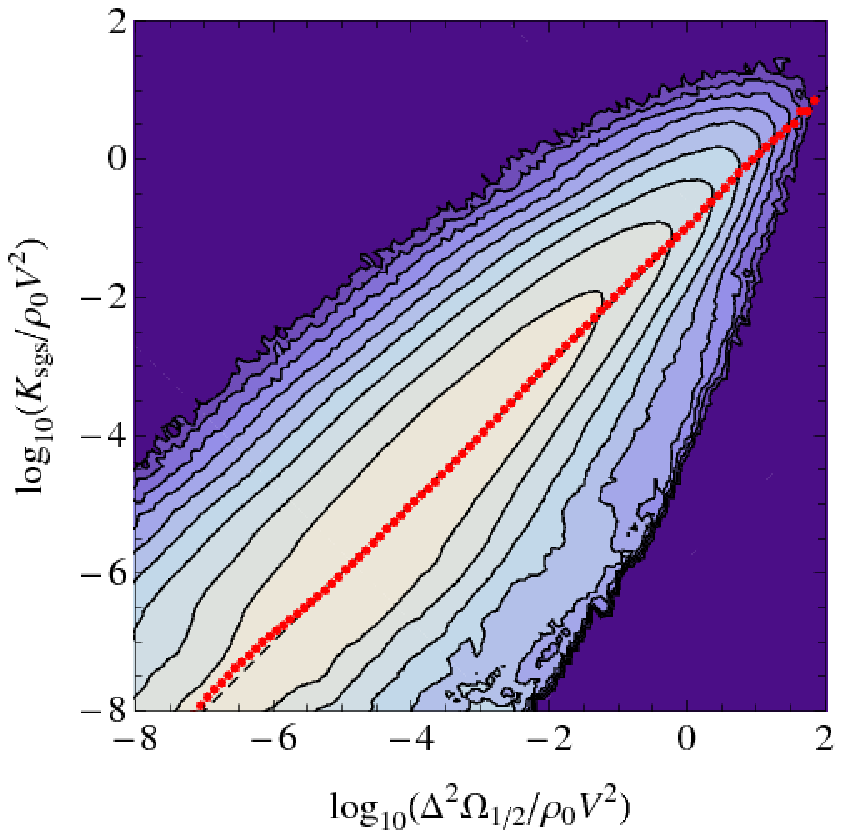}}}
\caption{Correlation diagrams of the SGS turbulence energy vs.\ the denstrophy, normalized by the
cutoff scale $\Delta$, for $512^{3}$ LES with solenoidal and compressive forcing. The contours are logarithmic. The average relation between both quantities is indicated by the dotted lines, and the dashed line shows the relation
$K_{\mathrm{sgs}} \sim 0.1\Delta^{2}\Omega_{1/2}$.}
\label{fig:corrl_energy}
\end{figure*}

As an example, Fig.~\ref{fig:ksgs} shows a visualization of $K_{\mathrm{sgs}}$ prepared from an LES with $512^{3}$ grid cells. The parameters of this simulation were chosen to match the ILES with solenoidal forcing in Section~\ref{sc:test_single}. The RMS Mach number of the flow is about $5.5$ in the statistically stationary regime. In the reddish regions, $K_{\mathrm{sgs}}$ is higher than the spatial mean, while it is lower in the bluish regions. For comparison, Fig~\ref{fig:denstrophy} shows the local denstrophy $\Omega_{1/2}=\frac{1}{2}\left|\vec{\nabla}\times\left(\varrho^{1/2}\vec{u}\right)\right|^2$, which is an indicator of compressible turbulent velocity fluctuations \citep{KritNor07}. It appears that high SGS turbulence energy is concentrated in regions of intense denstrophy. On average, $K_{\mathrm{sgs}} \sim 0.1\Delta^{2}\Omega_{1/2}$ for large denstrophy values, as one can see in the correlation diagram of $K_{\mathrm{sgs}}$ vs.\ $\Delta^{2}\Omega_{1/2}$ in Fig.~\ref{fig:corrl_energy} (a). The same relation is found for compressive forcing (see panel (b) of Fig.~\ref{fig:corrl_energy}). Nevertheless, the local values of $K_{\mathrm{sgs}}$ and $\Delta^{2}\Omega_{1/2}$ deviate substantially from the average relation. This is a consequence of the various processes contributing to the SGS dynamics, which are not fully encompassed by the derivative of the resolved velocity field. For this reason, derived quantities such as the rate of strain or the denstrophy are only of limited utility to estimate effects of turbulence on unresolved length scales. Since $P_{\mathrm{sgs}}=\frac{2}{3}K_{\mathrm{sgs}}$, this applies also to the turbulent pressure.

\begin{figure*}
\centering
\mbox{\subfigure[solenoidal]{\includegraphics[width=80mm]{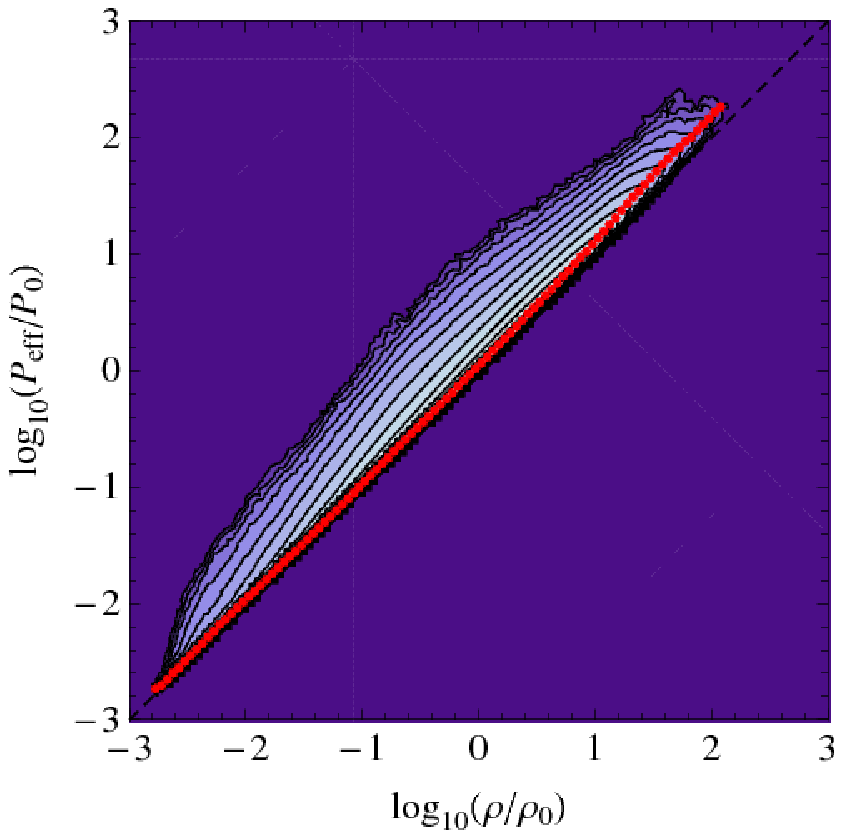}}\quad
\subfigure[compressive]{\includegraphics[width=80mm]{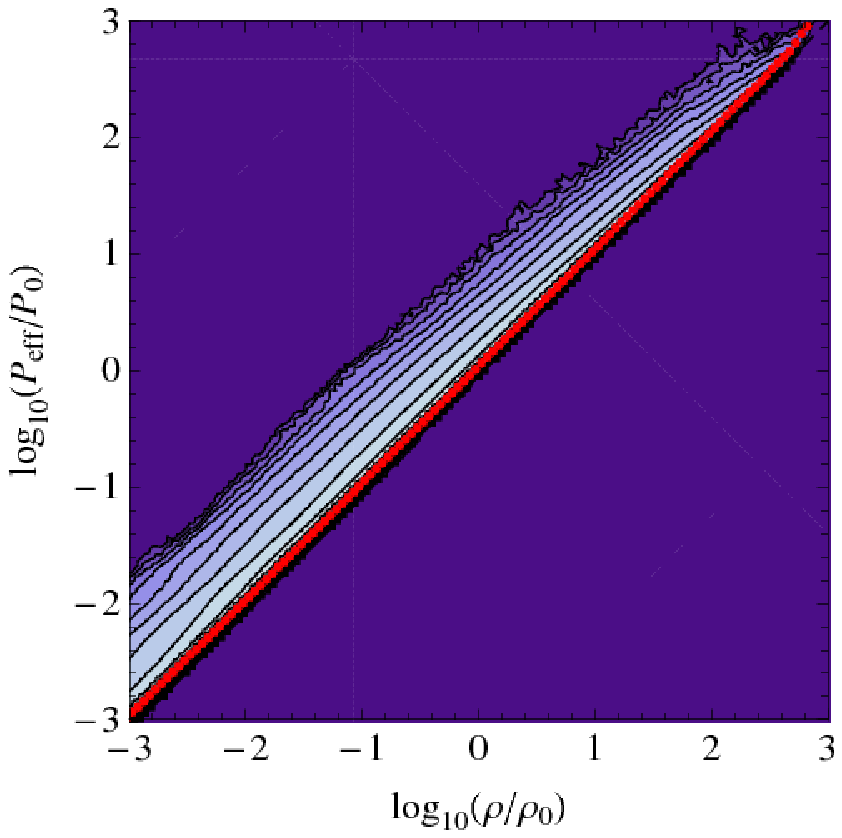}}}
\caption{Phase diagrams of the effective pressure defined by Eq.~(\ref{eq:press_eff}) vs.\ the  mass density for $512^{3}$ LES with solenoidal and compressive forcing. The contours are logarithmic. The averages of the SGS turbulence energy for particular values of the denstrophy are indicated by the dotted lines.}
\label{fig:press_diag}
\end{figure*}

The phase diagrams of the effective pressure~(\ref{eq:press_eff}) vs. the mass density are plotted in
Fig.~\ref{fig:press_diag} for both LES. One can see that the average of the effective pressure for a given mass density closely follows the isothermal relation $P\propto \rho$. This is because the mean turbulent pressure is small compared to the thermal pressure for the resolution $\Delta=L/256$ (see Section~\ref{sc:cutoff}). Locally, however, the intermittency of turbulent velocity fluctuations can give rise to an effective pressure that exceeds the thermal pressure by one order of magnitude. For this reason, the contribution of the turbulent pressure $P_{\mathrm{sgs}}$ is locally not negligible. This effect becomes stronger as the cutoff scale $\Delta$ increases in comparison to the integral scale of turbulence.

In Sect.~\ref{sc:euler_sgs}, we argue that the viscous stress term in the filtered momentum equation~(\ref{eq:momt_flt}) vanishes in the limit of infinite Reynolds number and the rate of energy dissipation on the grid scale, $\epsilon$, is determined by the SGS turbulence energy (see Eq.~\ref{eq:diss_sgs}). In contrast, an extrapolation of the expression for the microscopic dissipation rate on length scales $\ell\sim\eta$ to the grid scale $\Delta$ was proposed by \cite{PanPad09a}:
\begin{equation}
  \label{eq:diss_num}
  \varrho\epsilon = \varrho\mathrm{\nu}_{\Delta}|S^{\ast}|^{2}.
\end{equation}
The grid-scale viscosity $\nu_{\Delta}=\mathrm{const}.$ in the above expression is treated as a constant coefficient that is determined by the mean numerical dissipation of PPM. Defining the compressible Reynolds number of the resolved flow by $\mathrm{Re_{\Delta}}=2L^{2}\langle|S^{\ast}|^{2}\rangle/u_{\mathrm{rms}}^{2}$, where $u_{\mathrm{rms}}$ is the root mean square velocity\footnote{See \citet{SchmFeder09a}. Here,
we replace $\omega_{\rm rms}^{2}$ by $\langle|S^{\ast}|^{2}\rangle = \langle\omega^{2}+\frac{4}{3}d^{2}\rangle$ for consistency with Eq.~(\ref{eq:diss_num}).}, the viscosity can be evaluated from $\nu_{\mathrm{\Delta}}=VL/\mathrm{Re_{\Delta}}$. The problem with this approach is that the viscosity on the grid scale, which corresponds to the SGS eddy-viscosity, 
cannot be assumed to be constant. 

\begin{figure*}[t]
\centering
\mbox{\subfigure[solenoidal]{\includegraphics[width=80mm]{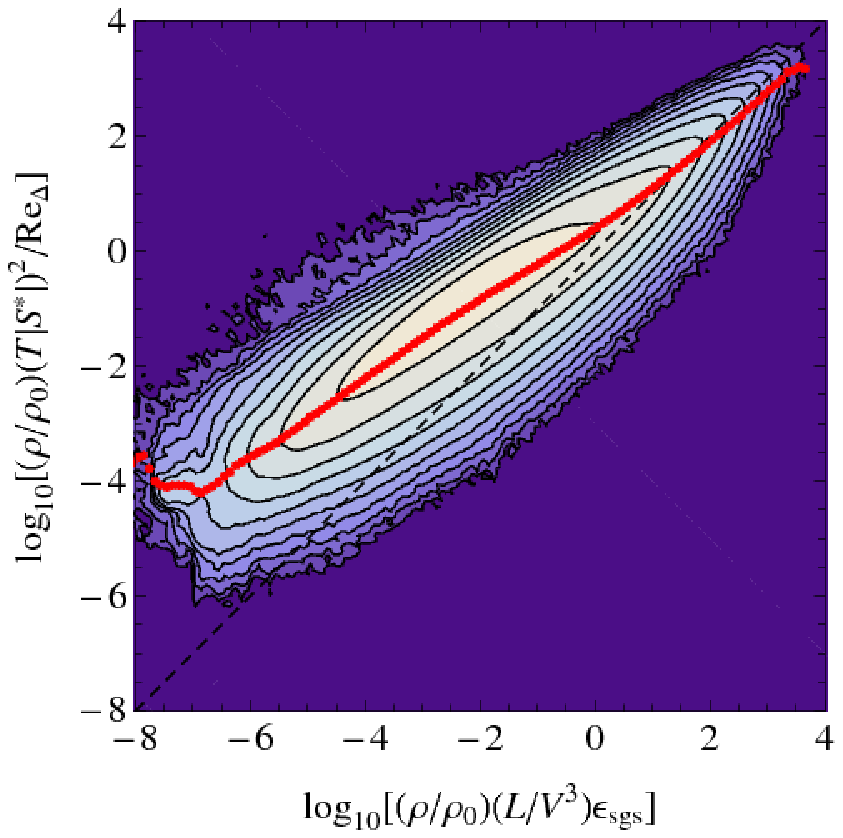}}\quad
\subfigure[compressive]{\includegraphics[width=80mm]{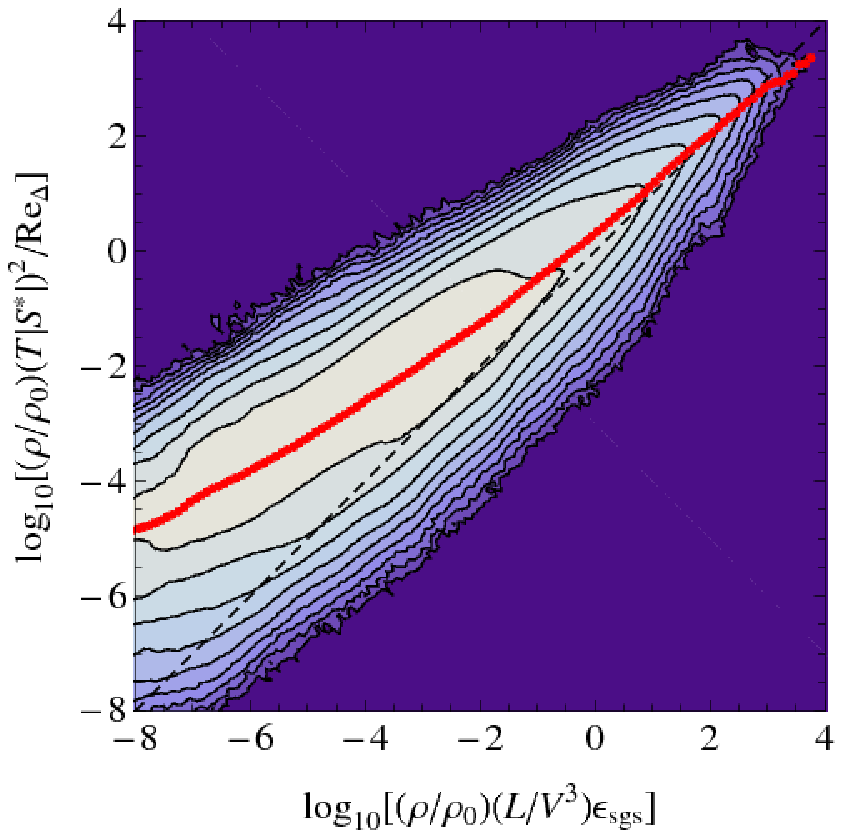}}}
\caption{Correlation diagrams of the normalized rate of energy dissipation defined by Eq.~(\ref{eq:diss_num}), where $\nu_{\Delta}$ assumes a constant value that is given by the numerical
Reynolds number, vs.\ the rate of energy dissipation~(\ref{eq:diss_sgs}) that is predicted by the SGS model. The averages of expression~ (\ref{eq:diss_num}) for given values of $\varrho\epsilon_{\mathrm{sgs}}$ are indicated by the dotted lines.}
\label{fig:corrl_diss}
\end{figure*}

Neglecting diffusion, compressibility and the non-linear term in the SGS turbulence stress~(\ref{eq:tau_nonlin}), the equilibrium between production and dissipation of SGS turbulence energy in  Eq.~(\ref{eq:energy_sgs2}) implies $K_{\mathrm{sgs}}\sim (C_{1}/C_{\epsilon})\varrho\Delta^{2}|S^{\ast}|^{2}$. Hence, $\epsilon\sim(\Delta/C_{\epsilon})^{2}(C_1|S^{\ast}|)^{3}$ according to Eq.~(\ref{eq:diss_sgs}). We emphasize that a relation of the form $\epsilon\sim\Delta^{2}|S^{\ast}|^{3}$ follows from any common SGS model under the assumption of local equilibrium \citep{Sagaut}. Comparing to Eq.~(\ref{eq:diss_num}), we see that that $\nu_{\Delta}\sim\Delta^{2}|S^{\ast}|$, which is not a constant. This is a consequence of the fact that $\varrho\epsilon\neq\sigma_{ij}u_{i,j}\propto  |S^{\ast}|^{2}$, as explained in Sect~\ref{sc:euler_sgs}. The discrepancy becomes apparent in Fig.~\ref{fig:corrl_diss}, which shows the correlation diagrams of the rate of energy dissipation calculated via Eq.~(\ref{eq:diss_num}) vs.\ $\epsilon$ following
from the SGS model. Toward low values of $\epsilon$, we find an average relation close to $|S^{\ast}|^{2}\propto\epsilon^{2/3}$, which is just the relation that follows from the above estimate of the equilibrium dissipation rate. This behavior is reasonable because the contribution of the non-linear term in the closure~(\ref{eq:tau_nonlin}), which is neglected in the estimate, is relatively small for low values of $K_{\mathrm{sgs}}$ (corresponding to low energy dissipation). Moreover, the unresolved velocity fluctuations tend to be small compared to the speed of sound in this limit, which corresponds to low compressibility.  Consequently, the results from the LES support the theoretical arguments against Eq.~(\ref{eq:diss_num}) as an approximation to the dissipation rate on the grid scale. Although Eqs.~(\ref{eq:diss_sgs}) and~(\ref{eq:diss_num}) yield about the same \emph{mean} dissipation rate, the former determines the \emph{local} rate of energy dissipation on the footing of a physically well motivated scale-separation of fluid dynamics, while the latter is based on a putative analogy between the numerical and the microscopic viscosity.

\subsection{Dependence on the cutoff scale}
\label{sc:cutoff}

The scaling of the turbulent velocity fluctuations in supersonic hydrodynamic turbulence has been inferred from energy spectrum functions and structure functions \citep[e.~g.,][]{KritNor07,SchmFeder08,SchmFeder09a,FederRom10,PriceFed10}. The pure velocity scaling in the supersonic regime is stiffer than Kolmogorov scaling, and it appears that the scaling exponent depends on the forcing. For example, \citet{FederRom10} find indices of the turbulence energy spectra $\beta=-1.86\pm 0.05$ and $-1.94\pm 0.05$ for solenoidal and compressive forcing, respectively. For incompressible turbulence, $\beta=-5/3$. Velocity variables with fractional mass-weighing, in particular $\rho^{1/3}u$, exhibit similar scaling laws, which can be interpreted as an indication of universality \citep{KritNor07,SchmFeder08}.

\begin{figure*}[t]
\centering
\begin{tabular}{cc}
\includegraphics[width=80mm]{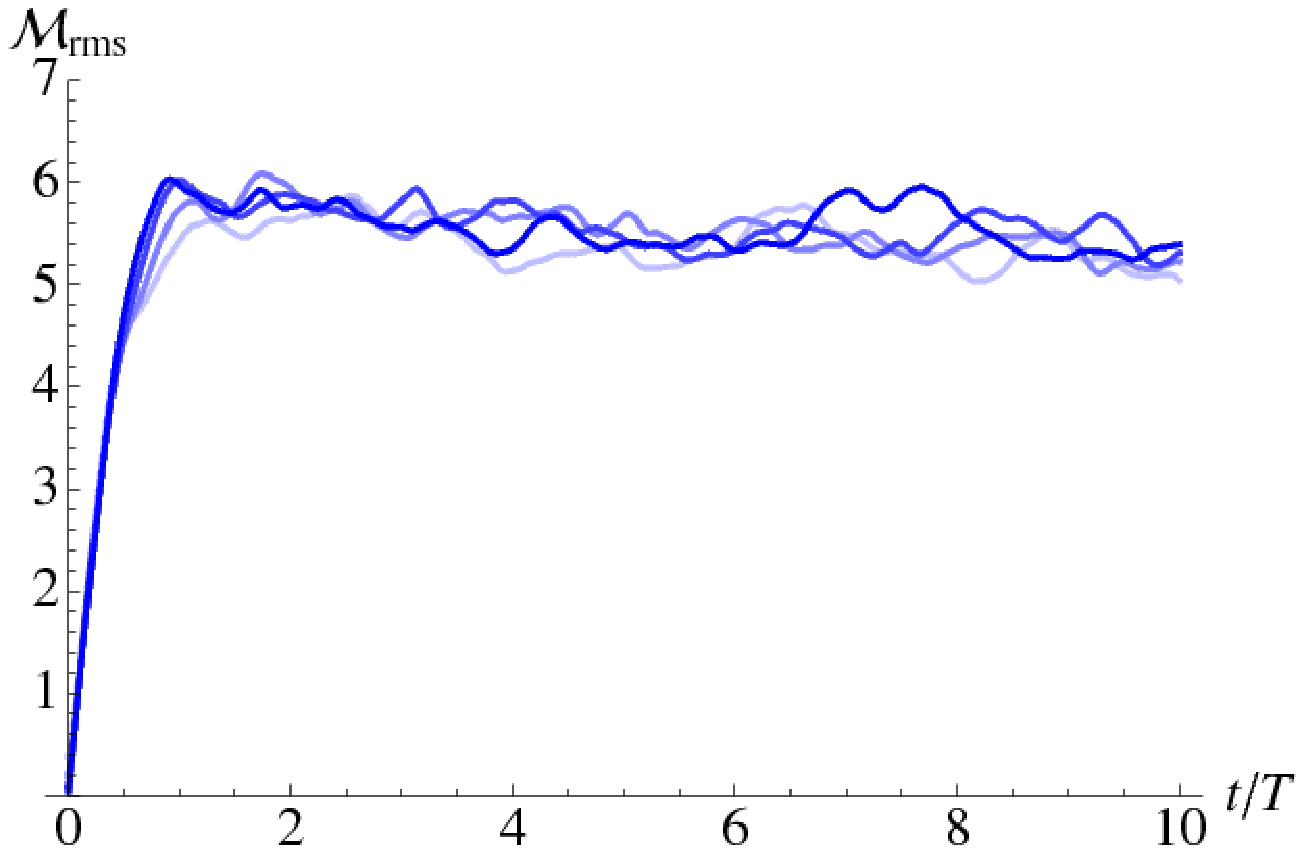} &
\includegraphics[width=80mm]{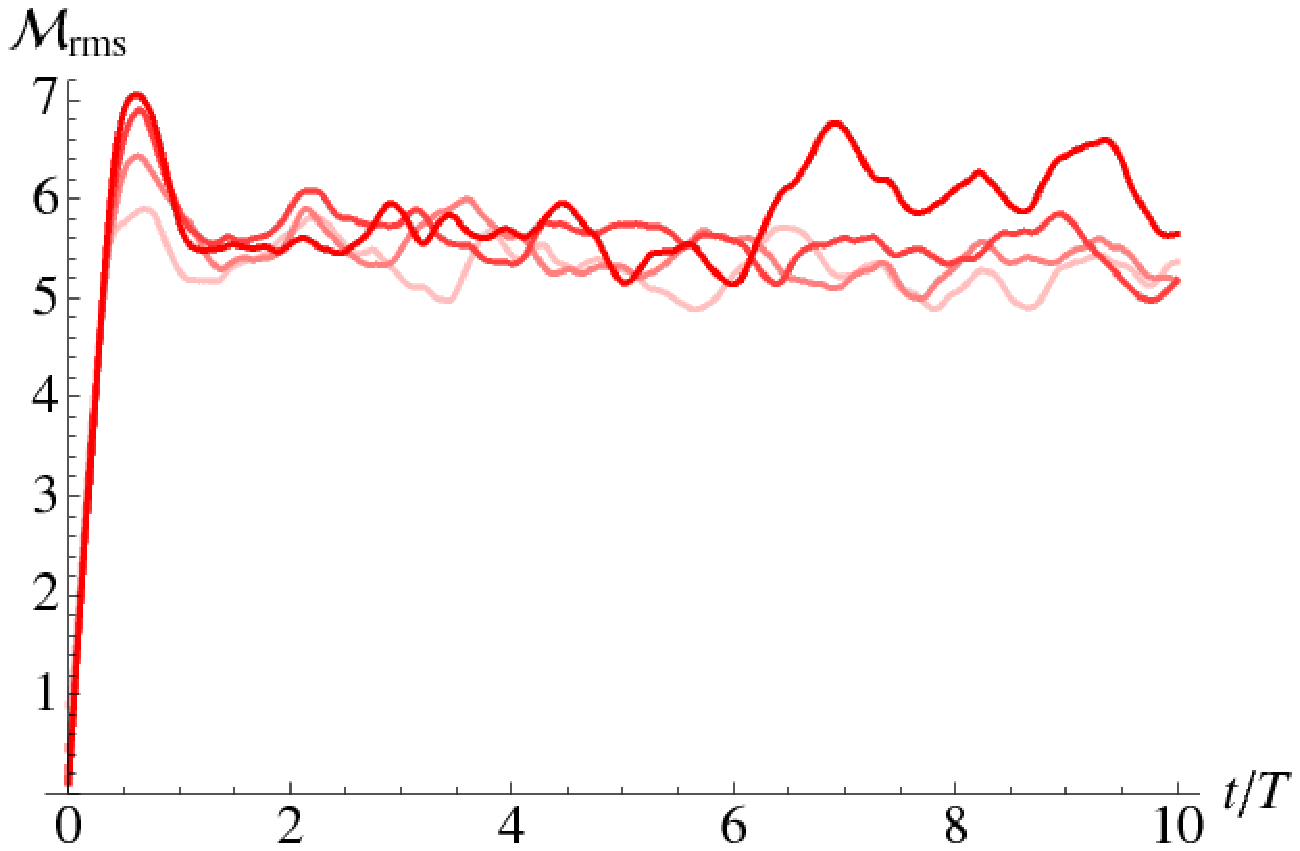}\\
\includegraphics[width=80mm]{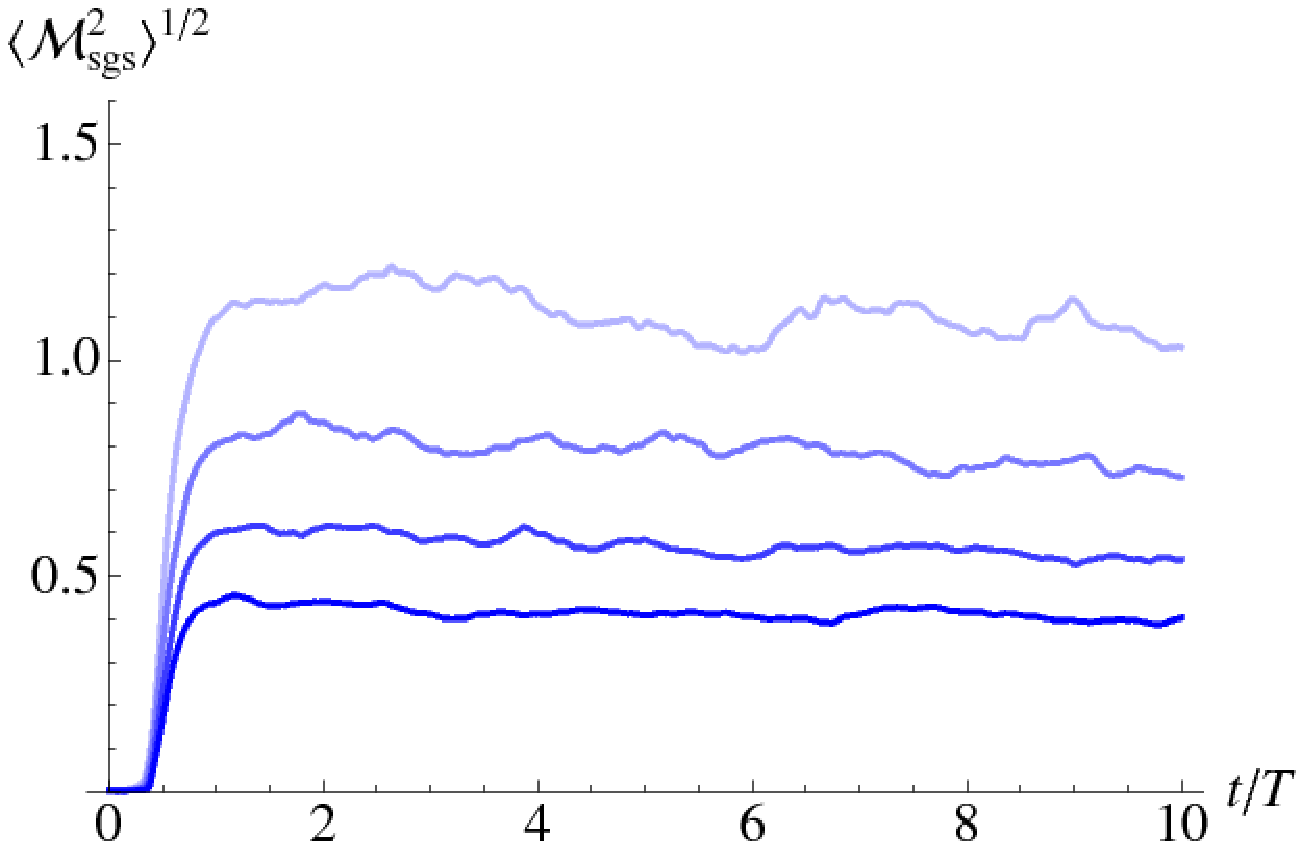} &
\includegraphics[width=80mm]{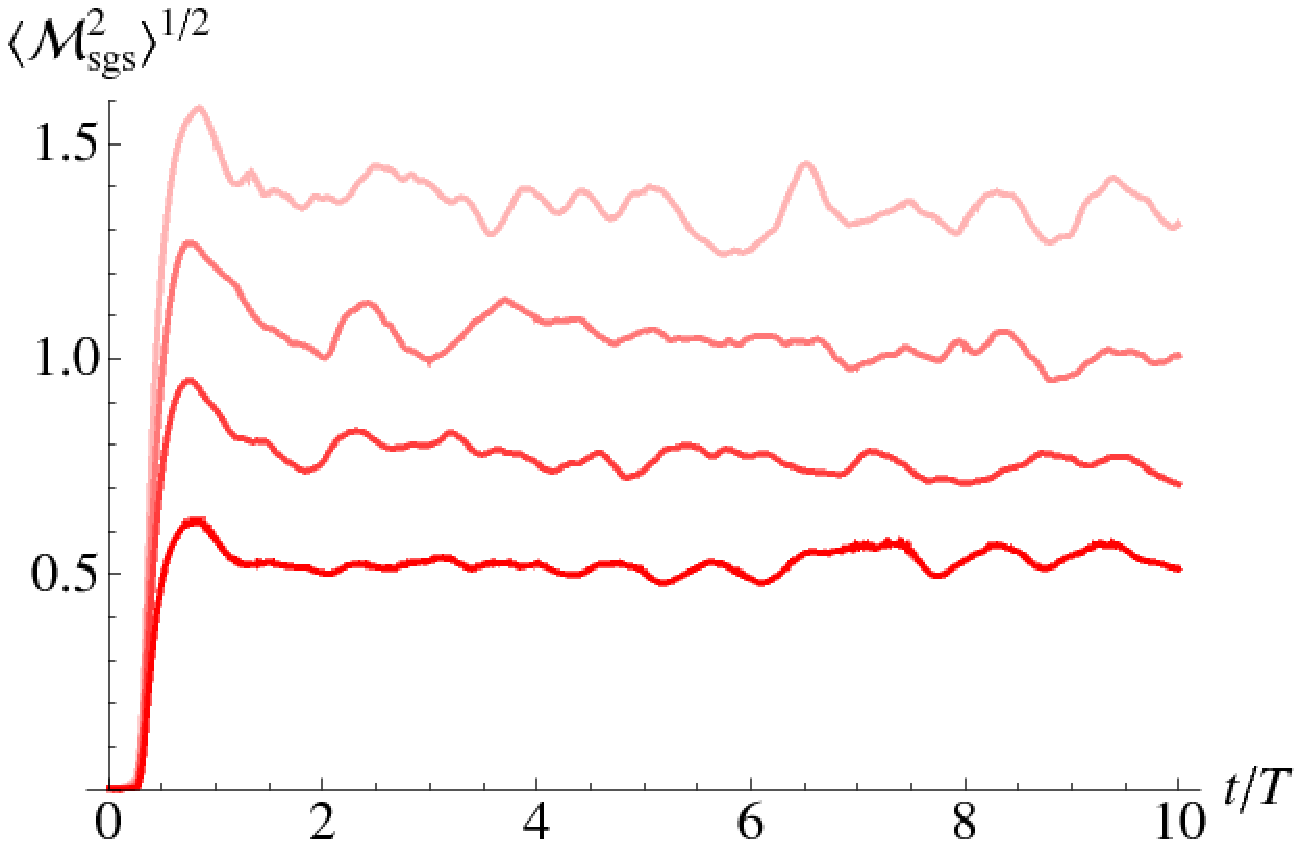}\\
\end{tabular}
\caption{Temporal evolution of the RMS Mach number (top) and the
	mean SGS turbulence Mach number (bottom) for solenoidal (left column) and compressive forcing (right column). The cutoff length $\Delta$ decreases from $L/32$ (light colour) to $L/256$ (full colour).}
\label{fig:stat_res_mach}
\end{figure*}

\begin{figure*}[t]
\subfigure[ SGS turbulence Mach number]{\includegraphics[width=80mm]{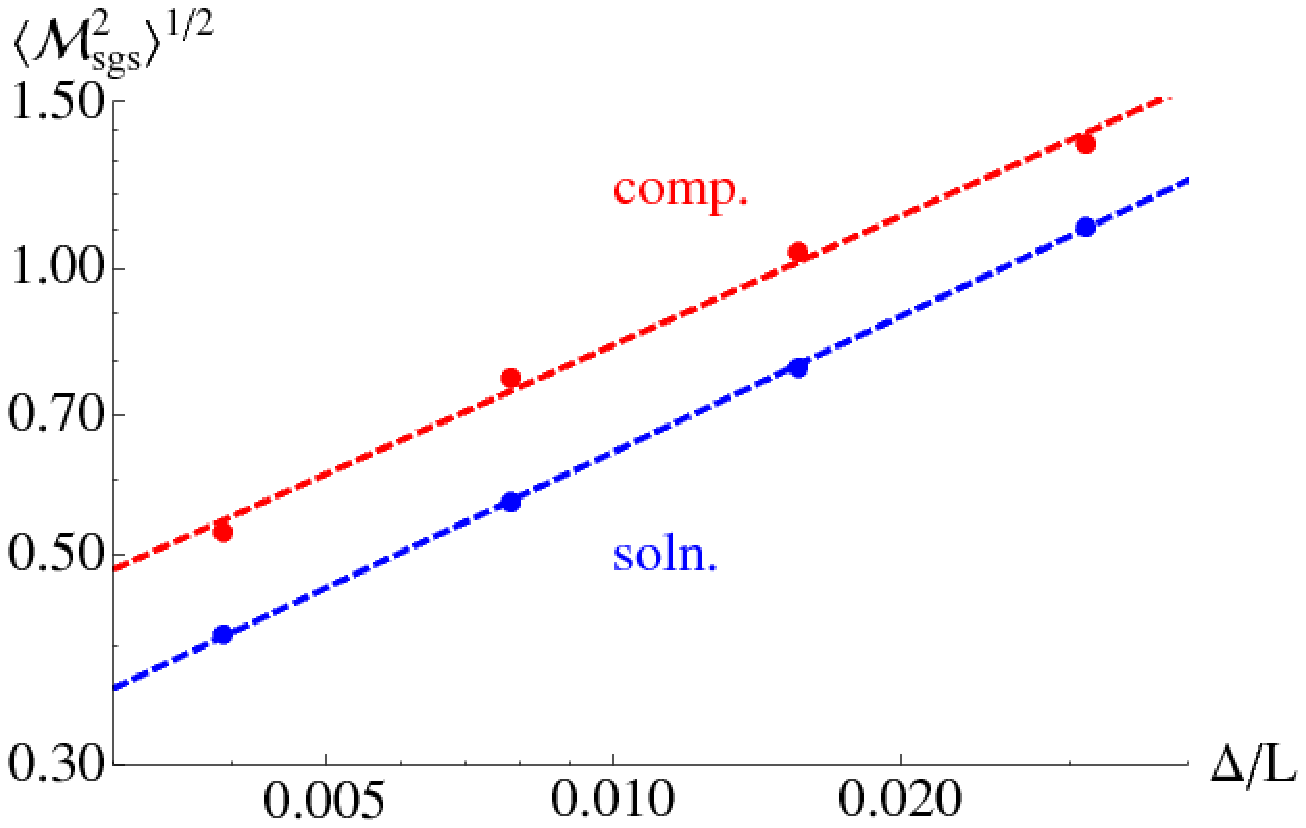}}
\subfigure[ SGS turbulence energy]{\includegraphics[width=80mm]{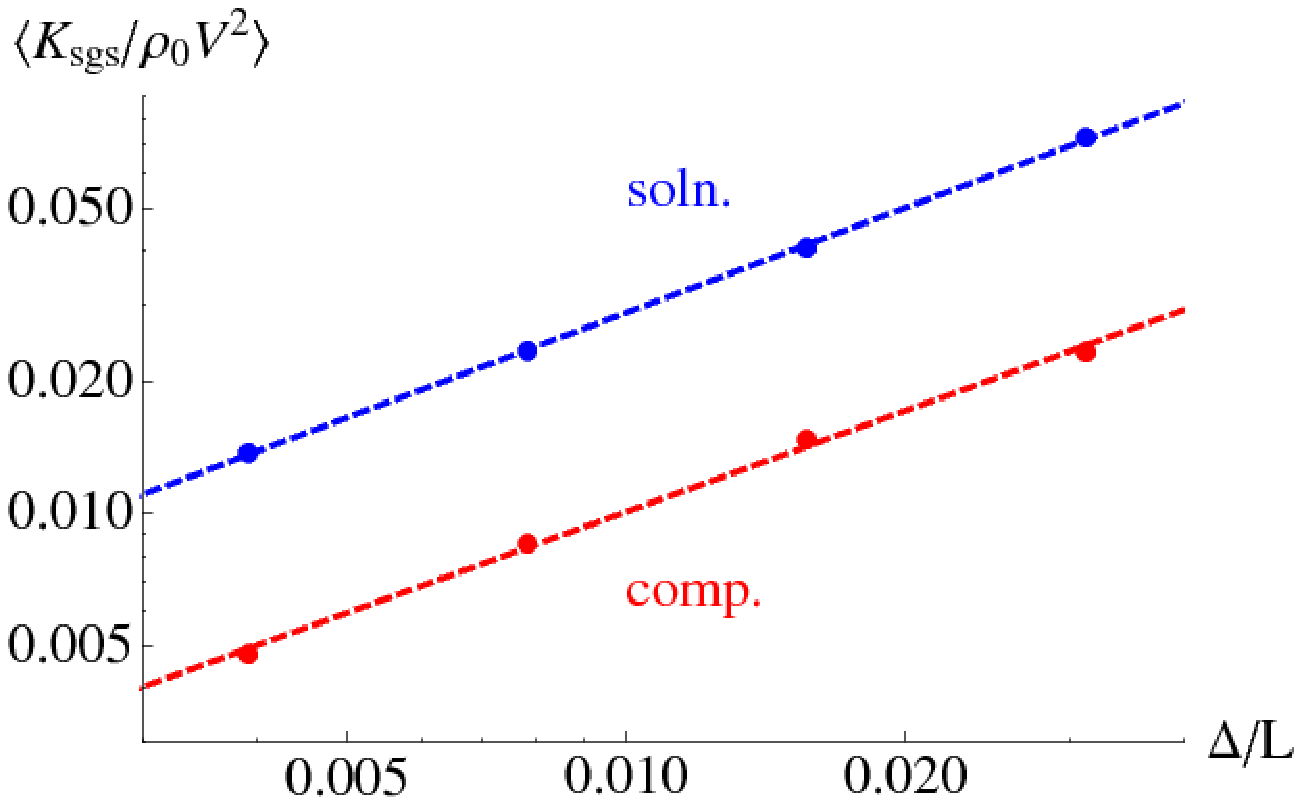}}\\
\caption{Scaling laws for the mean SGS turbulence Mach number (a) and energy (b) as functions
of the numerical resolution $\Delta$.}
\label{fig:scaling}
\end{figure*}

The SGS turbulence energy is given by the fluctuations of the velocity and density fields on length scales $\ell\lesssim\Delta$, as defined by Eqs.~(\ref{eq:tau_sgs_def}) and~(\ref{eq:trace_tau}). However, there is no obvious relation to the known scaling laws for turbulence, because the decomposition of the fluid dynamical variables cannot be related to the two-point statistics (structure functions) or the Fourier modes (energy spectra) in a straight-forward manner. To determine the scaling of SGS turbulence as a function of $\Delta$, we run several LES with $\Delta$ ranging from $L/256$ to $L/32$.
The mean values of the RMS Mach number and the SGS turbulence Mach number are plotted as functions of time in Fig.~\ref{fig:stat_res_mach}. The flow approaches a statistically stationary state after about 2 integral time scales \citep{SchmFeder09a,FederRom10}, for which $\mathcal{M}_{\rm rms}$ settles at values between 5 and 6. The temporal variation of $\mathcal{M}_{\rm rms}$ is caused by the stochastic forcing. As expected, $\langle\mathcal{M}_{\rm sgs}^2\rangle^{1/2}$ decreases with the cutoff scale. Averaging the spatial means from $t=2T$ to $10T$, we find the time-averaged mean values listed in Table~\ref{table:stat_res}. As one can see in Fig.~\ref{fig:scaling} (a), the time averages of $\langle\mathcal{M}_{\rm sgs}^2\rangle^{1/2}$ closely follow power laws, 
\begin{equation}
	\langle\mathcal{M}_{\rm sgs}^2\rangle^{1/2} \propto \Delta^{\alpha_{\mathcal{M}}},
\end{equation}
with $\alpha_{\mathcal{M}}=0.475\pm0.004$  for solenoidal and $0.451\pm0.026$ for compressive forcing.

\begin{figure*}[tbp]
\centering
\begin{tabular}{cc}
\includegraphics[width=80mm]{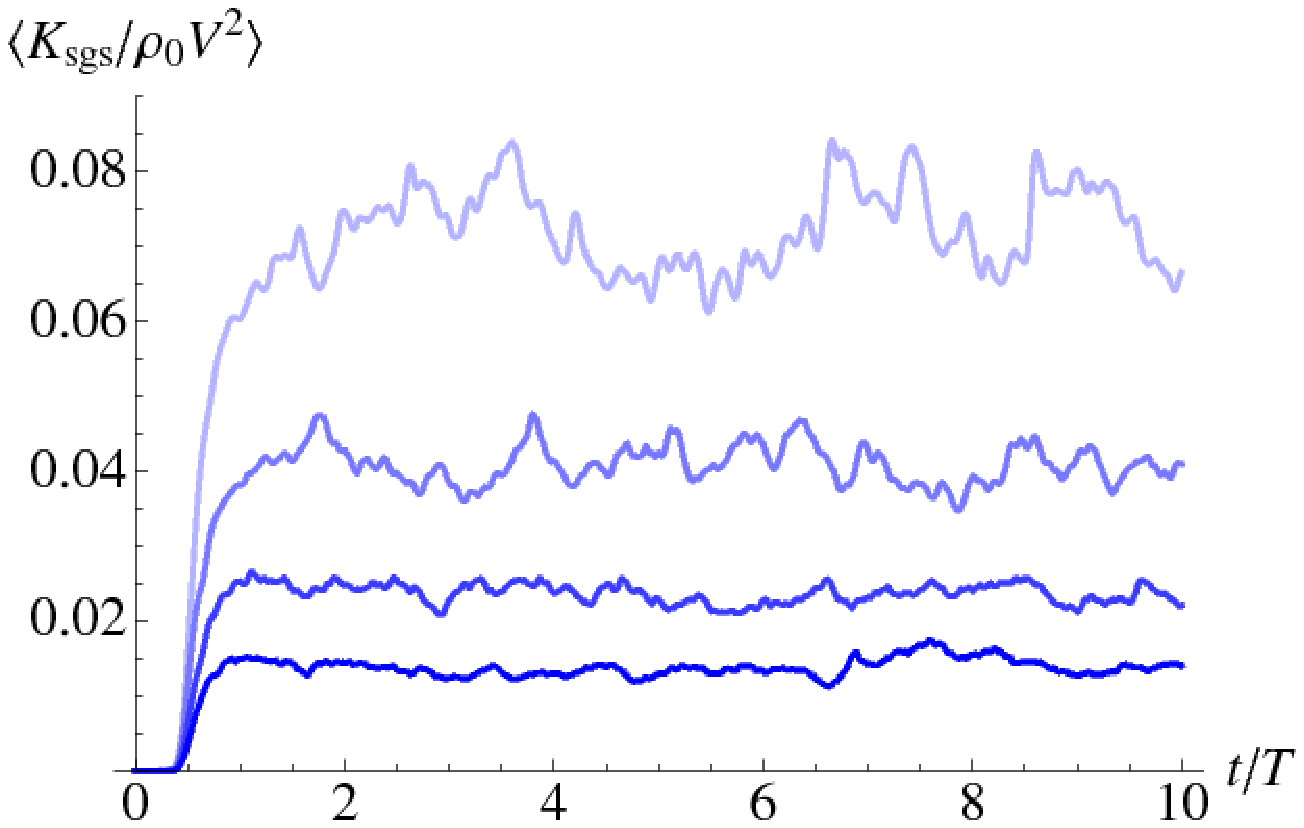} &
\includegraphics[width=80mm]{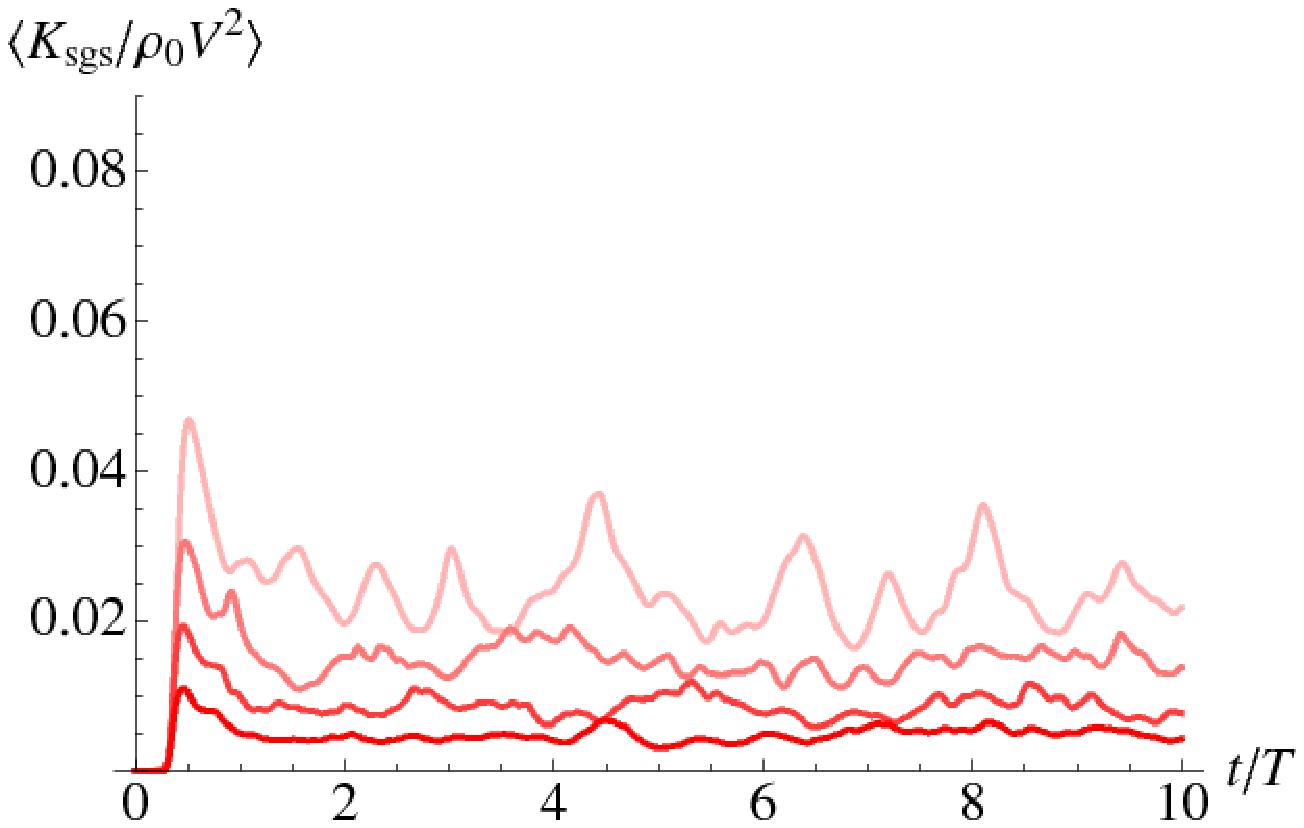}\\
\includegraphics[width=80mm]{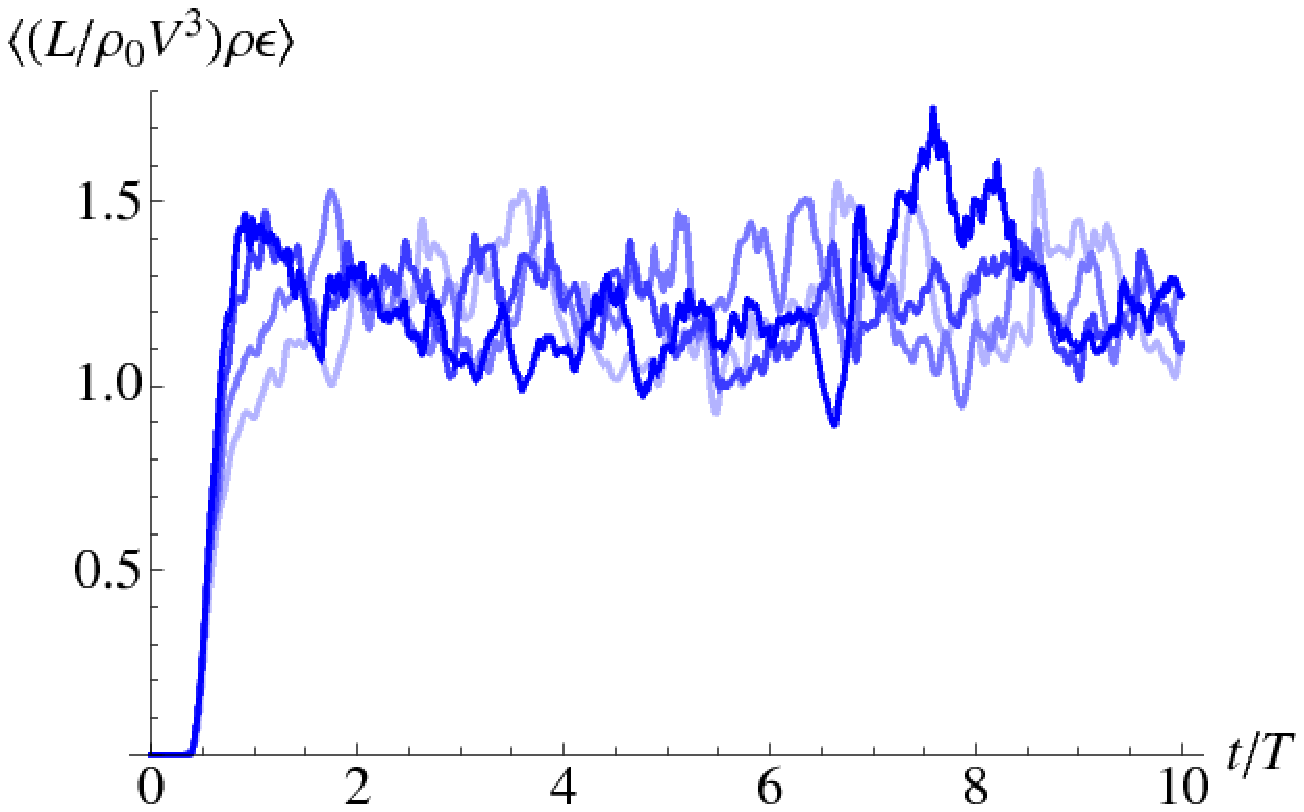} &
\includegraphics[width=80mm]{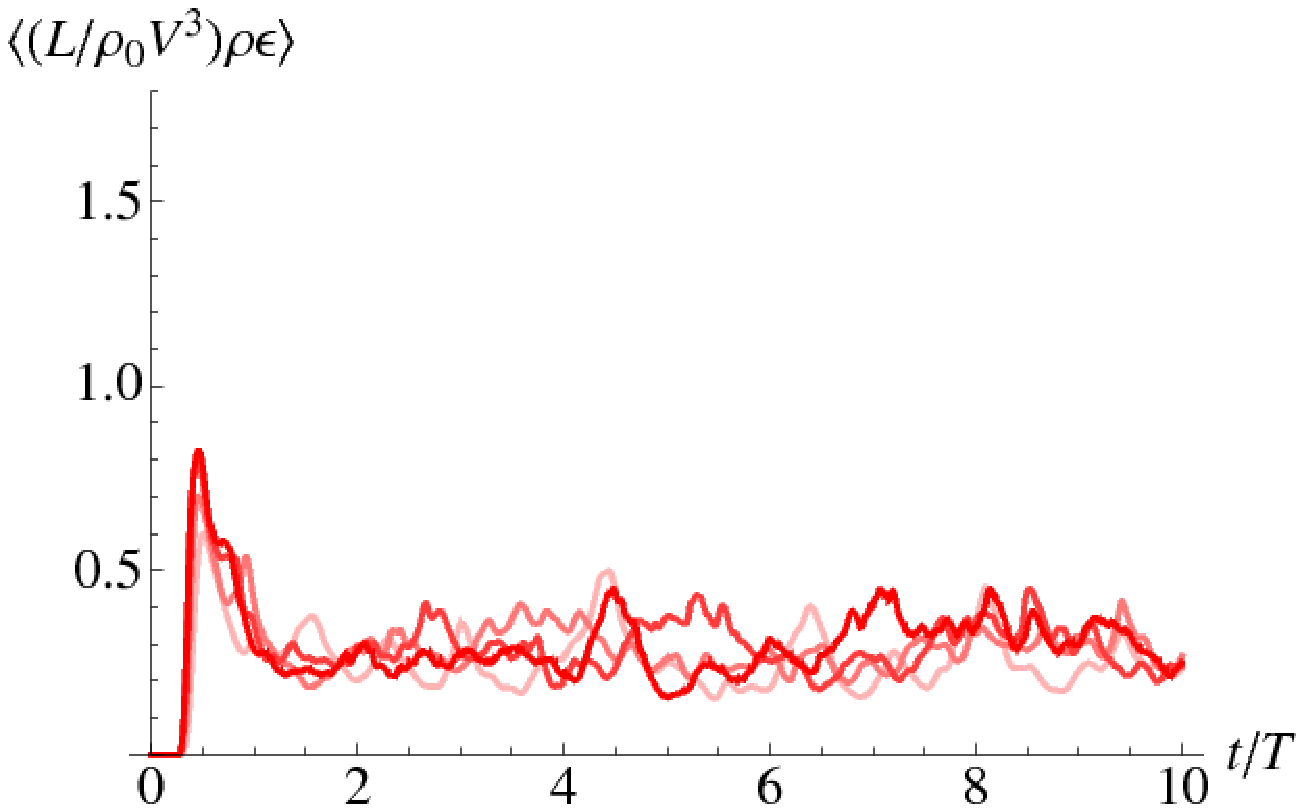}\\
\end{tabular}
\caption{Temporal evolution of the SGS turbulence energy (top) and the dissipation rate (bottom) for solenoidal (left column) and compressive forcing (right column). The cutoff length $\Delta$ decreases from $L/32$ (light colour) to $L/256$ (full colour).}
\label{fig:stat_res_energy}
\end{figure*}

The behaviour of the mean SGS turbulence energy is similar, although the intermittent fluctuations of $\langle K_{\rm sgs}\rangle$ are more pronounced in comparison to the mean SGS turbulence Mach number (see top panels of Fig.~\ref{fig:stat_res_energy}). The higher degree of intermittency stems from the mass density that is included in $K_{\rm sgs}$. For compressive forcing, $\langle K_{\rm sgs}\rangle$ is systematically lower in comparison to the LES with solenoidal forcing. This indicates that the total amount of energy in the turbulent structures on a given length scale is smaller in the compressive forcing case. The ratio of the mean values of $K_{\rm sgs}$ for compressive and solenoidal forcing in Table~\ref{table:stat_res} approximately agree with the ratio $0.38$ that is inferred from the filtered high-resolution data. On the other hand, the scaling laws 
\begin{equation}
	\langle K_{\rm sgs}\rangle \propto \Delta^{\alpha_{K}},
\end{equation}
are nearly the same for solenodial and compressive forcing (see Fig.~\ref{fig:scaling} (b)). We find
the slopes $\alpha_{K}=0.799\pm0.009$ and $0.769\pm0.029$, which agree within the error bars. This result is remarkable, because it suggests that the scaling properties of turbulence on small length scales are independent of the forcing.

\begin{table*}[t]
\caption{Time-averaged spatial mean values of various quantities and their standard deviations from the averages for different numerical resolutions.}
\label{table:stat_res}      
\centering                         
\begin{tabular}{rlcccc}        
\hline\hline                
N & $\Delta/L$ & $\mathcal{M}_{\rm rms}$ & $\langle\mathcal{M}_{\rm sgs}^2\rangle^{1/2}$ & 
$\langle K_{\rm sgs}\rangle/(\rho_0 V^2)$ & $(L/V^3)\langle\epsilon\rangle$  \\  
\hline                        
\multicolumn{6}{c}{solenodial forcing ($\zeta=1$)}\\
\hline
64  & 1/32 & 5.38 & $1.107 \pm 0.053$ & $0.0726 \pm 0.0055$ & $1.236 \pm 0.141$ \\
128 & 1/64 & 5.50 & $0.787 \pm 0.030$ & $0.0407 \pm 0.0025$ & $1.230 \pm 0.113$ \\
256 & 1/128 & 5.55 & $0.578 \pm 0.022$ & $0.0236 \pm 0.0013$ & $1.213 \pm 0.098$\\
512 & 1/256 & 5.52 & $0.412 \pm 0.012$ & $0.0138 \pm 0.0012$ & $1.219 \pm 0.159$\\
\hline                             
\multicolumn{6}{c}{compressive forcing ($\zeta=0$)}\\
\hline
64  & 1/32 & 5.29 & $1.353 \pm 0.049$ & $0.0235 \pm 0.0044$ & $0.253 \pm 0.072$\\
128 & 1/64 & 5.43 & $1.040 \pm 0.041$ & $0.0148 \pm 0.0018$ & $0.286 \pm 0.047$\\
256 & 1/128 & 5.57 & $0.767 \pm 0.029$ & $0.0086 \pm 0.0013$ & $0.293 \pm 0.061$\\
512 & 1/256 & 5.86 & $0.528 \pm 0.023$ & $0.0048 \pm 0.0008$ & $0.292 \pm 0.067$\\
\hline                                   
\end{tabular}
\end{table*}

As can be seen in the bottom panels of Fig.~\ref{fig:stat_res_energy}, the above scaling law of $\langle K_{\rm sgs}\rangle$ results in a mean dissipation rate $\langle\rho\epsilon\rangle$ that is independent of the cutoff scale, which is an essential property of the energy dissipation predicted by
the SGS model. The time-averaged mean values are listed in Table~\ref{table:stat_res}.
The significantly lower mean dissipation rate in the case of compressive forcing is consistent with the energy spectra of $\rho^{1/3}u$ \citep[see Fig.~A.1 in][]{FederRom10}. This mass-weighted velocity variable is related to the energy dissipation rate \citep{KritNor07}.
Moreover, Fig.~\ref{fig:adf_forcing} (right panel) shows that the growth of the mean internal energy in time becomes smaller as the weighing parameter $\zeta$ decreases from 1 (solenodial forcing) to 0 (compressive forcing). After subtracting the contribution from numerically resolved compression
effects \citep[see][]{SchmHille06}, we find that, independent of the cutoff length $\Delta$, about
3/4 of the change of the internal energy stems from SGS turbulence energy dissipation. The remainder is caused by numerical dissipation. This does not imply that the total rate of energy dissipation is much higher in LES compared to ILES, because the total energy dissipation is always determined by the energy injection due to the forcing. In conclusion, the greater part of kinetic energy is dissipated through the
SGS turbulence energy reservoir at a scale-free rate.

To quantify the relative importance of large values of $\mathcal{M}_{\rm sgs}$, we determine the volume fractions of cells with an SGS turbulence Mach number greater than a particular value. This fraction is given by $1-\mathrm{cdf}(\mathcal{M}_{\rm sgs})$, where $\mathrm{cdf}(\mathcal{M}_{\rm sgs})$ is the cumulative distribution function of $\mathcal{M}_{\rm sgs}$. In Figure~\ref{fig:adf_res}, the resulting functions are plotted for the LES with different cutoff lengths. As expected, the fraction with $\mathcal{M}_{\rm sgs}>1$ decreases with the cutoff length $\Delta$. However, the tails toward high $\mathcal{M}_{\rm sgs}$ demonstrate that even at relatively high resolution there are supersonic velocity fluctuations on unresolved length scales, and the corresponding turbulent pressure decreases only little with the cutoff scale.

For the lowest-resolution LES, we can compare the distribution of $\mathcal{M}_{\rm sgs}$ to the distribution inferred from the corresponding filtered $1024^{3}$ data (see Section~\ref{sc:test_single}). The filter length $\Delta_{\rm G}=16\Delta=L/32$ is equivalent to the cutoff length in the $64^{3}$ LES. Choosing yet a lower resolution of the LES, corresponding to a larger filter length for the ILES, turned out not to be feasible. Even for the LES with $\Delta/L=1/32$, the forcing range and the range of length scales that are directly affected by numerical dissipation overlap. For the filtering of the ILES, on the other hand, the filter length cannot be lowered (corresponding to a higher resolution of the LES), because the dynamical range of fluctuations between the grid scale and the filter length would become insufficient and the numerical smoothing would be too strong. Nevertheless, Fig.~\ref{fig:adf_res} (a) demonstrates that the distributions agree remarkably well is the case of solenoidal forcing. For compressive forcing, there are larger discrepancies.
However, given that Gaussian filtering corresponds only roughly to the implicit filter in an LES and that the SGS model is based on various approximations, the match is quite satisfactory. The larger deviations in the case of compressive forcing suggest that it is not possible to calibrate the SGS model coefficients in such a way that an optimal match is obtained both for solenoidal and for compressive forcing at the same time. 
The different shape of the distribution that is obtained from the high-resolution simulation with compressive forcing points toward a missing physical effect such as the pressure-dilatation, which is entirely neglected in our SGS model.
Anyhow, purely compressive forcing is a limiting case. In nature, some mixture of solenoidal and compressive forcing is more likely to occur. In Fig.~\ref{fig:adf_forcing} (right panel), we compare the distributions of $\mathcal{M}_{\rm sgs}$ for force fields with $\zeta$ varying from 1 (solenoidal) to 0 (compressive). High SGS turbulence Mach numbers become more frequent as the contribution of compressive forcing modes increases.

\subsection{Dependence on the Mach number}

\begin{figure*}[t]
\centering
\begin{tabular}{cc}
\includegraphics[width=80mm]{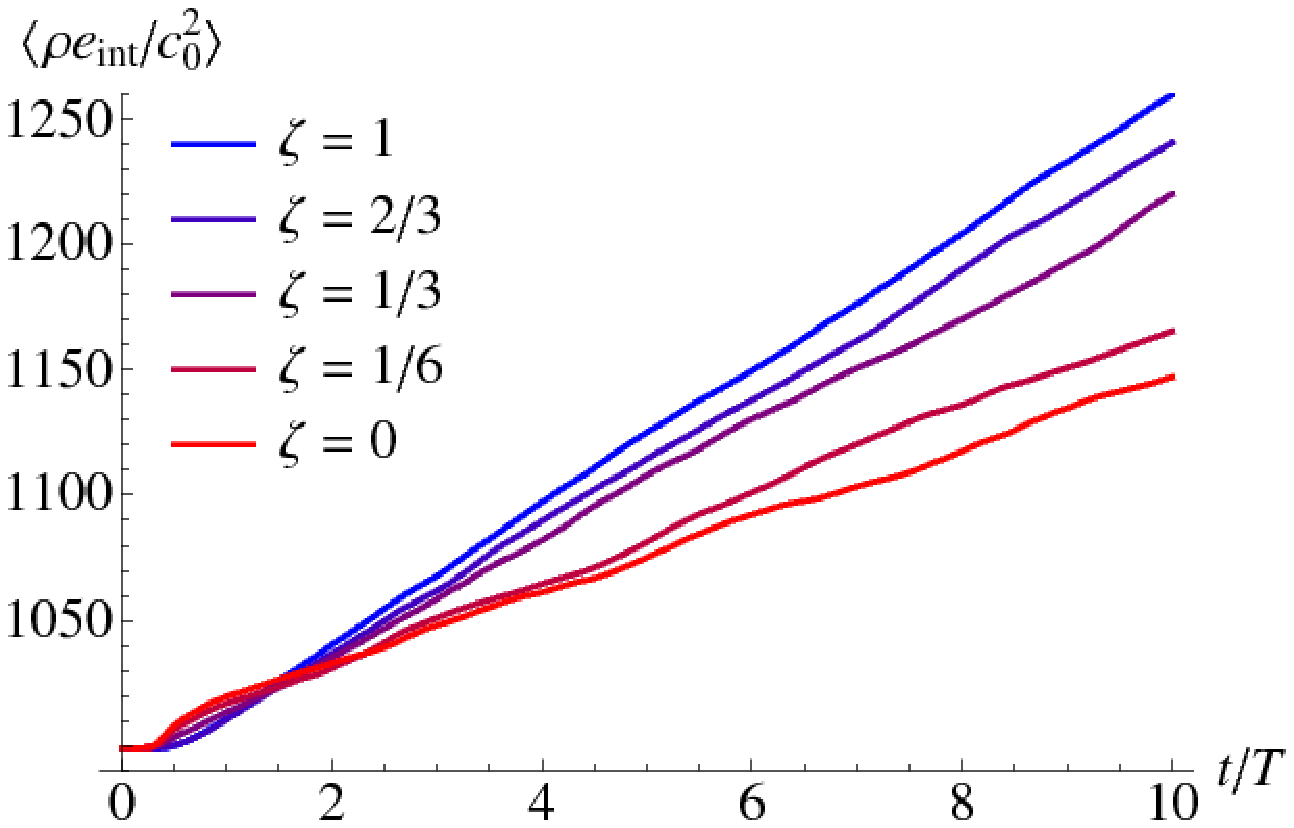} &
\includegraphics[width=80mm]{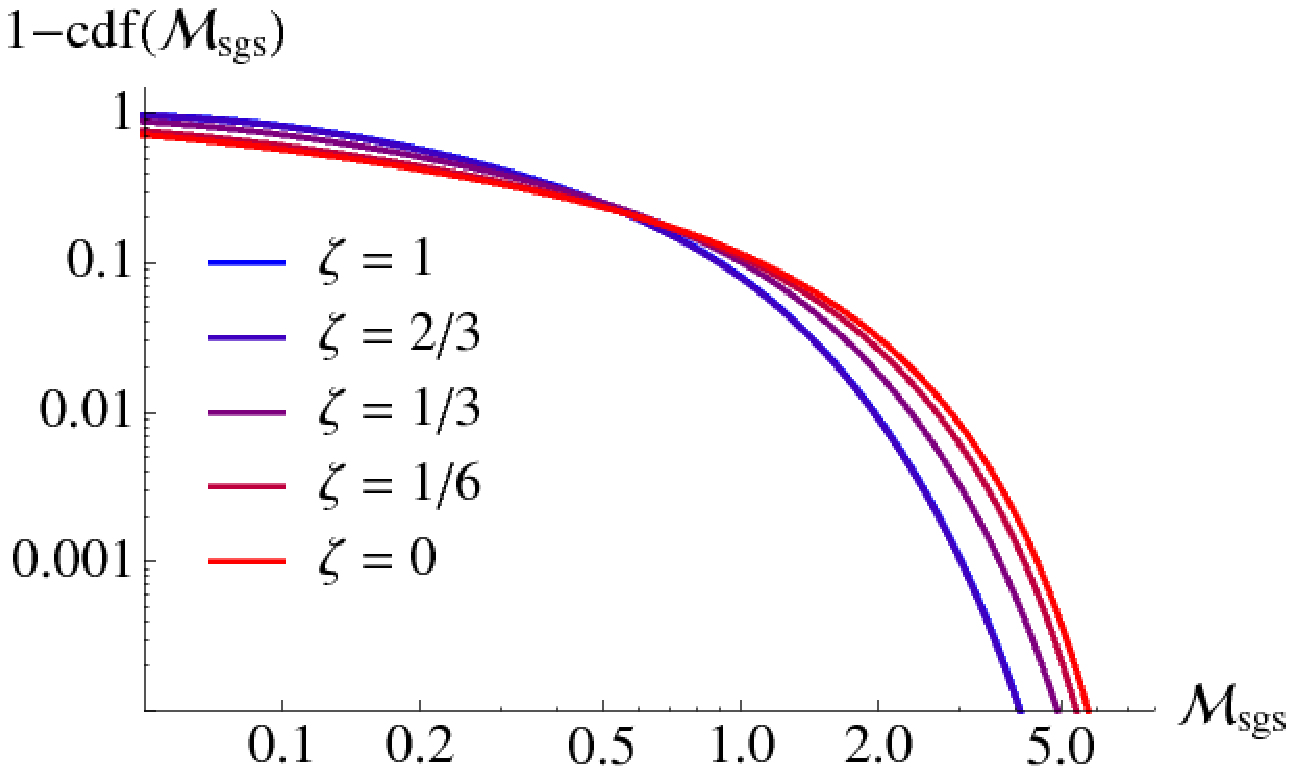}\\
\end{tabular}
\caption{Left panel: Time evolution of the mean internal energy for forcing with varying $\zeta$ and
about the same RMS Mach number. Right panel: Volume fractions of zones, in which the SGS turbulence Mach number is greater than a certain value for the same forcing parameters as in the left panel (the
curves for $\zeta=1$ and $\zeta=2/3$ almost coincide).}
\label{fig:adf_forcing}
\end{figure*}

\begin{figure*}[t]
\centering
\mbox{\subfigure[ solenoidal]{\includegraphics[width=80mm]{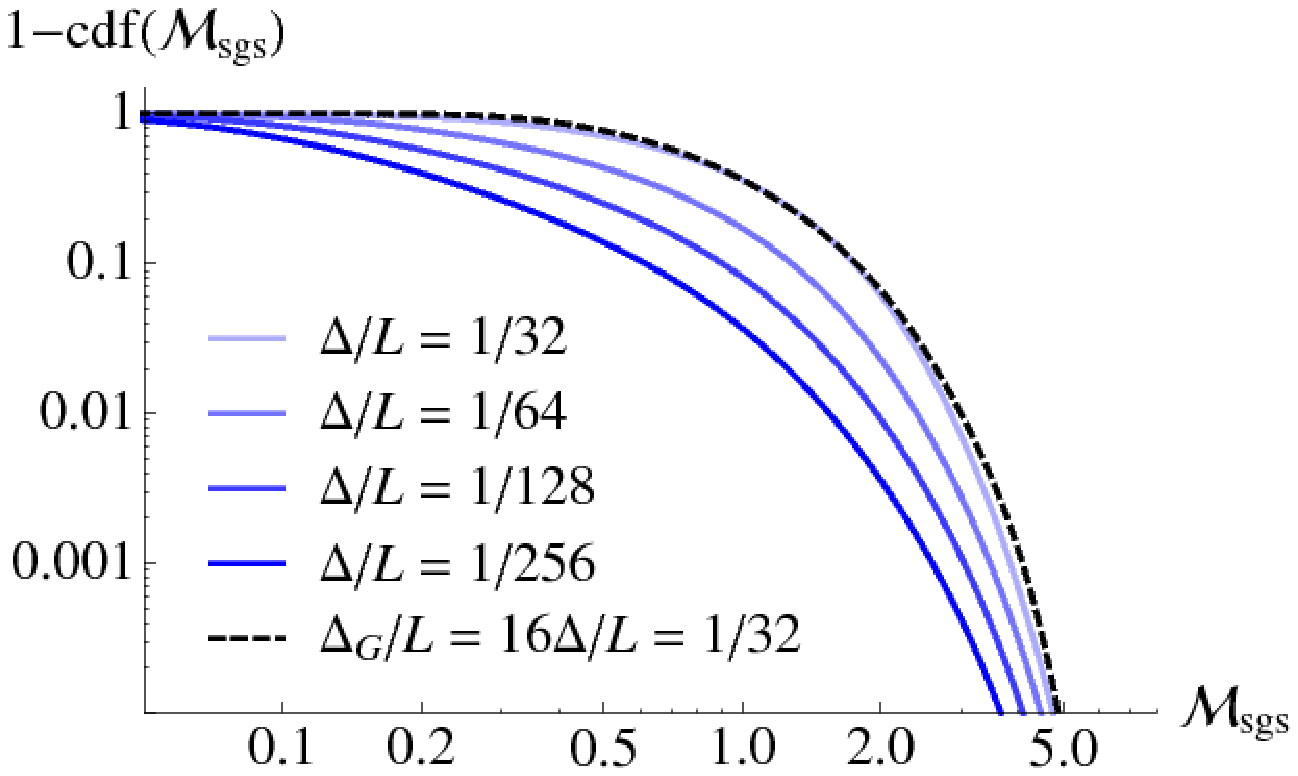}}\quad
\subfigure[ compressive]{\includegraphics[width=80mm]{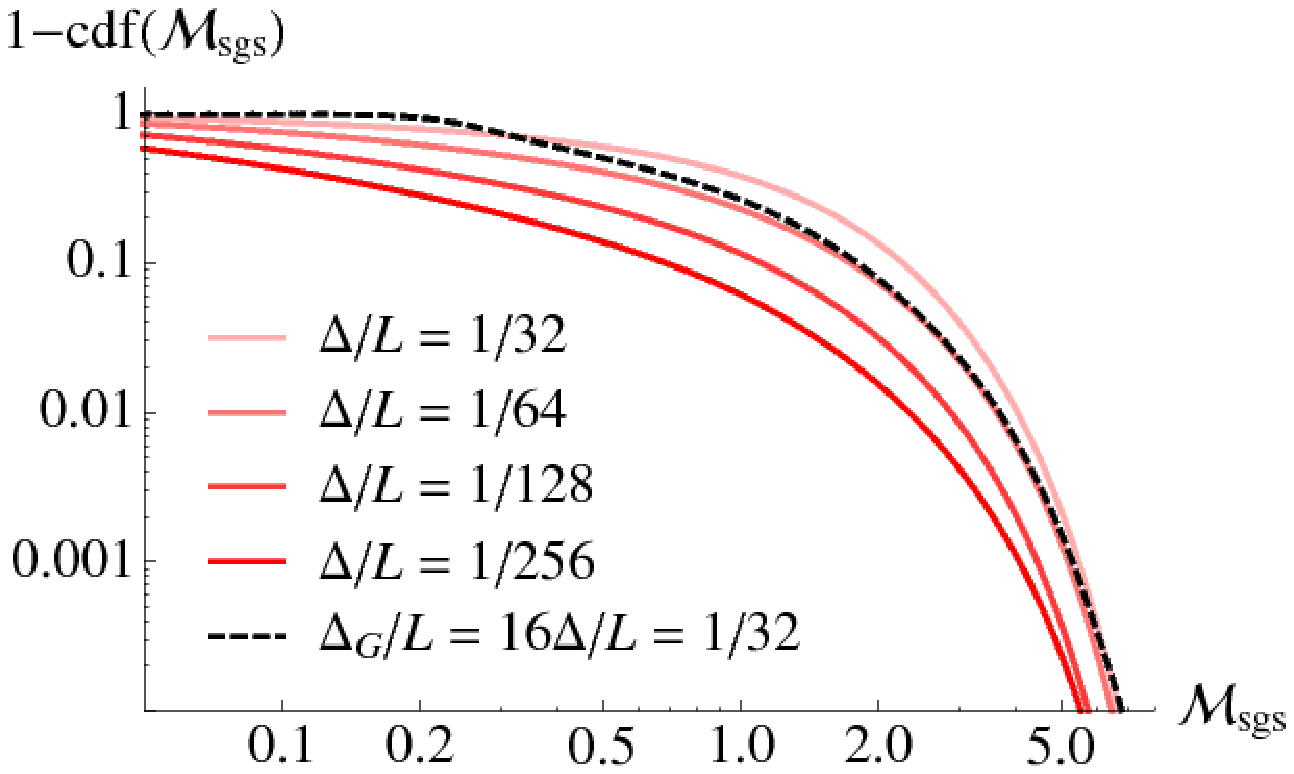}}}\\
\caption{Volume fractions of zones, in which the SGS turbulence Mach number is greater than
a certain value for different numerical resolutions. The dashed lines follows from the filtering of $1024^{3}$ ILES data with filter length $16\Delta$. This corresponds to the $64^{3}$ LES, for which $\Delta/L=1/32$.}
\label{fig:adf_res}
\end{figure*}

At fixed resolution, the SGS turbulence energy increases with the resolved kinetic energy of the flow. For isothermal turbulence, this also implies an increase of the SGS turbulence Mach number with rising RMS Mach number. To investigate this dependence, we varied the magnitude of mixed forcing with $\zeta=2/3$ and $\zeta=1/3$. In the case $\zeta=1/2$, a forcing field with two solenoidal and one longitudinal degrees of freedom is obtained. Fig.~8 in \citet{FederRom10} demonstrates that the ratio of the energy that is contained in transversal and longitudinal modes approaches a constant value for $\zeta>1/2$ \citep[see also][]{KritUst10}. Correspondingly, the cumulative distributions of $\mathcal{M}_{\rm sgs}$ plotted in Fig.~\ref{fig:adf_forcing} (b) show that there is almost no difference between forcing with $\zeta=2/3$ and purely solenoidal forcing ($\zeta=1$). On the other hand, one can see that there is a noticeable influence of compressive modes for $\zeta=1/3$, but the distribution differs from the purely compressive case ($\zeta=0$).

\begin{figure*}[t]
\centering
\begin{tabular}{cc}
\includegraphics[width=80mm]{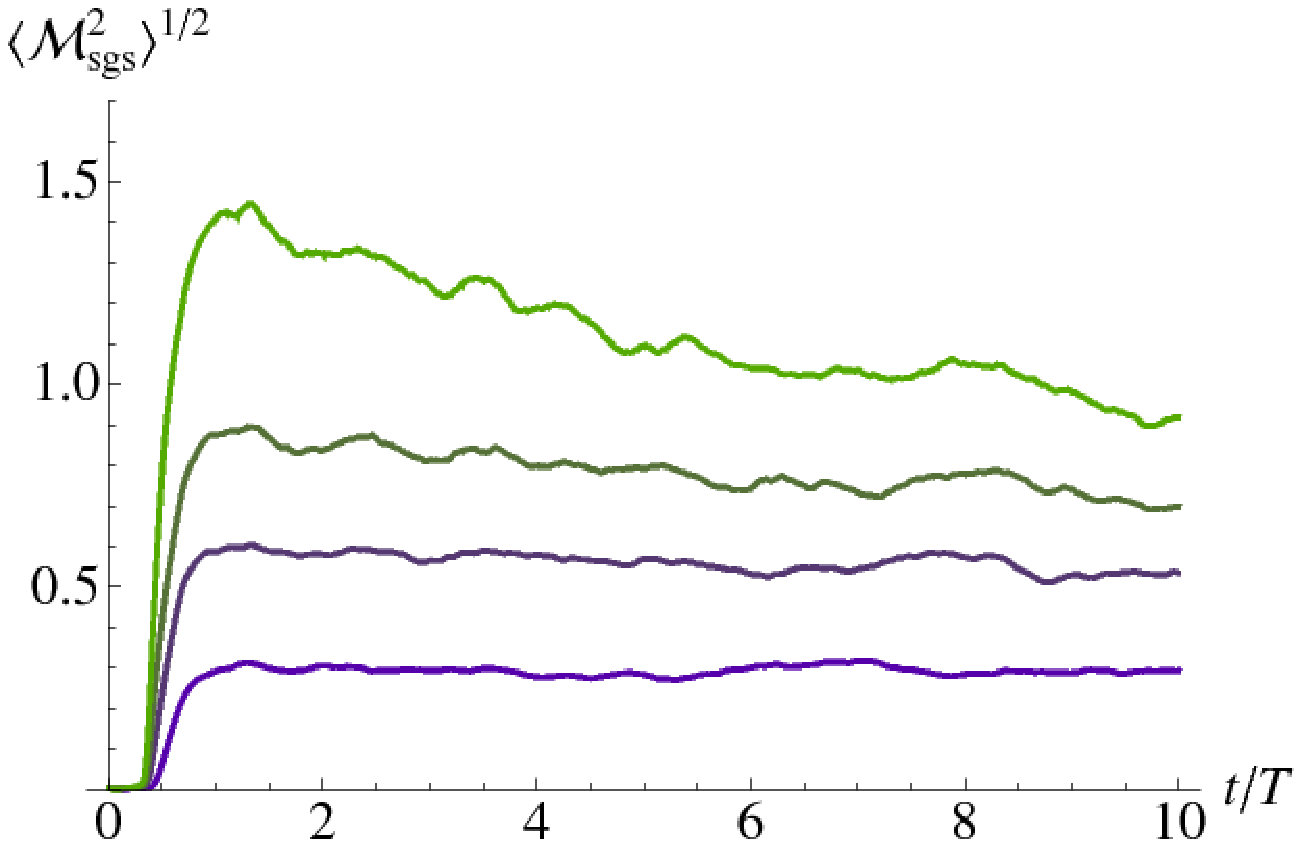} &
\includegraphics[width=80mm]{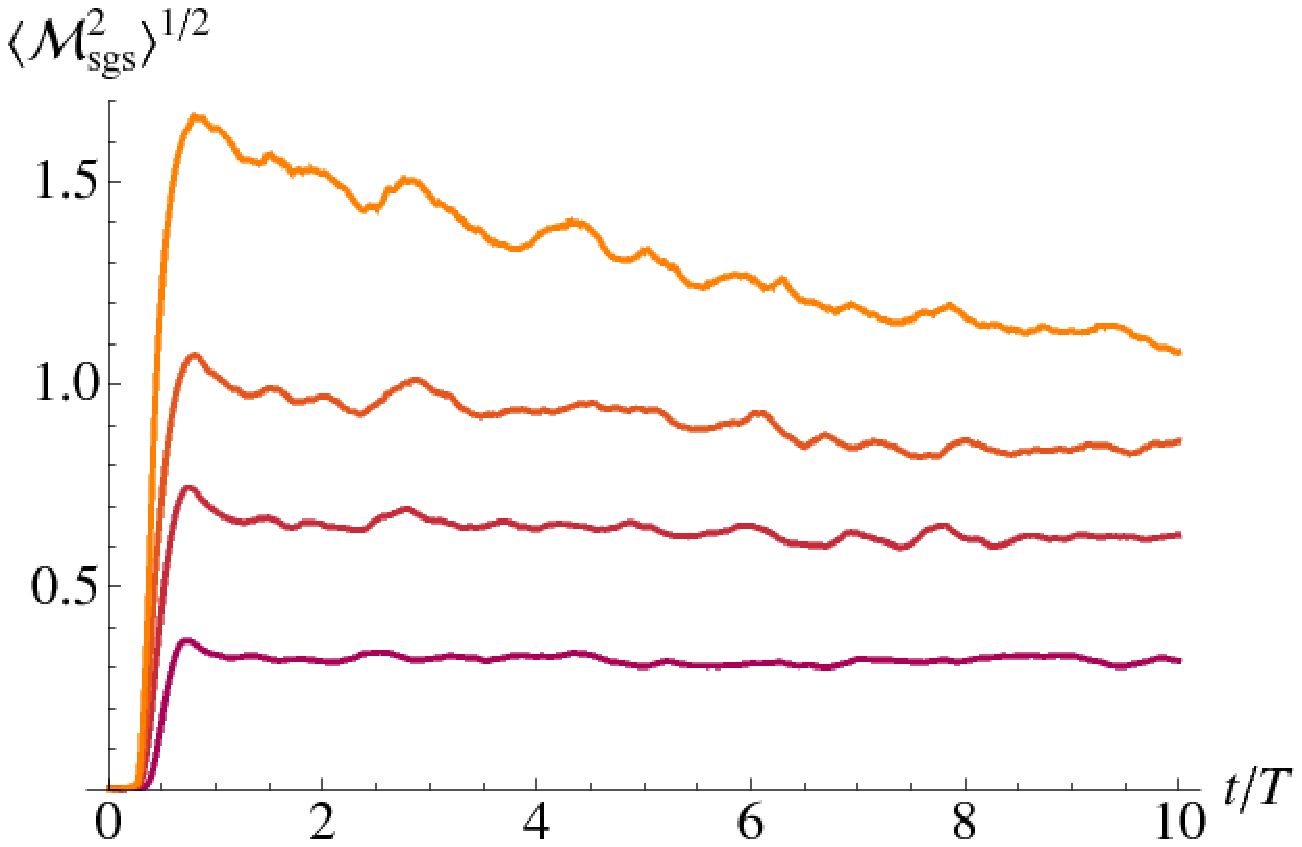}\\
\includegraphics[width=80mm]{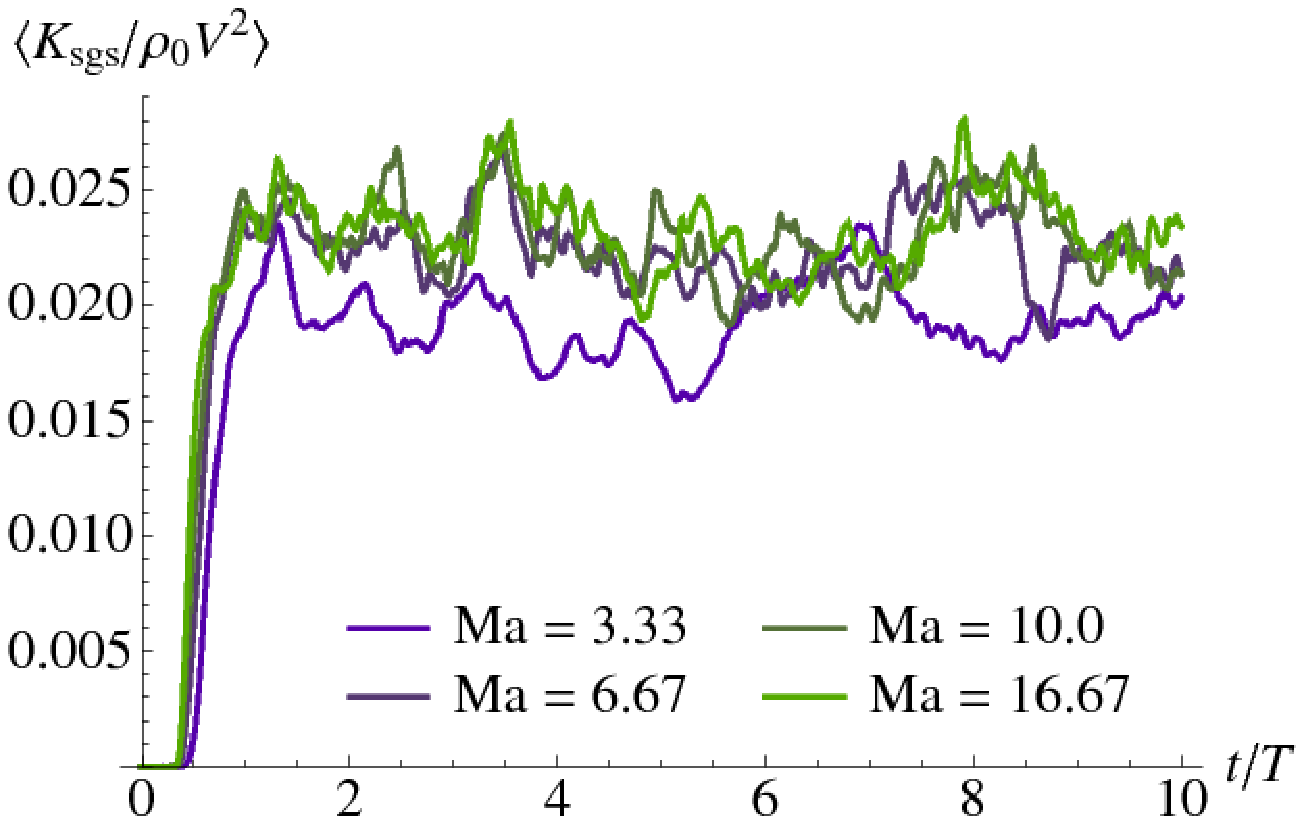} &
\includegraphics[width=80mm]{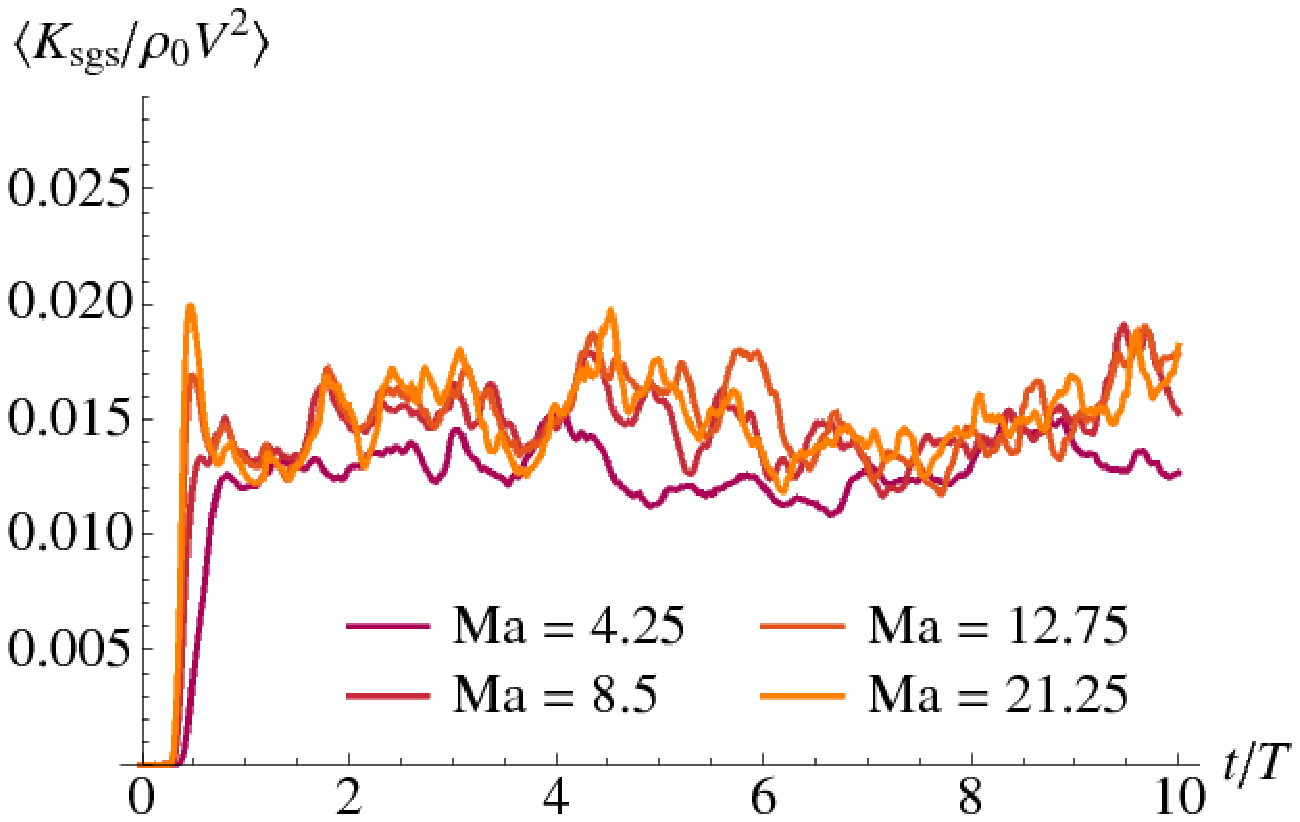}\\
\includegraphics[width=80mm]{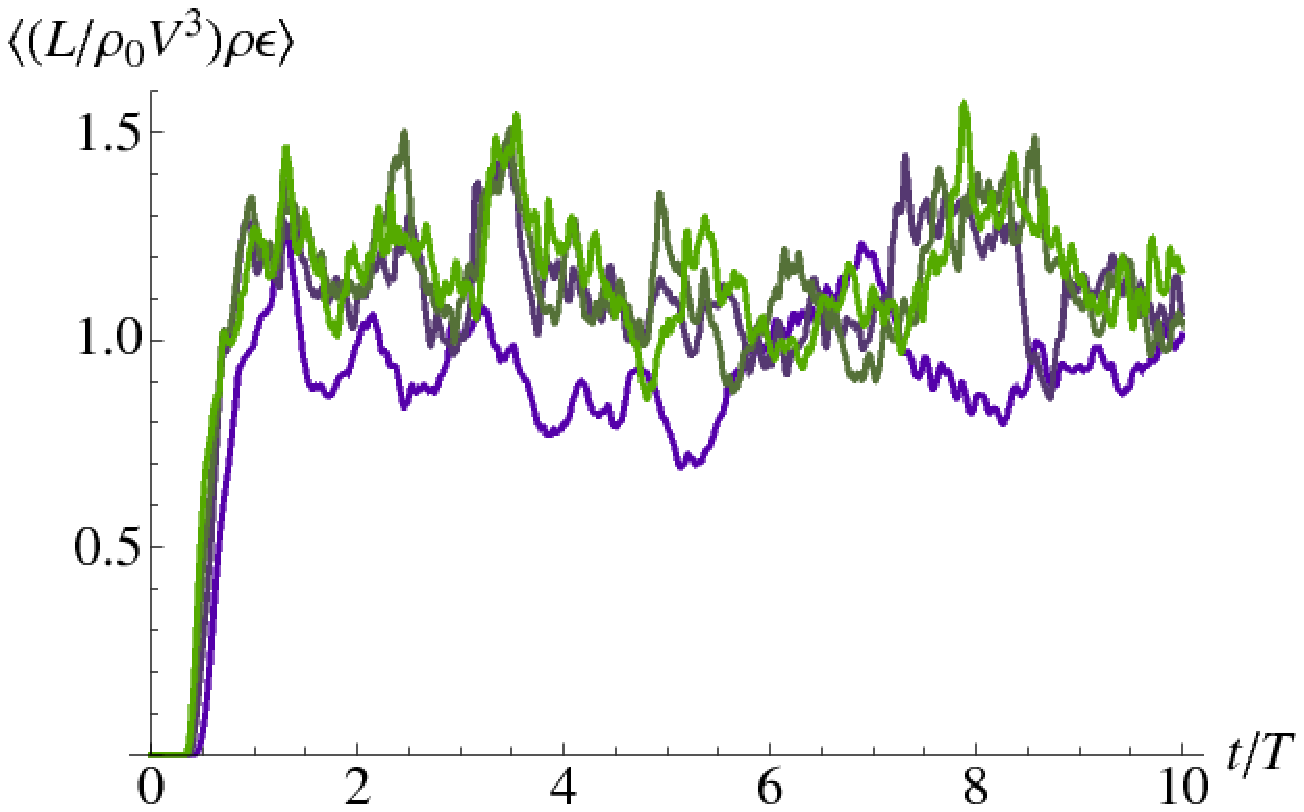} &
\includegraphics[width=80mm]{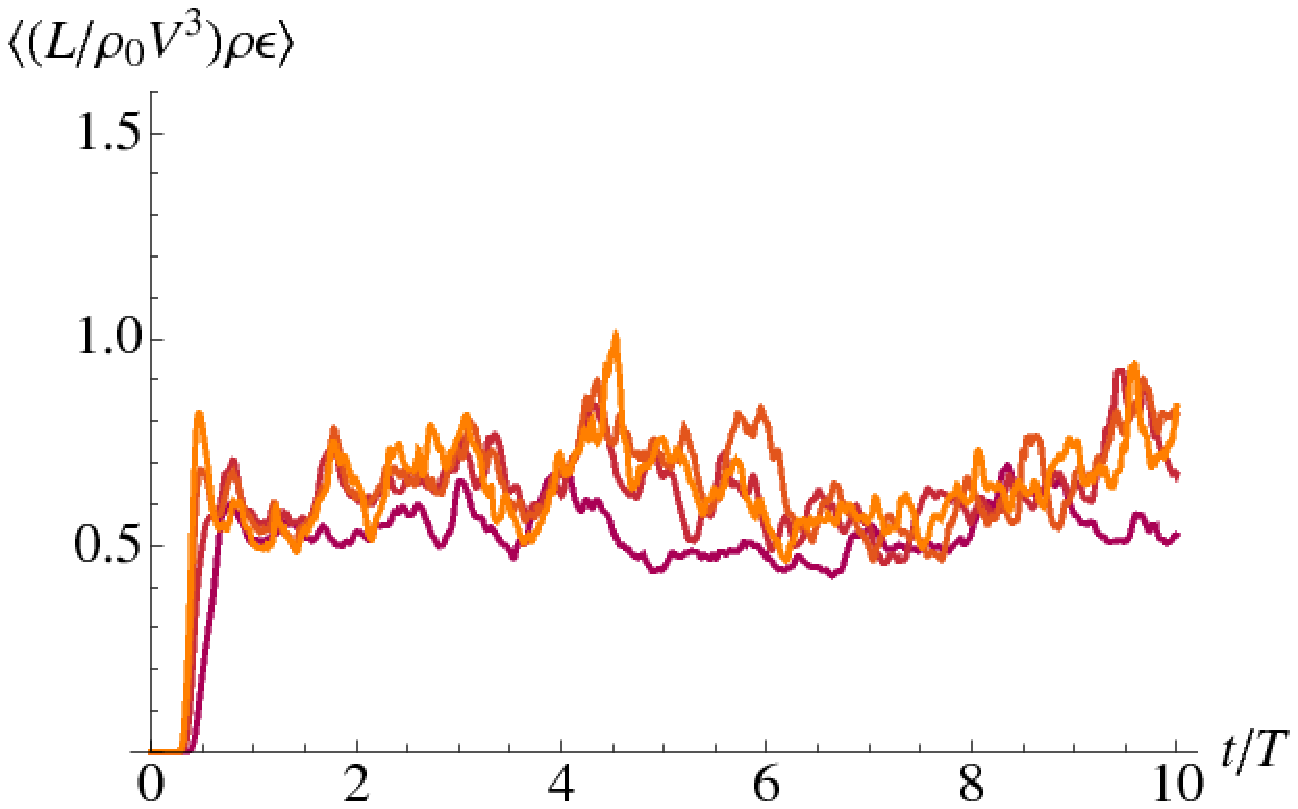}\\
\end{tabular}
\caption{Temporal evolution of the SGS turbulence Mach number (top) and energy (middle) and the dissipation rate (bottom) for $\zeta=2/3$ (left column) and $\zeta=1/3$ (right column). The different lines in each plot correspond to different forcing magnitudes (see Table~\ref{table:stat_mach}), which
are specified by the values of the characteristic Mach number $\mathrm{Ma}=V/c_{0}$ ($c_{0}$ is the initial speed of sound).}
\label{fig:stat_mach}
\end{figure*}

The dependence of the RMS SGS turbulence Mach number on the forcing amplitude is shown in the top panels of Fig.~\ref{fig:stat_mach} for $256^{3}$ LES. The corresponding characteristic Mach numbers and time-averaged statistics are listed in Table~\ref{table:stat_mach}. Independent of the Mach number, the ratio of $\langle\mathcal{M}_{\rm sgs}^2\rangle^{1/2}$ to $\mathcal{M}_{\rm rms}$ is nearly equal for $\zeta=2/3$ and $1/3$. This ratio is 0.102 for purely solenoidal forcing ($\zeta=1$), and 0.138 for purely compressive forcing ($\zeta=0$). Consequently, $\langle\mathcal{M}_{\rm sgs}^2\rangle^{1/2}/\mathcal{M}_{\rm rms}$ is mostly determined by the grid scale $\Delta$, except for small $\zeta$. As one can see in the middle panels of Fig.~\ref{fig:stat_mach}, the normalized mean SGS turbulence energy, $\langle K_{\rm sgs}\rangle/\rho_{0}V^{2}$, is  about the same for the different Mach numbers, with a weak trend to decrease toward low Mach numbers (see also Table~\ref{table:stat_mach}). The same behavior is found for the mean dissipation rate (Fig.~\ref{fig:stat_mach}, bottom panels), which further supports the validity of the SGS model in the supersonic regime. Following the trend discussed in Sect~\ref{sc:cutoff}, there is clearly an influence of the mixture of solenoidal and compressive modes in the forcing.

\begin{figure*}[t]
\centering
\mbox{\subfigure[ $\zeta=2/3$]{\includegraphics[width=80mm]{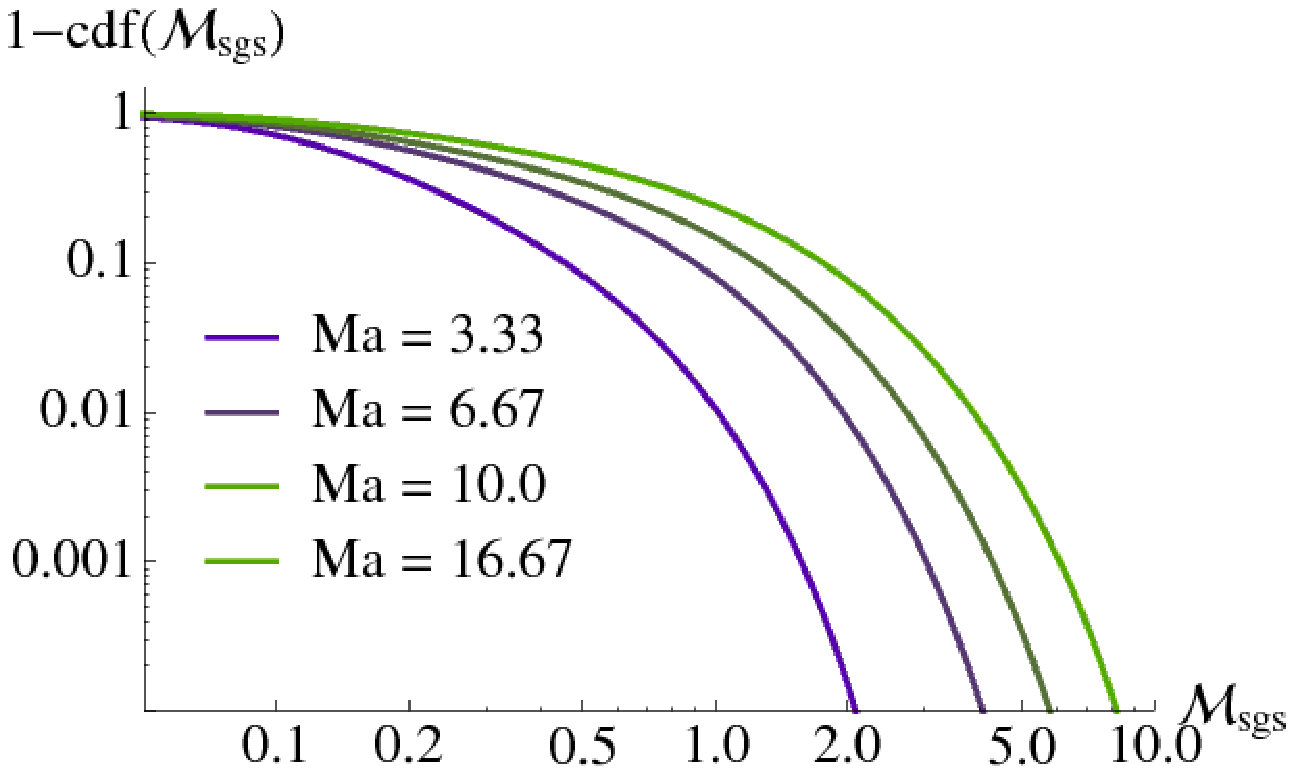}}\quad
\subfigure[ $\zeta=1/3$]{\includegraphics[width=80mm]{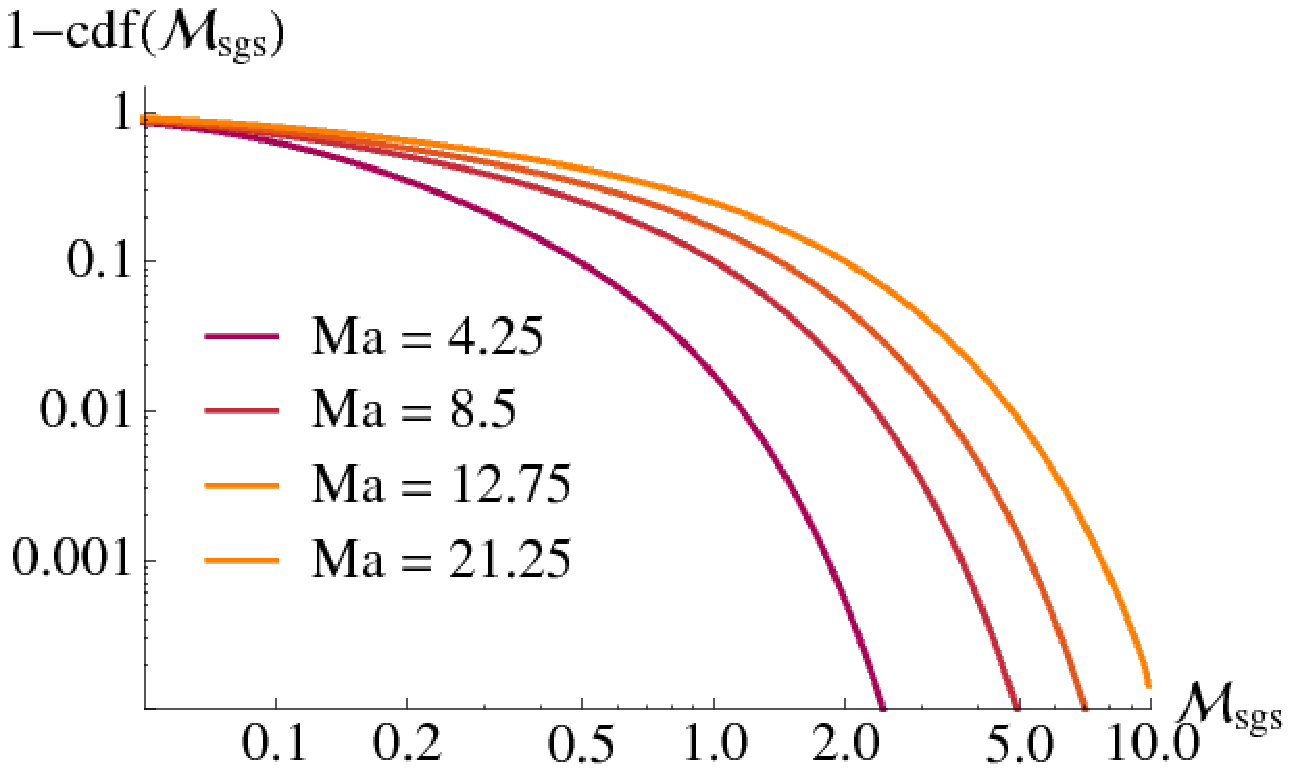}}}\\
\caption{Volume fractions of zones, in which the SGS turbulence Mach number is greater than
a certain value for different RMS Mach numbers and mixtures of solenoidal and compressive modes,
as in Fig.~\ref{fig:stat_mach}.}
\label{fig:adf_mach}
\end{figure*}

Fig.~\ref{fig:adf_mach} shows the distributions of $\mathcal{M}_{\rm sgs}$ for $\zeta=2/3$ (a) and $1/3$ (b), as explained in Section~\ref{sc:cutoff}. For both forcing types, the volume fractions with $\mathcal{M}_{\rm sgs}>1$ increase with the forcing magnitude. For RMS Mach numbers greater than $5$, supersonic turbulent velocity fluctuations at the cutoff scale fill more then $10\,\%$ of the total volume. If the ratio between the integral scale of turbulence and the cutoff scale is smaller, the volume filling factor increases further.

\begin{table*}[t]
\caption{Time-averaged spatial mean values of various quantities and their standard deviations from the averages for different forcing magnitudes
(defined by $\mathrm{Ma}=V/c_{0}$) and mixtures of solenoidal and compressive modes.}
\label{table:stat_mach}      
\centering                         
\begin{tabular}{rrcccc}        
\hline\hline                
$\mathrm{Ma}$ & $\mathcal{M}_{\rm rms}$ & $\langle\mathcal{M}_{\rm sgs}^2\rangle^{1/2}$ & 
$\langle\mathcal{M}_{\rm sgs}^2\rangle^{1/2}/\mathcal{M}_{\rm rms}$ &
$\langle K_{\rm sgs}\rangle/\rho_0 V^2$ & $(L/V^3)\langle\epsilon\rangle$  \\  
\hline                        
\multicolumn{6}{c}{$\zeta=2/3$}\\
\hline
3.33 & 2.91 & $0.293 \pm 0.010$ & 0.101 & $0.0193 \pm 0.0016$ & $0.933 \pm 0.109$ \\
6.67 & 5.38 & $0.559 \pm 0.021$ & 0.104 & $0.0223 \pm 0.0016$ & $1.125 \pm 0.123$\\
10.00 & 7.54 & $0.778 \pm 0.044$ & 0.103 & $0.0229 \pm 0.0018$ & $1.150 \pm 0.142$\\
16.67 & 10.49 & $1.097 \pm 0.117$ & 0.105 & $0.0232 \pm 0.0017$ & $1.168 \pm 0.130$\\
\hline                             
\multicolumn{6}{c}{$\zeta=1/3$}\\
\hline
4.25 & 2.96 & $0.320 \pm 0.009$ & 0.108 & $0.0129 \pm 0.0011$ & $0.535 \pm 0.064$\\
8.50  & 5.73 & $0.636 \pm 0.020$ & 0.111 & $0.0149 \pm 0.0015$ & $0.641 \pm 0.089$\\
12.75 & 8.14 & $0.896 \pm 0.052$ & 0.110 & $0.0153 \pm 0.0017$ & $0.666 \pm 0.101$\\
21.25 & 11.81 & $1.268 \pm 0.123$ & 0.107 & $0.0152 \pm 0.0015$ & $0.654 \pm 0.094$\\
\hline                                   
\end{tabular}
\end{table*}

\section{Conclusion}

Formulating a mixed closure for the flux of energy from the numerically resolved to the unresolved scales, we have generalized the subgrid-scale (SGS) turbulence energy model to the regime of highly compressible turbulence. This closure is based on ideas of \citet{WoodPort06} and features a non-linear term in addition to a linear eddy-viscosity term. In general, the turbulence energy cascade is an important source of SGS turbulence energy production in turbulent flows, and it should not be neglected even if other sub-resolution sources are emphasized in particular astrophysical applications \citep[e.~g., ][]{ScannBruegg08,Joung09}. Our proposed closure for the transfer of energy by the turbulent cascade complements theses models. 
We verified this closure by means of explicit filtering of high-resolution data from various simulations of supersonic isothermal and adiabatic turbulence \citep{SchmNie07,SchmFeder09a,FederRom10}. Tests in large eddy simulations of forced supersonic turbulence show that the SGS model meets several important requirements that should be satisfied by any sound SGS model:
\begin{itemize}
\item For statistically stationary turbulence, an equilibrium between the production and the dissipation of SGS turbulence is reached. The mean SGS turbulence energy depends on the grid scale via a power law.
\item The SGS turbulence Mach number, which specifies the importance of the turbulent pressure on the grid scale relative to the thermal pressure of the gas, depends linearly on the RMS Mach number of the resolved turbulence.
\item The SGS turbulence energy dissipation is independent of the grid scale.
\end{itemize}
Forced turbulence simulations in a periodic box are most suitable to test the properties listed above, because of the well defined statistics of the isotropic, homogeneous and stationary turbulence that is produced. In addition to these properties, we found a dependence of the SGS turbulence energy and the rate of energy dissipation on the mixture of solenoidal (divergence-free) and compressive (rotation-free) forcing modes. However, the scaling laws for the SGS turbulence energy are very similar for solenoidal and compressive forcing. This is an indication that the SGS model describes the dynamics in the inertial subrange, although the length scales close to the cutoff scale are affected by numerical dissipation. The differences in the mean values result from the substantial differences in the turbulent flow structure on larger scales \citep[see][]{FederKless08,FedKless09,FederRom10}.

The implementation of the SGS model into a fluid-dynamical code such as Enzo is rather straightforward. The SGS turbulence energy can be treated as a passive scalar with various source terms. To evaluate the mixed closure for the turbulence energy cascade, derivatives of the resolved velocity field are computed by means of centered differences. Care must be taken to ensure energy conservation, particularly, if the net change of the SGS turbulence energy in a certain grid cell exhausts the available energy over a time step or, vice-versa, if too much energy is drained from the resolved scales. However, these are exceptions that can be handled numerically. The effective increase in computing time is less than $10\,\%$. Moreover, the hydrodynamic coupling of the SGS model to the resolved flow introduces a reduction of the bottleneck effect in the turbulence energy spectra \citep{WoodPort06,Schm09}. 

In our large eddy simulations, we find correlations between resolved flow quantities such as the rate of strain or the denstrophy and SGS quantities, but the scatter is large. This is problematic if one intends to estimate unresolved flow properties on the basis of such correlations. In particular, this applies to the calculation of the dissipation rate from a constant numerical viscosity and the rate of strain, as proposed by \citet{PanPad09a}. We have shown that the assumption of a constant numerical dissipation coefficient is inconsistent with the equilibrium relation between the dissipation rate and the rate of strain on the grid scale in the limit of large Reynolds numbers. This relation, which follows from the SGS turbulence energy model with the linear eddy-viscosity closure (and also from the Smagorinsky model), is verified by our LES data for low turbulence intensity, while deviations become apparent for strong turbulent dissipation. This can be understood as a consequence of the non-linear term in the closure for the turbulence energy flux. Also the estimate of turbulent pressure effects on the basis of the rate of strain and the vorticity of the resolved flow that is put forward by \citet{ZhuFeng10} is incomplete because they do not distinguish between the contribution from the resolved flow and from the subgrid scales. The predictions from both approaches with regard to turbulence in the intergalactic medium are compared in ongoing work \citep{IapSchm10}.

From the probability distributions of the SGS turbulence Mach number, it follows that the turbulent pressure locally exceeds the thermal pressure even at moderate RMS Mach numbers and for relatively small grid scales. Since the grid scale in contemporary galactic disk simulations \citep[e.~g.,][]{AgerLake09,TaskTan09} is close to molecular cloud scales (a few pc), unresolved supersonic velocity fluctuations are quite likely and the turbulent pressure plays an important role. This has implications for the treatment of collapsing gas regions. The criterion for gravitational stability, which influences the grid resolution in adaptive mesh refinement simulations and controls the production of sink particles to capture the collapsing gas, is usually based on the thermal Jeans mass (among other criteria; see \citealt{FederrathBanerjeeClarkKlessen2010}). To account for the effects of turbulence below the grid scale, we suggest to include the turbulent pressure in the definition of the Jeans mass, in analogy to the magnetic pressure in self-gravitating MHD turbulence.

To model the fragmentation below the grid scale in more detail, a possible approach is based on the
assumption that the local star formation efficiency is regulated by turbulence on the grid scale. 
Then the star formation rate can be parameterized in terms of the turbulent Mach number
that is calculated from the SGS model \citep[see][]{KrumKee05,PadNord09}. On the other hand, star formation acts back on the SGS turbulence energy via stellar feedback. As suggested by \citet{Joung09}, a stellar feedback term can be included into the SGS turbulence energy equation. Statistically, we have
\[
	\frac{\dd}{\dd t} K_{\mathrm{sgs}}\sim
	\Sigma + \Sigma_{\star} - \rho\epsilon,
\]
where $\Sigma_{\star}\propto \rho e_{\star}/\tau_{\rm ff}$ accounts for the energy injection per unit mass, $e_{\star}$, by supernovae. The associated time scale is the free-fall time scale $\tau_{\rm ff}=[3\pi/(32G\rho)]^{1/2}$, which is the fundamental time scale of star formation. Neglecting the fluctuations of the gas density and setting the mean production rate $\Sigma \sim V^{2}/T$, where $V$ and $T$ are the typical velocity and the turn-over time scale, respectively, of the resolved turbulent flow, it follows that the turbulent pressure in equilibrium is of the order
\begin{equation}
	\label{eq:energy_eq_fb}
	P_{\rm sgs} \sim \rho\Delta^{2/3}
	\left[\frac{V^{2}}{T} + C_{\star}\frac{e_{\star}}{\tau_{\rm ff}}\right]^{2/3}.
\end{equation}
Kolmogorov scaling becomes manifest in the factor $\Delta^{2/3}$ in Eq.~(\ref{eq:energy_eq_fb}). For highly compressible turbulence, however, the scaling of the SGS turbulence energy deviates from the Kolmogorov law (see Sect.~\ref{sc:cutoff}). Depending on the ratios $V^{2}/e_{\star}$ and $T/\tau_{\rm ff}$, the production of SGS turbulence energy by the turbulent cascade or by supernovae dominates. The model of \citet{Joung09} follows in the limit \mbox{$(V^{2}/e_{\star})(\tau_{\rm ff}/T)\ll 1$}. In general, shear instabilities, gravitational instabilities, cooling instabilities, etc.\ above the grid scale feed energy to smaller scales.
For disk galaxies, a simple estimate can be obtained from the velocity dispersion of atomic hydrogen, which is about $10\,\mathrm{km\,s^{-1}}$. \citet{AgerLake09} show that gravitational instabilities grow on length scales ranging from $0.1$ to about $2\,\mathrm{kpc}$. Setting $V=10^{6}\,\mathrm{cm\,s^{-1}}$ and assuming $L>0.1\,\mathrm{kpc}\approx 3\times 10^{20}\,\mathrm{cm}$, the turbulence energy flux to smaller length scales is $V^{2}/T=V^{3}/L\lesssim 0.003\,\mathrm{erg\,g^{-1}\,s^{-1}}$. On the other hand, $e_{\star}\approx 4\times 10^{48}\,\eta\,\mathrm{erg}\,M_{\sun}^{-1}\approx 2\times 10^{15}\,\mathrm{erg\,g^{-1}}$ and $C_{\star}\approx 0.025$ imply $\Sigma_{\star}\sim 0.03\eta\, (n/1\,\mathrm{cm^{-3}})^{1/2}\,\mathrm{erg\,g^{-1}\,s^{-1}}$ \citep[see][]{Joung09}. Since the efficiency of the energy transfer from
supernova blast waves to the interstellar gas is roughly $\eta\simeq 0.1$ \citep{LowKless04}, the energy injection by stellar feedback is comparable to the turbulence energy flux for atomic hydrogen with density $n\sim 1\mathrm{cm^{-3}}$. The dependence of $\Sigma_{\star}$ on the gas
density implies a greater contribution form supernova feedback in the cold gas phase, but the above estimate does not account for the intermittency of turbulent velocity fluctuations, which entails large deviations from the mean. As a consequence, the assumption of \citet{Joung09} to consider the energy injection by supernova as the main source of the turbulent pressure is marginally fulfilled in cosmological simulations, in which the internal structure of galaxies is very poorly resolved. In galactic-scale simulations with high resolution, on the other hand, turbulence is not uniformly produced, and including the turbulence energy cascade improves the description of numerically unresolved processes. In particular, it will be useful to attempt a further generalization of the SGS model to multi-phase turbulence. A very simple ansatz has recently been presented by \citet{MurMona10}. A complete SGS model that accounts for a warm and a cold gas phase is presently under development \citep{BraunSchm10}.

Including stellar feedback and cooling into our SGS model will be of further utility for the numerical treatment of turbulence in the intergalactic medium \citep[see][]{SpringHern03}, where turbulence is produced by different processes \citep[see, for instance,][]{co99,ssh06,rkc08,IapiAdam08,IapSchm10}. \citet{OppDav09} show that a significant amount of the line broadening of \ion{O}{VI} in cosmological simulations stems from numerically unresolved turbulence. They apply a heuristic model in the postprocessing of the simulation data. Using an SGS model, on the other hand, the effect on the line broadening can be computed on the fly. Moreover, metals are mixed into the intergalactic medium by turbulence that is driven by galactic outflows. SGS turbulence enhances the turbulent mixing. Following an approach that is quite similar to the treatment of stellar feedback in galaxy simulations, our SGS model can be used in combination with phenomenological models for supernova-driven outflows \citep{Joung09,EvoFerra10}. In both cases, the use of adaptive mesh refinement is mandatory to achieve a sufficient dynamical range. Therefore, an essential objective for future work will be to incorporate the new closure for the highly compressible turbulent cascade into 
fluid mechanics with adaptively refined large eddy simulations (FEARLESS; \citealt{MaierIap09}).

\begin{acknowledgements}
We thank Jens Niemeyer and Ralf Klessen for valuable discussions. Computations described in this work were performed using the Enzo code developed by the Laboratory for Computational Astrophysics at the University of California in San Diego (http://lca.ucsd.edu). The computational resources were provided by the HLRBII project h0972 at the Leibniz Supercomputer Centre in Garching, Germany. 
CF acknowledges funding from the Landesstiftung Baden-W\"urrtemberg via their program International Collaboration II (grant P-LS-SPII/18), from the German Bundesministerium f\"ur Bildung und Forschung via the ASTRONET project STAR FORMAT (grant 05A09VHA), from the International Max Planck Research School for Astronomy and Cosmic Physics (\textsc{imprs-a}) and from the Heidelberg Graduate School of Fundamental Physics (\textsc{hgsfp}), which is funded by the Excellence Initiative of the Deutsche Forschungsgemeinschaft (\textsc{dfg}) \textsc{gsc} 129/1. Furthermore, CF received funding from the European Research Council under the European Community's Seventh Framework Programme (FP7/2007-2013 Grant Agreement no. 247060) for the research presented in this work.
\end{acknowledgements}

\bibliographystyle{aa}
\bibliography{ComprSGSModel}

\begin{thebibliography}{53}
\expandafter\ifx\csname natexlab\endcsname\relax\def\natexlab#1{#1}\fi

\bibitem[{{Agertz} {et~al.}(2009){Agertz}, {Lake}, {Teyssier}, {Moore},
  {Mayer}, \& {Romeo}}]{AgerLake09}
{Agertz}, O., {Lake}, G., {Teyssier}, R., {et~al.} 2009, \mnras, 392, 294

\bibitem[{{Agertz} {et~al.}(2010){Agertz}, {Teyssier}, \& {Moore}}]{AgerTey10}
{Agertz}, O., {Teyssier}, R., \& {Moore}, B. 2010, \mnras, 1527

\bibitem[{{Benzi} {et~al.}(2008){Benzi}, {Biferale}, {Fisher}, {Kadanoff},
  {Lamb}, \& {Toschi}}]{BenzBif08}
{Benzi}, R., {Biferale}, L., {Fisher}, R.~T., {et~al.} 2008, Physical Review
  Letters, 100, 234503

\bibitem[{{Bonazzola} {et~al.}(1987){Bonazzola}, {Heyvaerts}, {Falgarone},
  {Perault}, \& {Puget}}]{BonaHey87}
{Bonazzola}, S., {Heyvaerts}, J., {Falgarone}, E., {Perault}, M., \& {Puget},
  J.~L. 1987, \aap, 172, 293

\bibitem[{{Bonazzola} {et~al.}(1992){Bonazzola}, {Perault}, {Puget},
  {Heyvaerts}, {Falgarone}, \& {Panis}}]{BonaPer92}
{Bonazzola}, S., {Perault}, M., {Puget}, J.~L., {et~al.} 1992, J.\ Fluid Mech.,
  245, 1

\bibitem[{{Braun} \& {Schmidt}(2010)}]{BraunSchm10}
{Braun}, H. \& {Schmidt}, W. 2010, in preparation

\bibitem[{{Burkert} {et~al.}(2009){Burkert}, {Genzel}, {Bouche}, {Cresci}, \&
  {Khochfar et al.}}]{BurkGenz09}
{Burkert}, A., {Genzel}, R., {Bouche}, N., {Cresci}, G., \& {Khochfar et al.},
  S. 2009, E-print, arXiv:0907.4777

\bibitem[{{Cen} \& {Ostriker}(1999)}]{co99}
{Cen}, R. \& {Ostriker}, J.~P. 1999, \apj, 514, 1

\bibitem[{{Chandrasekhar}(1951)}]{Chandra51}
{Chandrasekhar}, S. 1951, Royal Society of London Proceedings Series A, 210, 26

\bibitem[{Colella \& Woodward(1984)}]{ColWood84}
Colella, P. \& Woodward, P.~R. 1984, J.\ Comp.\ Physics, 54, 174

\bibitem[{{Dobbs} {et~al.}(2008){Dobbs}, {Glover}, {Clark}, \&
  {Klessen}}]{DobbsGlov08}
{Dobbs}, C.~L., {Glover}, S.~C.~O., {Clark}, P.~C., \& {Klessen}, R.~S. 2008,
  \mnras, 389, 1097

\bibitem[{{Evoli} \& {Ferrara}(2010)}]{EvoFerra10}
{Evoli}, C. \& {Ferrara}, A. 2010, submitted to \mnras

\bibitem[{{Federrath} {et~al.}(2010{\natexlab{a}}){Federrath}, {Banerjee},
  {Clark}, \& {Klessen}}]{FederrathBanerjeeClarkKlessen2010}
{Federrath}, C., {Banerjee}, R., {Clark}, P.~C., \& {Klessen}, R.~S.
  2010{\natexlab{a}}, \apj, 713, 269

\bibitem[{{Federrath} {et~al.}(2008){Federrath}, {Klessen}, \&
  {Schmidt}}]{FederKless08}
{Federrath}, C., {Klessen}, R.~S., \& {Schmidt}, W. 2008, \apjl, 688, L79

\bibitem[{{Federrath} {et~al.}(2009){Federrath}, {Klessen}, \&
  {Schmidt}}]{FedKless09}
{Federrath}, C., {Klessen}, R.~S., \& {Schmidt}, W. 2009, \apj, 692, 364

\bibitem[{{Federrath} {et~al.}(2010{\natexlab{b}}){Federrath}, {Roman-Duval},
  {Klessen}, {Schmidt}, \& {Mac Low}}]{FederRom10}
{Federrath}, C., {Roman-Duval}, J., {Klessen}, R.~S., {Schmidt}, W., \& {Mac
  Low}, M. 2010{\natexlab{b}}, \aap, 512, A81+

\bibitem[{Frisch(1995)}]{Frisch}
Frisch, U. 1995, {Turbulence} (Cambridge University Press)

\bibitem[{{Fureby} {et~al.}(1997){Fureby}, {Tabor}, {Weller}, \&
  {Gosman}}]{FurTab97}
{Fureby}, C., {Tabor}, G., {Weller}, H.~G., \& {Gosman}, A.~D. 1997, Phys.\
  Fluids, 9, 3578

\bibitem[{Germano(1992)}]{Germano92}
Germano, M. 1992, J.\ Fluid Mech., 238, 325

\bibitem[{{Iapichino} {et~al.}(2008){Iapichino}, {Adamek}, {Schmidt}, \&
  {Niemeyer}}]{IapiAdam08}
{Iapichino}, L., {Adamek}, J., {Schmidt}, W., \& {Niemeyer}, J.~C. 2008,
  \mnras, 388, 1079

\bibitem[{{Iapichino} {et~al.}(2010){Iapichino}, {Schmidt}, \&
  {Niemeyer}}]{IapSchm10}
{Iapichino}, L., {Schmidt}, W., \& {Niemeyer}, J.~C. 2010, submitted to \mnras

\bibitem[{{Ishihara} {et~al.}(2009){Ishihara}, {Gotoh}, \& {Kaneda}}]{IshiGo10}
{Ishihara}, T., {Gotoh}, T., \& {Kaneda}, Y. 2009, Annual Review of Fluid
  Mechanics, 41, 165

\bibitem[{{Joung} \& {Mac Low}(2006)}]{Joung06}
{Joung}, M.~K.~R. \& {Mac Low}, M. 2006, \apj, 653, 1266

\bibitem[{{Joung} {et~al.}(2009){Joung}, {Mac Low}, \& {Bryan}}]{Joung09}
{Joung}, M.~R., {Mac Low}, M.-M., \& {Bryan}, G.~L. 2009, \apj, 704, 137

\bibitem[{{Kritsuk} {et~al.}(2007){Kritsuk}, {Norman}, {Padoan}, \&
  {Wagner}}]{KritNor07}
{Kritsuk}, A.~G., {Norman}, M.~L., {Padoan}, P., \& {Wagner}, R. 2007, ApJ,
  665, 416

\bibitem[{{Kritsuk} {et~al.}(2010){Kritsuk}, {Ustyugov}, {Norman}, \&
  {Padoan}}]{KritUst10}
{Kritsuk}, A.~G., {Ustyugov}, S.~D., {Norman}, M.~L., \& {Padoan}, P. 2010, in
  {Astronomical Society of the Pacifc Conference Series}, Vol. 429, {Numerical
  modeling of space plasma fows, ASTRONUM 2009}, ed. {N.~V.~Pogorelov,
  E.~Audit, and G.~P.~Zink}, 15--21

\bibitem[{{Krumholz} \& {McKee}(2005)}]{KrumKee05}
{Krumholz}, M.~R. \& {McKee}, C.~F. 2005, ApJ, 630, 250

\bibitem[{{Mac Low} \& {Klessen}(2004)}]{LowKless04}
{Mac Low}, M.-M. \& {Klessen}, R.~S. 2004, Reviews of Modern Physics, 76, 125

\bibitem[{{Maier} {et~al.}(2009){Maier}, {Iapichino}, {Schmidt}, \&
  {Niemeyer}}]{MaierIap09}
{Maier}, A., {Iapichino}, L., {Schmidt}, W., \& {Niemeyer}, J.~C. 2009, \apj,
  707, 40

\bibitem[{{Murante} {et~al.}(2010){Murante}, {Monaco}, {Giovalli}, {Borgani},
  \& {Diaferio}}]{MurMona10}
{Murante}, G., {Monaco}, P., {Giovalli}, M., {Borgani}, S., \& {Diaferio}, A.
  2010, \mnras, 405, 1491

\bibitem[{{Oppenheimer} \& {Dav{\'e}}(2009)}]{OppDav09}
{Oppenheimer}, B.~D. \& {Dav{\'e}}, R. 2009, \mnras, 395, 1875

\bibitem[{{Padoan} \& {Nordlund}(2009)}]{PadNord09}
{Padoan}, P. \& {Nordlund}, A. 2009, ArXiv e-print 0907.0248

\bibitem[{{Pan} {et~al.}(2009){Pan}, {Padoan}, \& {Kritsuk}}]{PanPad09a}
{Pan}, L., {Padoan}, P., \& {Kritsuk}, A.~G. 2009, Physical Review Letters,
  102, 034501

\bibitem[{{Price} \& {Federrath}(2010)}]{PriceFed10}
{Price}, D.~J. \& {Federrath}, C. 2010, \mnras, 406, 1659

\bibitem[{{R\"opke} \& {Schmidt}(2009)}]{RoepSchm09}
{R\"opke}, F.~K. \& {Schmidt}, W. 2009, in Lecture Notes in Physics, Vol. 756,
  Interdisciplinary Aspects of Turbulence, ed. {W.~Hillebrandt, \& F. Kupka},
  255--+

\bibitem[{Ryu {et~al.}(2008)Ryu, Kang, Cho, \& Das}]{rkc08}
Ryu, D., Kang, H., Cho, J., \& Das, S. 2008, Science, 320, 909

\bibitem[{Sagaut(2006)}]{Sagaut}
Sagaut, P. 2006, {Large eddy simulation for incompressible flows: An
  introduction} (Berlin: Springer-Verlag)

\bibitem[{Sarkar(1992)}]{Sarkar1992}
Sarkar, S. 1992, Physics of Fluids, 4, 2674

\bibitem[{{Scannapieco} \& {Br{\"u}ggen}(2008)}]{ScannBruegg08}
{Scannapieco}, E. \& {Br{\"u}ggen}, M. 2008, \apj, 686, 927

\bibitem[{{Schmidt}(2010)}]{Schm09}
{Schmidt}, W. 2010, in {Astronomical Society of the Pacifc Conference Series},
  Vol. 429, {Numerical modeling of space plasma fows, ASTRONUM 2009}, ed.
  {N.~V.~Pogorelov, E.~Audit, and G.~P.~Zink}, 45--50

\bibitem[{{Schmidt} {et~al.}(2009){Schmidt}, {Federrath}, {Hupp}, {Kern}, \&
  {Niemeyer}}]{SchmFeder09a}
{Schmidt}, W., {Federrath}, C., {Hupp}, M., {Kern}, S., \& {Niemeyer}, J.~C.
  2009, \aap, 494, 127

\bibitem[{{Schmidt} {et~al.}(2008){Schmidt}, {Federrath}, \&
  {Klessen}}]{SchmFeder08}
{Schmidt}, W., {Federrath}, C., \& {Klessen}, R. 2008, \prl, 101, 194505

\bibitem[{Schmidt {et~al.}(2006)Schmidt, Hillebrandt, \&
  Niemeyer}]{SchmHille06}
Schmidt, W., Hillebrandt, W., \& Niemeyer, J.~C. 2006, {Comp. Fluids.}, 35, 353

\bibitem[{{Schmidt} {et~al.}(2006{\natexlab{a}}){Schmidt}, {Niemeyer}, \&
  {Hillebrandt}}]{SchmNie06b}
{Schmidt}, W., {Niemeyer}, J.~C., \& {Hillebrandt}, W. 2006{\natexlab{a}},
  \aap, 450, 265

\bibitem[{{Schmidt} {et~al.}(2006{\natexlab{b}}){Schmidt}, {Niemeyer},
  {Hillebrandt}, \& {R{\"o}pke}}]{SchmNie06c}
{Schmidt}, W., {Niemeyer}, J.~C., {Hillebrandt}, W., \& {R{\"o}pke}, F.~K.
  2006{\natexlab{b}}, \aap, 450, 283

\bibitem[{{Schmidt} {et~al.}(2007){Schmidt}, {Niemeyer}, {Hupp}, {Federrath},
  \& {Maier}}]{SchmNie07}
{Schmidt}, W., {Niemeyer}, J.~C., {Hupp}, M., {Federrath}, C., \& {Maier}, A.
  2007, {A New Modelling Approach for Turbulent Astrophysical Flows}, {DECI
  project report, available from URL
  http://www.deisa.eu/science/deci/projects2005-2006/files/fearless\_report.pd%
f}

\bibitem[{{Springel} \& {Hernquist}(2003)}]{SpringHern03}
{Springel}, V. \& {Hernquist}, L. 2003, \mnras, 339, 289

\bibitem[{{Subramanian} {et~al.}(2006){Subramanian}, {Shukurov}, \&
  {Haugen}}]{ssh06}
{Subramanian}, K., {Shukurov}, A., \& {Haugen}, N.~E.~L. 2006, MNRAS, 366, 1437

\bibitem[{{Tasker} \& {Tan}(2009)}]{TaskTan09}
{Tasker}, E.~J. \& {Tan}, J.~C. 2009, \apj, 700, 358

\bibitem[{{Truelove} {et~al.}(1997){Truelove}, {Klein}, {McKee}, {Holliman},
  {Howell}, \& {Greenough}}]{TrueloveEtAl1997}
{Truelove}, J.~K., {Klein}, R.~I., {McKee}, C.~F., {et~al.} 1997, \apjl, 489,
  L179

\bibitem[{{Woodward} {et~al.}(2006){Woodward}, {Porter}, {Anderson}, {Fuchs},
  \& {Herwig}}]{WoodPort06}
{Woodward}, P.~R., {Porter}, D.~H., {Anderson}, S., {Fuchs}, T., \& {Herwig},
  F. 2006, Journal of Physics Conference Series, 46, 370

\bibitem[{{Woodward} {et~al.}(2001){Woodward}, {Porter}, {Sytine}, {Anderson},
  {Mirin}, {Curtis}, {Cohen}, {Dannevik}, {Dimits}, {Eliason}, {Winkler}, \&
  {Hodson}}]{PortWood01}
{Woodward}, P.~R., {Porter}, D.~H., {Sytine}, I., {et~al.} 2001, in
  {Computational Fluid Dynamics, Proceedings of the Fourth UNAM Supercomputing
  Conference Mexico City, June 2000}, ed. E.~{Ramos}, G.~{Cisneros},
  R.~{Fernandez-Flores}, \& A.~{Santillan-Gonzalez} ({World Scientific}), 3--15

\bibitem[{{Zhu} {et~al.}(2010){Zhu}, {Feng}, \& {Fang}}]{ZhuFeng10}
{Zhu}, W., {Feng}, L., \& {Fang}, L. 2010, \apj, 712, 1

\end{thebibliography}

\end{document}